\documentclass[a4paper,10pt]{article}
\usepackage[english]{babel}
\usepackage[utf8]{inputenc}
\usepackage[T1]{fontenc}
\usepackage{amsmath}
\usepackage{amssymb}
\usepackage{latexsym}
\usepackage[table]{xcolor}
\usepackage{pdflscape}
\usepackage[dvips]{graphicx}
\usepackage{epsfig}

\usepackage[affil-it]{authblk}
\usepackage{sidecap}
\usepackage{subcaption}
\usepackage{graphicx}
\usepackage{comment}
\usepackage{float}
\usepackage{csquotes}
\usepackage[style=numeric-comp,natbib=true,uniquename=init,giveninits,bibstyle=trad-plain,sorting=none, backend=biber, isbn=false, 
doi=false, url=false]{biblatex}
\begin{document}

\title{Intestinal villi and crypts density robustly maximizes nutrient absorption}

\author[a,b]{Martin Garic}
\author[c]{Rohan Vernekar}
\author[d]{Dacil Y Martin}
\author[d]{Stéphane Tanguy}
\author[c]{Clément de Loubens}
\author[a,1]{Claude Loverdo}

\affil[a]{Sorbonne Université, CNRS, Laboratoire Jean Perrin, LJP, F-75005 Paris, France}
\affil[b]{current address: LPENS, Département de Physique, École Normale Supérieure, PSL University, CNRS, Paris, France}
\affil[c]{Univ. Grenoble Alpes, CNRS, Grenoble INP, LRP, 38000 Grenoble, France}
\affil[d]{Univ. Grenoble Alpes, CNRS, UMR 5525, VetAgro Sup, Grenoble INP, TIMC, 38000 Grenoble, France}

\date{}

\maketitle
\footnotetext[1]{claude.loverdo@sorbonne-universite.fr}

\begin{abstract}
The villi and crypts of the gastrointestinal tract increase the effective surface area of the intestinal mucosa, potentially enhancing nutrient absorption. It is commonly assumed that this is their primary function, and that a higher villi density necessarily leads to improved absorption. However, when villi are packed too closely together, diffusion can be hindered, potentially offsetting this benefit. In this work, we investigate quantitatively the relationship between the density of these structures and the overall efficiency of absorption. In three different simplified geometries, approximating leaf-like villi, finger-like villi, and colonic crypts, we calculate analytically the concentration profile and the absorption flux, assuming that there is only diffusion between these structures while the lumen is well mixed. When plotting the absorption flux per unit of gut length as a function of the structures' density, we observe that there is a density maximizing absorption. We study numerically this optimum. We find that it is robust to the nutrient absorption properties: a geometry optimal for one nutrient is close to optimum for another nutrient. Physiological data from various animal species fall within this predicted optimal range, consistent with the hypothesis that structure density is shaped by selection for efficient nutrient uptake.
\end{abstract}

The gastrointestinal tract's main function is absorbing nutrients, critical for sustaining metabolic activities and overall health~\cite{reed_review_2009}. Propelled by the stomach into the small intestine, the chyme is transported by various patterns of muscle contractions~\cite{henderson_mechanism_1928,lentle_review_2015}. These patterns optimize enzymatic digestion, breaking down large nutrients into smaller ones that can be absorbed by the intestinal wall.

The inner surface of the small intestine is characterized by the presence of small projections (of the $0.1$--$1$~mm size range) known as villi. These structures are lined with epithelial cells through which nutrient molecules are transported towards nearby blood and lymph capillaries. Across different species, positions along the digestive tract and stages of development, villi exhibit morphological variations, ranging from elongated finger-shaped structures to leaflet formations~\cite{eltantaway_effect_2015}. Their size and density also vary from species to species~\cite{hilton_morphology_1902, walton_blueprint_2018}. They can be more or less long and more or less thin. After traversing the small intestine, the chyme continues its journey through the large intestine. This organ lacks villi but possesses crypts~\cite{n_i_contribution_1984}. Similar to villi, these crypts vary in depth, width, and density between species~\cite{prieto2019differences,helander1973morphological}. These structures may play a role in absorption across species.

A widely accepted idea is that villi increase the absorptive surface area, thereby enhancing nutrient uptake~\cite{walton_generation_2016,helander_surface_2014,vertzoni2019impact}. However, there exists a trade-off: increasing the density of villi (resp. crypts) increases the surface area for absorption; but this reduces the intervillous space (resp. crypt width), which decreases the amount of nutrients that enter these spaces, reducing absorption~\cite{strocchi_role_1993}. The question of the optimal distance between microstructures, or in other words their density, has not been studied quantitatively yet. If better absorption has been selected by evolution, structure geometry will be close to optimal. This study aims to test this evolutionary hypothesis and provide a theoretical framework for understanding the observed variations in villi and crypt structures across species~\cite{shyer2013villification}.

One potential obstacle to the selection of optimal geometries is that there are several nutrients, which properties differ. Nutrients vary in size so they have different diffusion coefficients. Their absorption properties on the epithelium are also different~\cite{fiori2022pharmacokinetics, lennernaas1998human}. One question is whether there is a single optimal geometry that is robust to nutrient properties, or if there are trade-offs. 


In this article, through analytical calculations, we examine the interplay between these structures' configurations and absorption efficiency. We explore a limiting case in which we assume the lumen is well-mixed, while there is only diffusive transport between villi and within crypts. We first present a simplified model for absorption between leaf-like villi, in which symmetries allow us to solve this system in 2D, similar to~\cite{winne1989effect, oliver1998surface}. We then study the case of finger-like villi using a 3D model. Thirdly, we conduct a similar study for colonic crypts. We investigate whether the optimal geometry is robust, i.e, similar for various nutrients with different diffusion and absorption properties. Finally, we compare our results for the optimal density with physiological data for various species.

\section{General Model}

In the present study, we assume that the villi are not moving relative to each other. In particular, this means that there is no fluid flow caused by their movement in the intervillous space. We also assume that convective mixing in the lumen does not extend to the intervilli space. While the extent of flow near the epithelial surface remains debated, several studies suggest the presence of an unstirred water layer thicker than the villi height~\cite{wilson_intestinal_1974, smithson_intestinal_1981, lentle_flow_2011,fagerholm_experimental_1995}, which would strongly limit convective transport between villi. In the colon, a dense mucus layer several hundred microns thick covers the epithelium~\cite{swidsinski_comparative_2007}, resulting in a predominantly diffusive transport. Although the mucus gel forms a dense fiber network~\cite{johansson2011composition,strugala2003colonic}, its mesh size is on the order of 100--200~nm~\cite{cone2009barrier}, much larger than small nutrients such as glucose or amino acids. These molecules therefore diffuse through the mucus mesh largely unhindered by steric interactions with the mucin network, so that diffusion within the mucus can be treated as diffusion in water. In this study we assume the lumen is well-mixed, while there is only diffusive transport between villi and within crypts.

Due to this localized diffusion regime, we model the system using the steady-state diffusion equation (also called the Laplace equation) where $C$ is the concentration profile close to these structures. The steady-state assumption is justified if diffusion time-scales are much shorter than transit time-scales in the gut. The longest villi are of order $h\sim 1$~mm so the diffusion time-scale for glucose (diffusion coefficient $D=6\times10^{-10}$ m$^2$/s~\cite{bashkatov2003glucose}) between villi is $T\approx h^2/(2D)\approx 833$ seconds which is less than 15 minutes, smaller than the residence time in the gut~\cite{kimura2002gastrointestinal}. Secondly, we assume that the length scales for longitudinal concentration gradients along the digestive tract are much larger than the typical size of the villi or crypt structures studied, and that mixing is efficient in the lumen. Thus the luminal concentration is fixed at a concentration $c_0$. This gives the following equations between the structures:
\begin{equation}
    \nabla ^2 C=0
    \label{eq:Laplace}
\end{equation}
\begin{equation}
    C(y=h)=c_0
    \label{eq:BC1}
\end{equation}
Lastly, we consider that the epithelial walls absorb imperfectly. For this, we use the Robin boundary condition:
\begin{equation}
    D(\vec n \cdot \nabla C) + kC=0
    \label{eq:BC2}
\end{equation}
Where $D$ is the molecular diffusion coefficient, $\vec n$ is the normal vector to the wall pointing outward, and $k$ is a positive absorptivity constant. When $k$ approaches zero, the condition tends towards reflection of nutrient particles by the structures' walls, whereas when $k$ tends towards infinity, it approaches perfect absorption.

\begin{figure}
    \centering
  \begin{minipage}{0.49 \linewidth}
    \includegraphics[width=\linewidth]{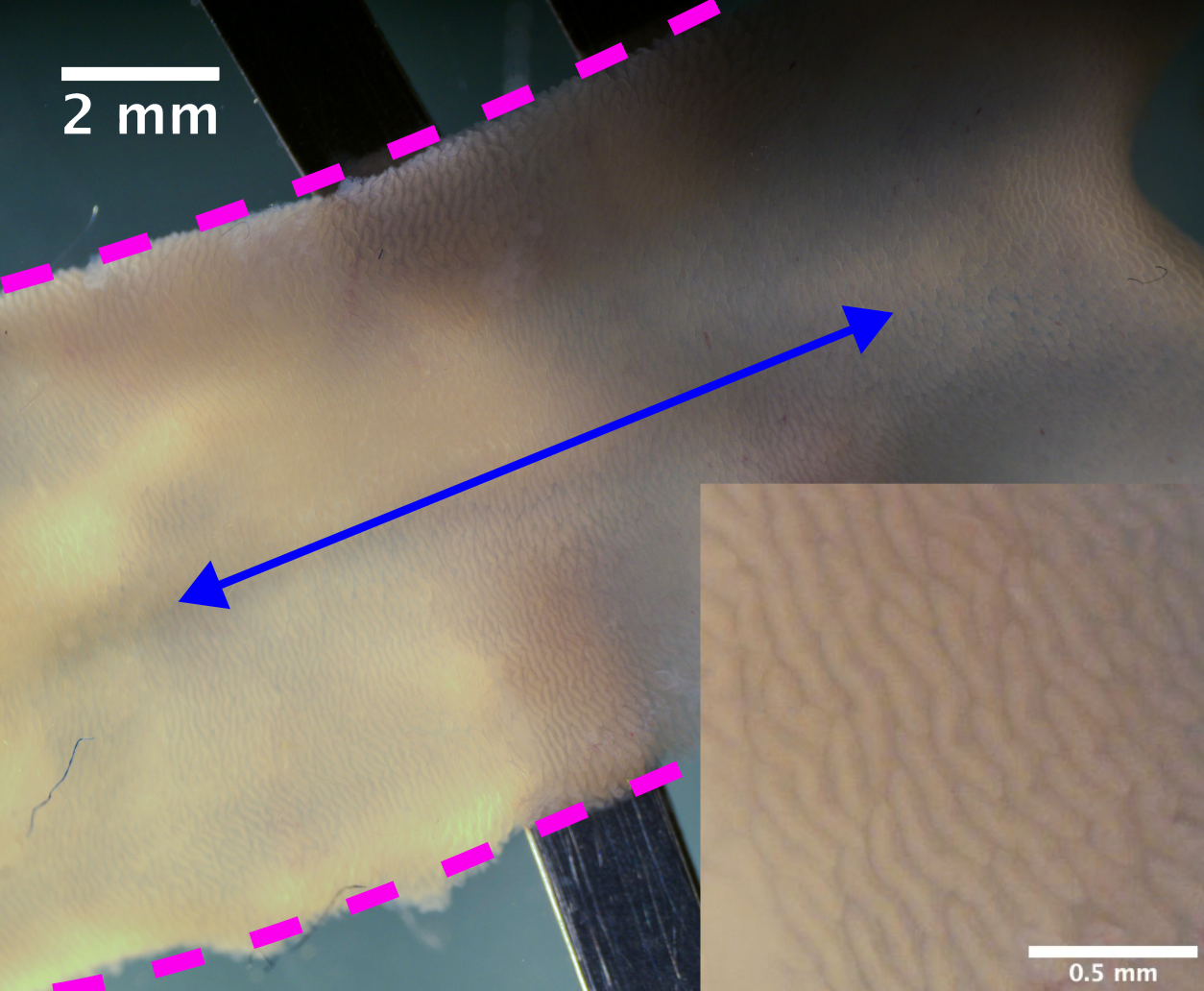}

  a. Inner surface of a rat ileum

  \end{minipage}
    \begin{minipage}{0.49 \linewidth}
    \includegraphics[width=\linewidth]{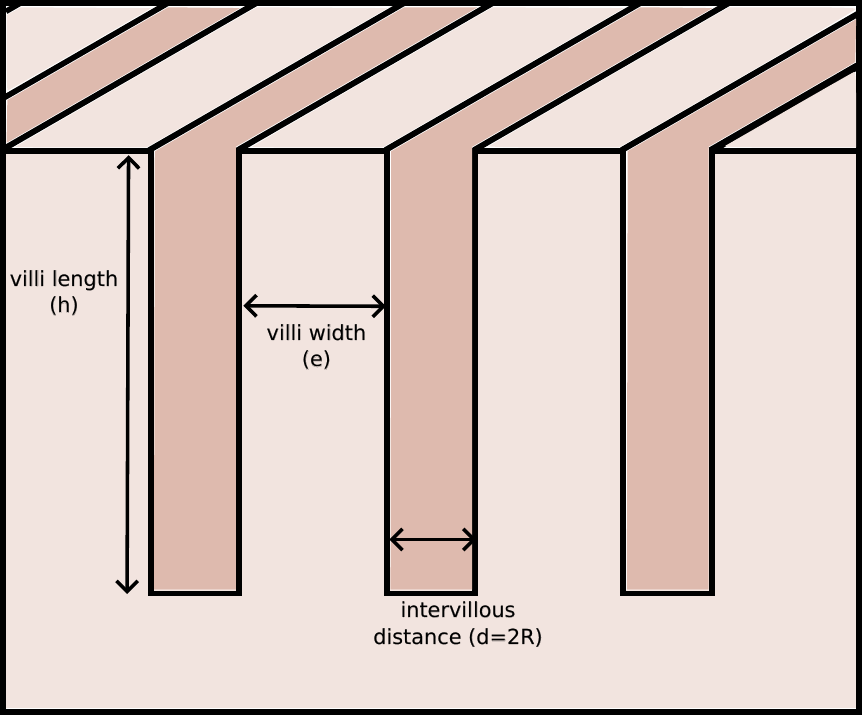}

  b. Sketch of leaf-like intestinal villi.

  \end{minipage}

  ~

  ~

   \begin{minipage}{0.49 \linewidth}
    \includegraphics[width=\linewidth]{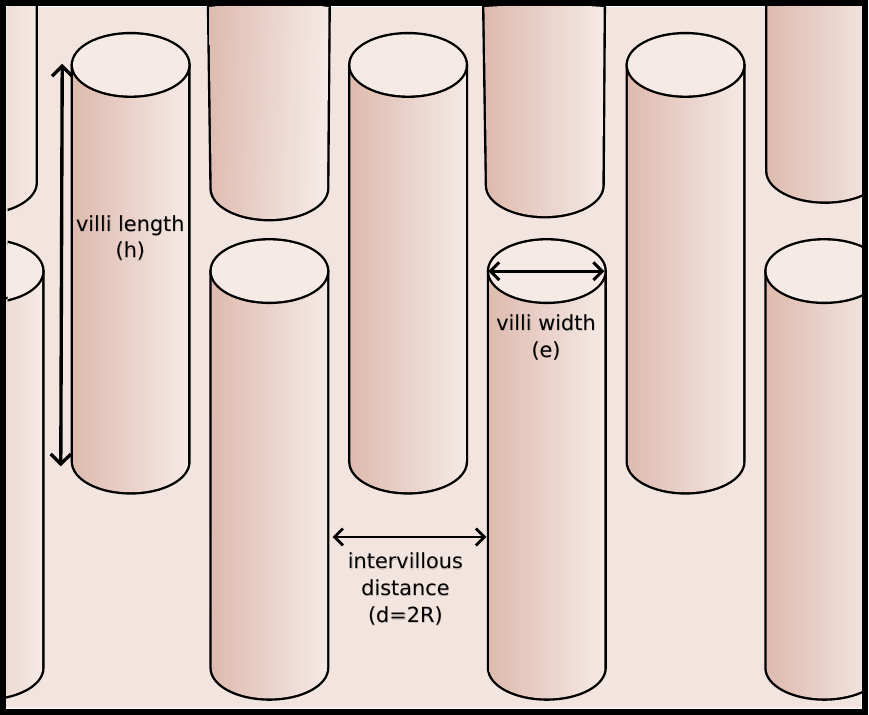}

  c. Sketch of finger-like intestinal villi.

  \end{minipage}
    \begin{minipage}{0.49 \linewidth}
    \includegraphics[width=\linewidth]{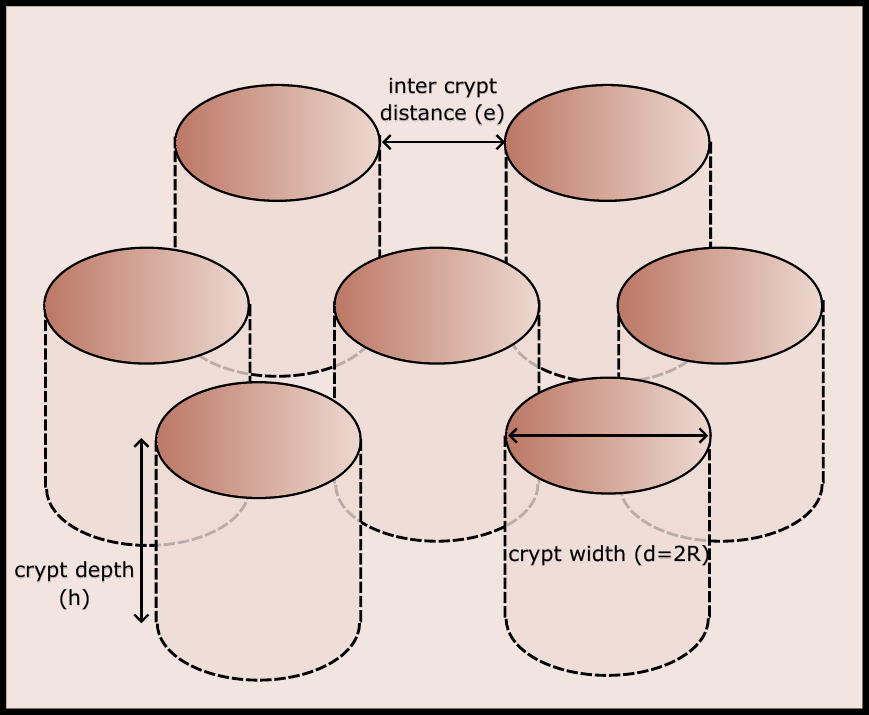}

  d. Sketch of colonic crypts.
  \end{minipage}

\caption{Structures on the inner surface of the gastrointestinal tract. (a) Experimental image of the inner surface of a rat ileum. The blue arrow represents the longitudinal axis, and the magenta dashed lines represent where the tube has been cut open for imaging. A zoom shows that in this sample, the villi are leaf-like. (b,c,d) Sketches of the three studied geometries in this paper. The parameters are not to scale. $h$ corresponds to the length or depth of the villi and crypts. $e$ corresponds to the width of the villi or the distance between crypts. $2R$ corresponds to the intervillous distance or the crypt diameter. The space characterized by $h$ and $R$ is the space where particles diffuse freely.}\label{fig:mainfig}
\end{figure}

The archetypal geometries of the structures studied in this article are presented in figure~\ref{fig:mainfig}. We calculate nutrient absorption for the three geometries: leaf-like villi, finger-like villi, and colonic crypts. Equation~\ref{eq:Laplace} is solved for each of the three geometries using equations~\ref{eq:BC1} and~\ref{eq:BC2} as boundary conditions. We find concentration profiles, quantify absorption for each geometry and discuss which geometries maximize absorption.

\section{Analytical calculations - Absorption for each geometry}

\subsection{Absorption flux for leaf-like villi}

We first look at absorption between two leaf-like villi, which is the simplest of the three cases. The symmetry allows us to reduce the system from 3 dimensions to 2 dimensions. As illustrated in figure~\ref{fig:mainfig}b, we consider villi as stationary rectangular projections, and the space between two villi as analogous to a well. In our system, $h$ is the villi height, $R$ is half of the intervillous distance, and $e$ is the villi width.

To non-dimensionalize the equations, we introduce dimensionless units: $\tilde x = x/R$, $\tilde y = y/h$, $\xi = R/h$, $\Theta = D/(kh)$, and $c = C/c_0$. Here, $\xi$ is the rescaled intervillous width, and $\Theta$ is the ratio of the absorption time scale to the diffusion time scale for a given nutrient, which we call the rescaled inverse surface absorptivity. Consequently, the equations~\ref{eq:Laplace}--\ref{eq:BC2} become:
\begin{equation}
    \frac{\partial^2c}{\partial \tilde x^2}+\xi^2\frac{\partial^2c}{\partial \tilde y^2}=0
    \label{eq:Laplace_leaf_main}
\end{equation}
\begin{equation}
    c(\tilde x, 1)=1
    \label{eq:BC1_leaf_main}
\end{equation}
\begin{equation}
    \Theta \vec n \left(\frac{1}{\xi}\frac{\partial c}{\partial \tilde x}\vec e_x + \frac{\partial c}{\partial \tilde y}\vec e_y\right)+c=0
    \label{eq:BC2_leaf_main}
\end{equation}
We solve equation~\ref{eq:Laplace_leaf_main} with the boundary conditions of equations~\ref{eq:BC1_leaf_main} and~\ref{eq:BC2_leaf_main} using separation of variables. We obtain the following solution (details in SI Section A):

\begin{equation}
    c(\tilde x, \tilde y) = \sum_{n=1}^\infty \frac{4 \sin(\mu_n)\cos(\mu_n \tilde x)\left(\Theta\frac{\mu_n}{\xi}\cosh\left(\frac{\mu_n}{\xi}\tilde y\right)+\sinh\left(\frac{\mu_n}{\xi}\tilde y\right)\right)}{\left(2\mu_n+\sin(2\mu_n)\right)\left(\Theta\frac{\mu_n}{\xi}\cosh\left(\frac{\mu_n}{\xi}\right)+\sinh\left(\frac{\mu_n}{\xi}\right)\right)}
\label{eq:concentration_leaf_main}
\end{equation}

where $\mu_n$ is the $\text{n}^\text{th}$ positive root of the equation:

\begin{equation}
    \Theta \frac{\mu_n}{\xi}= \frac{1}{\tan(\mu_n)}
    \label{eq:mu_leaf_main}
\end{equation}

\begin{figure}[h!]
\centering
\includegraphics[width=1\linewidth]{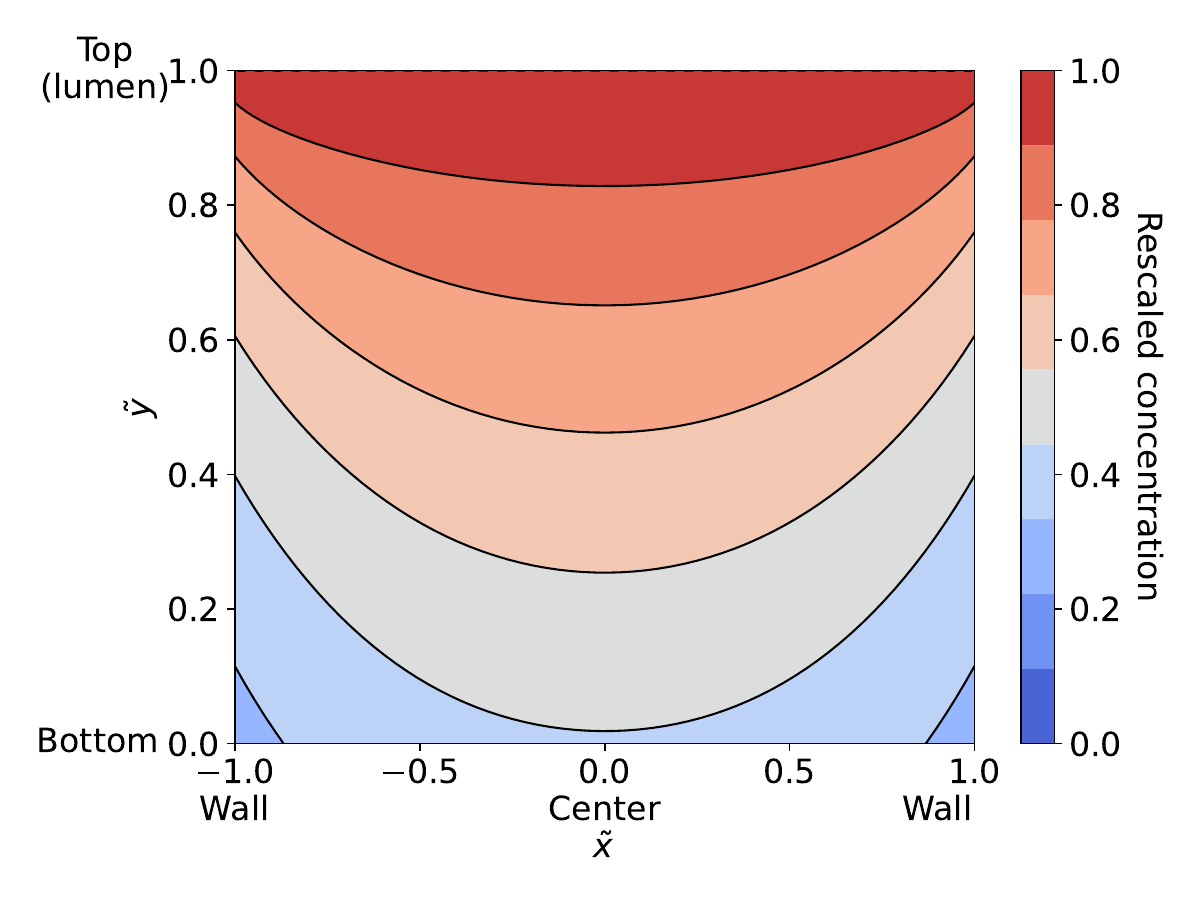}
\caption{Contour plot of the rescaled nutrient concentration within an intervillous space for the leaf-like villi geometry (equation~\ref{eq:concentration_leaf_main}). In this figure, the rescaled intervillous width $\xi=1$ and the rescaled inverse surface absorptivity $\Theta = 1$. The colors correspond to the rescaled concentration within this space (the lumen concentration is set to 1). Absorption between villi results in lower concentrations the lower the position. Left-right symmetry is a result of the symmetry of the system.}\label{fig:concentration_leaf_main}
\end{figure}

Figure~\ref{fig:concentration_leaf_main} illustrates the concentration distribution between two villi for $\Theta=1$ and $\xi=1$. Consistent with expectations, the concentration diminishes with depth between the villi, due to particle absorption by the walls. A reduction in $\Theta$ intensifies absorption, resulting in a steeper concentration gradient (SI figure S1). We recover the results from article~\cite{winne1989effect}, where similar assumptions and this geometry were used. In addition, we study the absorbing flux, taking into account absorption at the tip of the villi, and discuss the optimal spacing between them.

The rate of absorption is given by the concentration flux through the surface S of the villi walls:
\begin{equation}
    J = -D \iint_S \vec n \cdot \nabla C(x,y) d\vec S
    \label{eq:flux}
\end{equation}
This measure gives information on the efficiency of the intervillous space in absorbing particles diffusing in this space. A higher flux means more absorption. The expression of the dimensionless flux for one intervillous space is:
\begin{equation}
    \tilde J=\sum_{n=1}^\infty\frac{8\sin^2(\mu_n)\left(\Theta\frac{\mu_n}{\xi}\sinh\left(\frac{\mu_n}{\xi}\right)+\cosh\left(\frac{\mu_n}{\xi}\right)\right)}{\left(2\mu_n+\sin(2\mu_n)\right)\left(\Theta\frac{\mu_n}{\xi}\cosh\left(\frac{\mu_n}{\xi}\right)+\sinh\left(\frac{\mu_n}{\xi}\right)\right)}
\label{eq:flux_leaf_main}
\end{equation}
With $\tilde J=J/(Dc_0)$ the dimensionless flux. This calculation is done for one intervillous space. What matters is nutrient absorption per unit of intestinal tube length. To be able to compare different villi densities, we introduce the notion of flux per unit length (flux density) where the unit length is one period ($2R+e$) where $e$ is the villi width. In this case, the flux per unit length is:
\begin{equation}
    j=\frac{J+kc_0e}{2R+e}
\end{equation}
Where $kc_0e$ is the flux on the villi tip. In the absence of villi (flat gut walls), the flux per unit length would be $j=kc_0$ under the assumption of perfect mixing in the lumen. We use this information to non-dimensionalize this flux. Using $\tilde e=e/h$, the dimensionless flux per unit length becomes:
\begin{equation}\label{jleaf}
    \tilde j = \frac{\Theta\tilde J + \tilde e}{2\xi+\tilde e},
\end{equation}
where $\tilde j = j/(kc_0)$.

\subsection{Absorption flux for finger-like villi}
We now look at absorption between finger-like villi. We model the villi as evenly spaced cylinders of height $h$ on a triangular lattice (see figure~\ref{fig:hexagon_to_circle}) such that the radius of a villus is $e/2$ and the shortest distance between the sides of two villi is $2R$ (consistent with the 2D model as seen in figure~\ref{fig:mainfig}c). There are repeating hexagons centered on the villi at the nodes of the triangular lattice of side length $R+e/2$. To solve this problem using cylindrical coordinates, we approximate the hexagons as disks, keeping the surface of the disk equal to the surface of the hexagon (see figure S3). The dimensionless units for the system are: $\tilde r=r/R$, $\tilde z=z/h$, $\xi=R/h$, $\tilde e=e/h$, $\Theta=\frac{D}{kh}$, and $c=C/c_0$. Using cylindrical coordinates, the system is reduced to an axisymmetric one. Solving equations~\ref{eq:Laplace}--\ref{eq:BC2} by separation of variables gives an infinite series solution in Bessel functions $J_n$ and $Y_n$ of the first and second kind, from which equation~\ref{eq:flux} gives the following dimensionless flux $\tilde J=J/(Dc_0h)$ within an intervillous space (details in SI Section B):

\begin{multline}
    \tilde J = \sum_{n=1}^\infty \tilde e \pi\left(\int_{\frac{\tilde e}{2\xi}}^{1+\frac{\tilde e}{2\xi}}rf_n\left(r\right)dr\right)\left(-\frac{Y_1\left(\mu_n\left(1+\frac{\tilde e}{2\xi}\right)\right)}{J_1\left(\mu_n\left(1+\frac{\tilde e}{2\xi}\right)\right)}J_1\left(\mu_n \frac{\tilde e}{2\xi}\right)+Y_1\left(\mu_n \frac{\tilde e}{2\xi}\right)\right)\\
    \times\frac{\left(\Theta\frac{\mu_n}{\xi} \sinh\left(\frac{\mu_n}{\xi}\right) + \cosh\left(\frac{\mu_n}{\xi}\right)\right)}{\left(\int_{\frac{\tilde e}{2\xi}}^{1+\frac{\tilde e}{2\xi}}rf_n\left(r\right)^2dr\right)\left(\Theta \frac{\mu_n}{\xi}\cosh\left(\frac{\mu_n}{\xi}\right) + \sinh\left(\frac{\mu_n}{\xi}\right)\right)}
    \label{eq:flux_finger_main}
\end{multline}

with $\mu_n$ as the solution to the following transcendental equation:

\begin{equation}
    \Theta \frac{\mu_n}{\xi} = -\frac{Y_1\left(\mu_n\left(1+\frac{\tilde e}{2\xi}\right)\right)J_0\left(\mu_n\frac{\tilde e}{2\xi}\right)-J_1\left(\mu_n\left(1+\frac{\tilde e}{2\xi}\right)\right)Y_0\left(\mu_n\frac{\tilde e}{2\xi}\right)}{Y_1\left(\mu_n\left(1+\frac{\tilde e}{2\xi}\right)\right)J_1\left(\mu_n\frac{\tilde e}{2\xi}\right)-J_1\left(\mu_n\left(1+\frac{\tilde e}{2\xi}\right)\right)Y_1\left(\mu_n\frac{\tilde e}{2\xi}\right)}
\label{eq:mu_finger_main}
\end{equation}

and $f_n(\tilde r)$ is:

\begin{equation}
    f_n(\tilde r)=-\frac{Y_1\left(\mu_n\left(1+\frac{\tilde e}{2\xi}\right)\right)}{J_1\left(\mu_n\left(1+\frac{\tilde e}{2\xi}\right)\right)}J_0\left(\mu_n \tilde r\right)+Y_0\left(\mu_n \tilde r\right)
\end{equation}
where $\tilde J=J/(Dc_0h)$. In the previous equations, $J_n(x)$ and $Y_n(x)$ are Bessel functions of the first and second type, of order $n$. To account for villi density, we introduce the flux per unit surface (in comparison to the flux per unit length in the previous geometry), $j$, where the unit surface is a disk of radius $R+e/2$. In this case, the new dimensionless flux is:

\begin{equation}
    \tilde j = \frac{\Theta \tilde J+\pi(\frac{\tilde e}{2})^2}{\pi(\xi+\frac{\tilde e}{2})^2},
\end{equation}
where $\tilde j=j/(kc_0)$. We obtain a flux per unit surface that uses the same parameters as the flux per unit length for the case of leaf-like villi. This will allow us, in a later section, to compare the results between geometries.

\subsection{Absorption flux for colonic crypts}
Crypts can be modeled as 3D wells of radius $R$ and height $h$. For symmetry simplifications, we assume that the crypts are positioned on a triangular lattice such that the shortest distance between the sides of two crypts is $e$. Solving the system using equations~\ref{eq:Laplace}--\ref{eq:BC2} in cylindrical coordinates, by the same method as for finger-like villi, gives via equation~\ref{eq:flux} the dimensionless flux $\tilde J$ in a crypt as an infinite series in Bessel functions $J_n$ of the first kind. With hexagons centered on the crypts of side length $R+e/2$ (figure~\ref{fig:mainfig}d), the dimensionless flux per unit surface is (details in SI Section C):
\begin{equation}
    \tilde J = \sum_{n=1}^\infty \frac{4\pi J_1^2\left(\mu_n\right)\left(\Theta \frac{\mu_n}{\xi} \sinh\left(\frac{\mu_n}{\xi}\right)+\cosh\left(\frac{\mu_n}{\xi}\right)\right)}{\mu_n\left(J_0^2\left(\mu_n\right)+J_1^2\left(\mu_n\right)\right)\left(\Theta \frac{\mu_n}{\xi} \cosh\left(\frac{\mu_n}{\xi}\right)+\sinh\left(\frac{\mu_n}{\xi}\right)\right)}
    \label{eq:flux_crypt_main}
\end{equation}

with $\mu_n$ as the solution to the following transcendental equation:
\begin{equation}
    \Theta\frac{\mu}{\xi} = \frac{J_0\left(\mu\right)}{J_1\left(\mu\right)}
    \label{eq:mu_crypt_main}
\end{equation}

The dimensionless flux per unit surface is then:
\begin{equation}
    \tilde j = \frac{\Theta \tilde J + \frac{3\sqrt{3}}{2}(\xi+\tilde e/2)^2-\pi\xi^2}{\frac{3\sqrt{3}}{2}(\xi+\tilde e/2)^2}
\end{equation}

\section{Results - Optimal geometries}

\subsection{The rescaled intervilli distance $\xi$ is the parameter to optimize}

In the previous three subsections, absorption was quantified in the intestines for three archetypal geometries: leaf-like villi, finger-like villi, and crypts. To compare the three models effectively, similar dimensionless parameters were introduced: $\xi$ is the intervillous distance (resp. crypt radius) rescaled by its depth $h$, $\tilde e$ is the villi width (resp. inter-crypt distance) once again rescaled by depth $h$, and $\Theta$ is the rescaled inverse surface absorptivity $\Theta=D/kh$ with $D$ the nutrient diffusion coefficient and $k$ the surface absorption coefficient.

Our theoretical analysis shows that for a single intervillous space, it is beneficial to have an infinitely wide gap to maximize absorption. However, a smaller gap allows for a higher density of structures per unit of gut length. The density maximizing the absorption flux per unit of gut length comes from this trade-off.

The flux depends on several parameters. The smaller the value of the inverse rescaled absorptivity $\Theta$ is, the larger the absorption. For a given nutrient it will be fixed by the nutrient's diffusion coefficient and the epithelial cell absorption properties. The greater the value of $h$ is, the greater the absorption is. We assume $h$ is set by other constraints such as resources needed to build longer villi or diffusion times to the bottom of these structures. The smaller the villi width $e$, the better the absorption, as there can be more intervillous space per unit length. But $e$ needs to have some minimal value to accommodate blood and lymph circulation~\cite{ma2007immmunohistochemical}. The remaining parameter that can be varied to maximize absorption is the distance between villi, R, which corresponds to $\xi$ in dimensionless units.

\begin{figure}[h]
\centering
\includegraphics[width=1\linewidth]{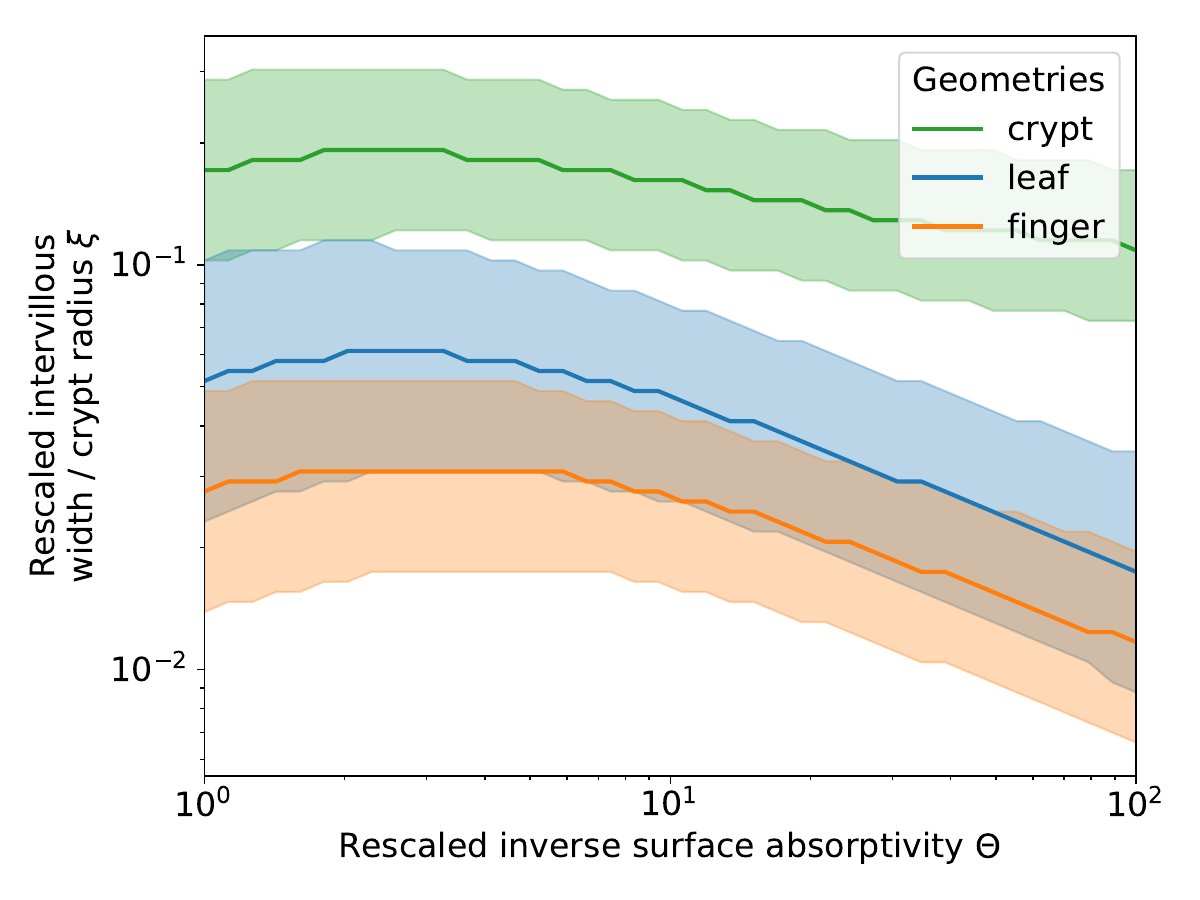}
\caption{Optimal rescaled intervillous width / crypt radius $\xi^*$ (line) and the range achieving $95\%$ of the maximal flux (shaded band), as a function of $\Theta$, at fixed $\tilde e=0.2$ (a typical value of the physiological data in Table~\ref{tab:physiological_data}), for the three geometries studied.}\label{fig:xi_vs_theta}
\end{figure}

Now that we have calculated the absorption in the three different geometries, we can maximize it as a function of $\xi$ by solving numerically the following equation:

\begin{equation}
    \frac{\partial j(\xi,\tilde e,\Theta)}{\partial \xi}=0
    \label{eq:optimize_xi}
\end{equation}

\subsection{Physiological range of $\Theta$}

The only parameter that is directly dependent on the nutrient characteristics is $\Theta$. The physiological range of $\Theta$ can be estimated by bracketing it between a well-absorbed and a poorly-absorbed molecule. Glucose and mannitol have similar diffusion coefficients, $D \approx 6\times10^{-10} m^2/s$~\cite{bashkatov2003glucose}. Absorption can be very different: glucose is taken up efficiently by the epithelium, whereas mannitol is largely unabsorbed and is commonly used as a low-permeability reference~\cite{fiori2022pharmacokinetics}. For glucose, an overestimate of the permeability is $k \approx 10^{-5} m/s$~\cite{lennernaas1998human} (overestimated because it is renormalized to the projected rather than the true epithelial surface); for mannitol on a flat surface, $k \approx 10^{-9} m/s$~\cite{watson2005interferon}. With villi heights h in the range 300–1000 $\mu$m, and since a higher $k$ gives a lower $\Theta$, these values yield an underestimate of $\Theta \approx 0.2$ for glucose and $\Theta \approx 400$ for mannitol. These two values are edge cases rather than typical ones. Most relevant nutrients are expected to share the same order of magnitude for $D$ and to have absorption constants between these two extremes, thus a realistic physiological range for $\Theta$ is between $1$ and $100$.

\subsection{Shape of the optimum - The optimum is robust to different nutrient properties}

\begin{figure}[h]
\includegraphics[width=1\linewidth]{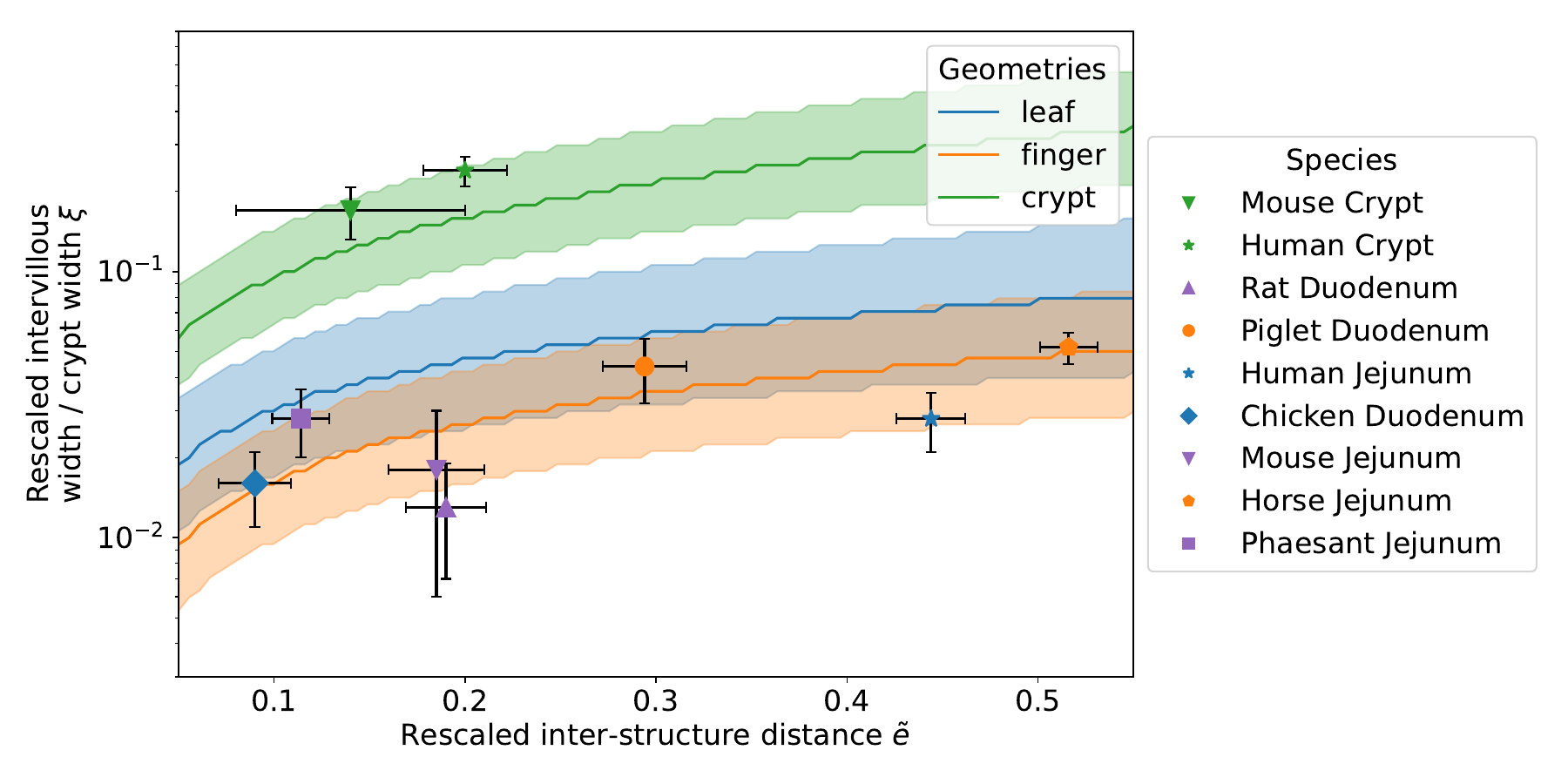}
\caption{Plot of the rescaled intervillous width maximizing absorption per unit of gut length as a function of the rescaled inter-structure distance for the three geometries introduced earlier. The shaded regions correspond to $95\%$ of the maximal absorption for each geometry. Data points show physiological data from various species (Table~\ref{tab:physiological_data}). The green markers correspond to crypts, the blue markers correspond to leaf-like villi, the orange markers correspond to finger-like villi and the purple markers to a mix of finger- and leaf-like villi. Here $\Theta=10$, in the middle of the physiological range for $\Theta$, but similar results are obtained for other physiological values of $\Theta$ in SI figure S10.}\label{fig:CompareAllAnalVC_Log}
\end{figure}

We now study numerically the optimal geometry for absorption. Equation~\ref{eq:optimize_xi} shows that $\xi$ maximizing absorption depends on $\Theta$ and $\tilde e$. The optimal $\xi$ and the range of $\xi$ where the absorption is within $95\%$ of the maximum is shown as a function of $\Theta$ in figure~\ref{fig:xi_vs_theta} and as a function of $\tilde e$ in figure~\ref{fig:CompareAllAnalVC_Log} for all three geometries. The two villi geometries (leaf-like and finger-like) have similar optima. The optimal $\xi$ for crypts is almost an order of magnitude higher.

As $\tilde e$ increases, the optimal value of $\xi$ increases (fig~\ref{fig:CompareAllAnalVC_Log}). Plotted on the log-log axes (SI figure~\ref{fig:CompareAllAnalVC_LogLog}), the $\xi$ seems to depend with close to $\sqrt{\xi}$, but the range of physiological $\tilde e$ is too small for a robust fit. The dependence is similar for the three geometries.

In most cases, a value of $\xi$ that is optimal for a given $\Theta$ is close to optimum (absorption within $95\%$ of the maximum) for another $\Theta$ in the physiological range (fig~\ref{fig:xi_vs_theta}). In other terms, a geometry optimal to absorb a given nutrient will be close to optimal for other nutrients. The optimum is robust, there is no significant trade-offs between nutrients for adaptation.

\begin{table*}[t!]
\centering
\begin{tabular}{ |l r||c|c|c|c|c|c|c|  }
 \hline
 \multicolumn{9}{|c|}{\textbf{Villi Measurements}} \\ 
 \hline
 Species & Section & Length ($h$) & Width ($e$) & Intervillous  & $\tilde e$ & $\xi$ & Villi & Source \\
 &  &  & &  space ($2R$) &  &  &shape & \\
 \hline
 Human    & J & $360\pm10$ & $160\pm5$ & $20\pm5$ & $0.44$ & $0.028$ & Leaf  & \cite{marsh_study_1969} \\
 Rat      & D & $390\pm34$ & $70\pm6$  & $10\pm5$ & $0.18$ & $0.013$ & Unclear  & \cite{seyyedin_histomorphometric_2017},\cite{fujimiya_serotonin-containing_1991} \\
 Mouse    & J & $405\pm8$  & $75\pm10$ & $15\pm10$& $0.18$ & $0.019$ & Both  & \cite{abbas_internal_1989} \\
 Piglet   & D & $340\pm15$ & $100\pm6$ & $30\pm8$ & $0.29$ & $0.044$ & Finger  & \cite{skrzypek_mechanisms_2018} \\
 Pheasant & J &$460\pm10$  & $52\pm7$  & $26\pm7$ & $0.11$ & $0.028$ & Unclear  & \cite{goodarzi_histology_2021} \\
 Horse& J&  &  &  & $0.52$ & $0.052$ & Finger  & \cite{roberts_mucosa_1974} \\
 Chicken  & D &$1000\pm200$& $90\pm6$& $32\pm8$  & $0.09$ & $0.016$ & Leaf  & \cite{samanya_histological_2002} \\
 \hline
 \multicolumn{9}{|c|}{\textbf{Crypt Measurements}} \\ 
 \hline
 Species && Depth ($h$) & Inter-crypt  &
 Crypt &
 $\tilde e$ &
 $\xi$
 &
 &
 Source\\
 &&  &  distance ($e$) &
 diameter ($2r$) &
 &
 &
 &
 \\
 \hline
 Human    && $500\pm50$ & $100\pm5$ & $120\pm10$ & $0.2$ & $0.24$ & & \cite{qi2008automated} \\
 \hline
 Mouse    && $350\pm50$ & $50\pm20$ & $60\pm10$ & $0.14$ & $0.17$ & &  \cite{farkas2015cryosectioning} \\
 \hline
\end{tabular}
\caption{Physiological data on villi and crypt length, width, and spacing. Every length is in $\mu m$. The second column corresponds to the section of the small intestine for villi, where D denotes Duodenum while J denotes Jejunum. The standard deviation around the mean lengths is also shown in the data. The shapes of the villi are shown in the penultimate column. When the shape is unknown due to the images only showing a 2D section of the villi, the description shows "Unclear". When the villi are finger- and leaf-like on a single image, the description shows "Both". Data is plotted in figure~\ref{fig:CompareAllAnalVC_Log} to show the connection with theoretical predictions.}
\label{tab:physiological_data}
\end{table*}

\subsection{Comparison with physiological data}

The objective of this study is to determine if the villi and crypt geometry is adapted to maximize absorption. We calculate the structure density that maximizes absorption through different types of structures. $\xi$ denotes the rescaled intervillous width or crypt radius, while $\tilde e$ represents the rescaled villi width or the distance between two crypts.

To confront our predictions to observed villi and crypt geometries, we surveyed the literature on gut physiology (see SI section D), collected measurements for various species (Table~\ref{tab:physiological_data}), and compared with our model predictions (figure~\ref{fig:CompareAllAnalVC_Log}). Most physiological geometries fall within the predicted region for $95\%$ of the maximal absorption. We captured the similarity between leaf-like and finger-like villi, as well as the fact that the crypt width is much larger than the distance between villi.

\section{Conclusion}

In this study, we examine the role of micro-structures, intestinal villi and colonic crypts, in maximizing intestinal absorption. While it is usually assumed that any increase in surface increases absorption~\cite{walton_generation_2016,helander_surface_2014,vertzoni2019impact}, we actually quantify a crucial trade-off: higher villi and crypt densities increase total absorption surface area, but also decrease nutrient penetration in a given intervillous space or in a given crypt~\cite{strocchi_role_1993}. It is actually an intermediate density that maximizes absorption.

Our main assumption is that nutrients are well mixed in the lumen, but that their transport within the intervillous spaces and colonic crypts occurs solely by diffusion. In three simplified geometries, representing leaf-like villi, finger-like villi, and crypts, we solve analytically the diffusion equation with semi-absorbing boundary conditions. We obtain the absorption flux per unit of gut length.

Our hypothesis is that villi/crypt geometry may have been selected to maximize absorption. The absorption flux per unit of gut length depends on the dimensionless parameters $\Theta$, $\tilde e$, and $\xi$. $\Theta$ is the rescaled inverse surface absorptivity and a decrease of $\Theta$ means that the epithelial surface would better absorb nutrients, so $\Theta$ should be as low as possible. $\tilde e$ and $\xi$ are two lengths rescaled by $h$ the villi height / crypt depth. A larger $h$ enables more absorption, but is likely limited by physiological factors that we did not explore. The diffusion time to the bottom of a structure may limit the usefulness of a deeper $h$. $\tilde e$ is $e/h$, with $e$ the thickness of the structure (villi width or distance between crypts). Smaller $\tilde e$ enables denser structures with the same nutrient penetration, so it increases absorption. But it has to have some finite value to be able to perform its physiological functions, like blood and lymph transport. Thus the parameter that may be optimized is $\xi=R/h$, with $R$ half the intervillous distance, or the crypt radius.

We found numerically that for a given $(\tilde e,\Theta)$, there is one value of $\xi$ maximizing absorption, which is strongly increased relative to a smooth surface. Crucially, the optimal value of $\xi$ depends only weakly on $\Theta$, thus the optimal geometry is similar for different nutrients. These results are robust to different nutrient properties. Indeed, since the optimal spacing depends only weakly on $\Theta$ over the physiological range, a geometry optimized for one nutrient remains close to optimal for others, despite their differing diffusion and absorption properties.

We measured physiological geometries for various animals using experimental images in the scientific literature. Most of the data falls within the region predicted to achieve 95\% of maximal absorption, and the model correctly predicts a larger $\xi$ for crypts than for villi. This is consistent with structure density having been tuned for absorption efficiency, though the breadth of the 95\% region means the data cannot by themselves rule out other constraints shaping the geometry.

For the physiological data, we relied on measurements from published experimental studies. A limitation of this approach is that tissue fixation for imaging can alter the native geometry of the samples, potentially affecting dimensional accuracy. Additionally, during digestion, physiological factors such as luminal pressure may cause dynamic changes in tissue shape that are not captured in fixed samples. However, experimental studies suggest that longitudinal distension is relatively limited under physiological conditions~\cite{holzheimer1989influence}. Despite these modifications, the flatness of the optimum suggests that such variations are unlikely to move measurements significantly away from the optimal range. 

We also used simplified geometries for the calculations. For instance, we assumed finger like villi to be regularly spaced, and  approximated the geometry as cylinders to solve the equations (see fig~\ref{fig:hexagon_to_circle}). This approximation should introduce very little error, except at high villi density. Furthermore, villi have often an intermediate shape between leaf-like and finger like, we calculated absorption for the two extreme geometries and found that the optima are relatively close.

The main hypothesis of this study is that the lumen is well-mixed and that there is only diffusion in the intervillous space and crypts. We also assumed that the structures are static. For the crypts, it is likely a good approximation. It has been proposed that villi, because they are on a surface which contracts and thus move them, enhance mixing and flow near the intestinal wall~\cite{lentle2013mucosal,lim2015flow,Rohan2025,strocchi_role_1993,wang2017three, puthumana2022steady, wang2010multiscale}. We also neglected secretion and absorption, in particular of water, which could result in a net flow that would modify transport. However, the extent of flow between these structures remains debated, especially because of the mucus presence. Several studies suggest the presence of an unstirred water layer thicker than the villi height~\cite{lentle_flow_2011,wilson_intestinal_1974,smithson_intestinal_1981, fagerholm_experimental_1995}. The gradient of concentration induced by luminal mixing may extend to the intervillous space for large rescaled intervillous widths / crypt widths $\xi$.
What we find is that the optimum is for relatively densily packed structures (optimal $\xi$ <0.07 for villi, <0.2 for crypts), thus in the regime where our approximation should be valid.

To conclude, while some studies, such as~\cite{strocchi_role_1993}, have suggested that overly dense structures may hinder diffusion, to our knowledge no quantitative model like ours had previously calculated the optimal geometry. These results support the hypothesis that villi and crypt density has evolved to maximize nutrient uptake~\cite{shyer2013villification} rather than being arbitrary. There exist diseases that modify villi and crypt geometries and that are associated with weaker nutrient absorption~\cite{shalimar2013mechanism}. There could be other biological parameters affecting absorption, but our results show that geometry itself plays an important role.

~

\textit{LRP is part of the LabEx Tec21 (ANR-11-LABX-0030) and of the PolyNat Carnot Institute (ANR-11-CARN-007-01). The authors thank Agence Nationale de la Recherche for its financial support of the project TransportGut, ANR-21-CE45-0015.
}

\medskip
\noindent\textbf{Ethics.} The imaged intestinal tissue (Appendix~\ref{sec:experimental_images}) was obtained from animals euthanized under a previously approved, unrelated study; no animals were used or euthanized specifically for this work. All animal care and procedures were conducted in strict compliance with the guidelines outlined by the European Union for the use of experimental animals (Directive 2010/63/EU).

\smallskip
\noindent\textbf{Author contributions.} All authors: conceptualization. M.G.: investigation. D.Y.M. and S.T.: microscopy images. C.d.L. and C.L.: supervision and funding acquisition. M.G.: writing original draft. All authors: writing, review and editing.

\smallskip
\noindent\textbf{Competing interests.} We declare we have no competing interests.

\newpage

\begin{center}
 \LARGE Supplementary material \normalsize
\end{center}

\renewcommand{\thefigure}{S\arabic{figure}}

\setcounter{figure}{0}

\setcounter{equation}{0}

\appendix

\section{Leaf-like geometry calculations}

Here we solve the Laplace equation in the leaf-like villi case. Starting from equation 1 of the main text and with the rescaled $\tilde x=x/R$, $\tilde y=y/h,c=C/c_0$, $\xi=R/h$ and $\Theta=D/(kh)$, the differential equation to solve is:

\begin{equation}
    \frac{\partial^2 c}{\partial \tilde x^2} + \xi^2\frac{\partial^2 c}{\partial \tilde y^2} = 0
    \label{eq:Laplace_leaf}
\end{equation}

\begin{equation}
    c(\tilde x, 1)=1
    \label{eq:BC1_leaf}
\end{equation}

\begin{equation}
    \Theta \left( \xi^{-1} \frac{\partial c}{\partial \tilde x} \vec e_x + \frac{\partial c}{\partial \tilde y}\vec e_y\right)\vec n + c = 0, (\tilde x, \tilde y) \in S,
    \label{eq:BC2_leaf}
\end{equation}

where $S$ is either $\tilde x=-1$, $\tilde x=1$ or $\tilde y=0$. The Laplace equation is solved using separation of variables. Writing the concentration field $c$ as the product of two functions gives:

\begin{equation}
    c(\tilde x,\tilde y) = f(\tilde x)g(\tilde y)
  \label{eq:separation_leaf}
\end{equation}

Substituting equation~\ref{eq:separation_leaf} into equation~\ref{eq:Laplace_leaf} yields:

\begin{equation}
    f''(\tilde x)g(\tilde y) + \xi^2f(\tilde x)g''(\tilde y) = 0
\end{equation}

\begin{equation}
    \frac{f''(\tilde x)}{f(\tilde x)} = -\xi^2\frac{g''(\tilde y)}{g(\tilde y)} = -\lambda
    \label{eq:lambda_leaf}
\end{equation}

Where $f'(\tilde x)$ is the derivative of $f(\tilde x)$ with respect to $\tilde x$, $g'(\tilde y)$ is the derivative of $g(\tilde y)$ with respect to $\tilde y$ and $\lambda$ is chosen positive; otherwise, non-physical solutions are found. We define $\lambda = \mu^2$.

Let us first solve for $f(\tilde x)$:

\begin{equation}
    f''(\tilde x)+\mu^2 f(\tilde x) = 0
\end{equation}

\begin{equation}
    f(\tilde x) = C \cos(\mu \tilde x)+D \sin(\mu \tilde x)
\end{equation}

The boundary equation from equation~\ref{eq:BC2_leaf} is then used to determine $C$ and $D$. The left-right symmetry implies that $D=0$. For an arbitrary non-zero $C$, the result is:

\begin{equation}
    \tan(\mu) = \frac{\xi}{\Theta \mu}
\label{eq:mu_leaf}
\end{equation}

There exists an infinite number of solutions $\mu_n$ to equation~\ref{eq:mu_leaf}, where $n$ is a positive integer. So, $f_n(\tilde x)$ is defined as:

\begin{equation}
    f_n(\tilde x) = \cos(\mu_n\tilde x)
\end{equation}

Where $\mu_n$ is the nth solution to equation~\ref{eq:mu_leaf} and $g_n$ such that $c_n(\tilde x,\tilde y)=f_n(\tilde x)g_n(\tilde y)$. Each solution $f_n(\tilde x)$ forms an orthogonal basis because integration over the space $[-1,1]$ of $\cos(\mu_n x)\cos(\mu_m x)=0$ if $n\neq m$. The full solution to the differential equation is a superposition of these solutions that form an orthogonal basis: $c(\tilde x,\tilde y)=\sum_{n=1}^\infty c_n(\tilde x, \tilde y)$. We now solve for $g_n(\tilde y)$ using equation~\ref{eq:lambda_leaf}:

\begin{equation}
    \xi^2g_n''(\tilde y)-\lambda g_n(\tilde y) = 0
\end{equation}

\begin{equation}
  g_n(\tilde y) = B_n \cosh\left(\frac{\mu_n}{\xi} \tilde y\right)+A_n \sinh\left(\frac{\mu_n}{\xi} \tilde y\right)
\end{equation}

The boundary conditions are then applied to find $A_n$ and $B_n$:

\begin{equation}
    -\frac{\Theta}{\xi} \mu_n A_n + B_n = 0
\end{equation}

\begin{equation}
    B_n = \Theta \frac{\mu_n}{\xi} A_n
\end{equation}

Assuming that $A_n\neq 0$, the resulting function $g_n(\tilde y)$ is:

\begin{equation}
    g_n(\tilde y) =\left( \Theta\frac{\mu_n}{\xi} \cosh(\frac{\mu_n}{\xi} \tilde y) + \sinh(\frac{\mu_n}{\xi} \tilde y)\right)A_n
\end{equation}

Then,

\begin{equation}
    c_n(\tilde x,\tilde y) = A_n \cos(\mu_n x)(\Theta\frac{\mu_n}{\xi} \cosh(\frac{\mu_n}{\xi} \tilde y) + \sinh(\frac{\mu_n}{\xi} \tilde y))
\end{equation}

And:

\begin{equation}
    c(\tilde x,\tilde y) = \sum_{n=1}^{n=\infty}A_n \cos(\mu_n x)\left(\Theta\frac{\mu_n}{\xi} \cosh\left(\frac{\mu_n}{\xi} \tilde y\right) + \sinh\left(\frac{\mu_n}{\xi} \tilde y\right)\right)
\end{equation}

To determine $A_n$, the final boundary condition $c(\tilde x,1)=1$ (equation~\ref{eq:BC1_leaf}) is applied:

\begin{equation}
    c(\tilde x,1) = 1=\sum_{n=1}^{n=\infty} A_n \cos(\mu_n x)\left(\Theta\frac{\mu_n}{\xi} \cosh\left(\frac{\mu_n}{\xi}\right) + \sinh\left(\frac{\mu_n}{\xi}\right)\right)
\end{equation}

Substituting $\tilde A_n = A_n(\Theta \frac{\mu_n}{\xi}\cosh(\frac{\mu_n}{\xi}) + \sinh(\frac{\mu_n}{\xi}))$ gives following equation:

\begin{equation}
    1 = \sum_{n=1}^{n=\infty}\tilde A_n \cos(\mu_n x)
\end{equation}

We recognize a generalized Fourier expansion, where $\cos(\mu_n x)$ constitutes an orthogonal set on the domain, the coefficient $\tilde A_n$ is expressed as:

\begin{equation}
    \tilde A_n = \frac{\int_{-1}^1\cos(\mu_n x)dx}{\int_{-1}^1\cos^2(\mu_n x)dx}=\frac{4\sin(\mu_n)}{2\mu_n+\sin(2\mu_n)}
\end{equation}

So the full expression of $c(\tilde x,\tilde y)$ is:

\begin{equation}
    c(\tilde x, \tilde y) = \sum_{n=1}^\infty \frac{4 \sin(\mu_n)\cos(\mu_n \tilde x)\left(\Theta\frac{\mu_n}{\xi}\cosh\left(\frac{\mu_n}{\xi}\tilde y\right)+\sinh\left(\frac{\mu_n}{\xi}\tilde y\right)\right)}{\left(2\mu_n+\sin(2\mu_n)\right)\left(\Theta\frac{\mu_n}{\xi}\cosh\left(\frac{\mu_n}{\xi}\right)+\sinh\left(\frac{\mu_n}{\xi}\right)\right)}
\label{eq:concentration_leaf}
\end{equation}

where $\mu_n$ is again the nth solution to equation~\ref{eq:mu_leaf}.

\begin{figure}[h]
\begin{subfigure}{.32\textwidth}
  \centering
  \includegraphics[width=.9\linewidth]{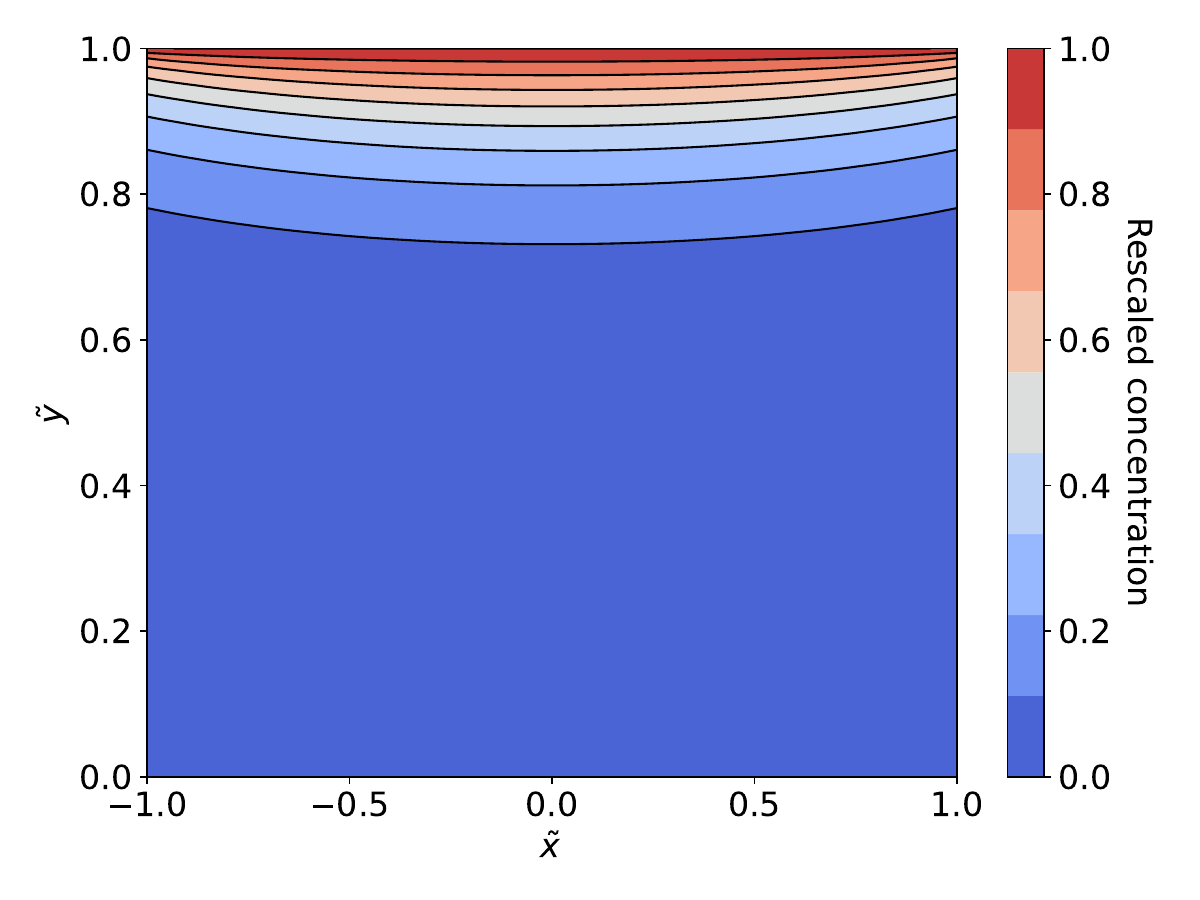}
  \caption{$\xi=0.1$ and $\Theta=0.1$}
\end{subfigure}%
\hfill
\begin{subfigure}{.32\textwidth}
  \centering
  \includegraphics[width=.9\linewidth]{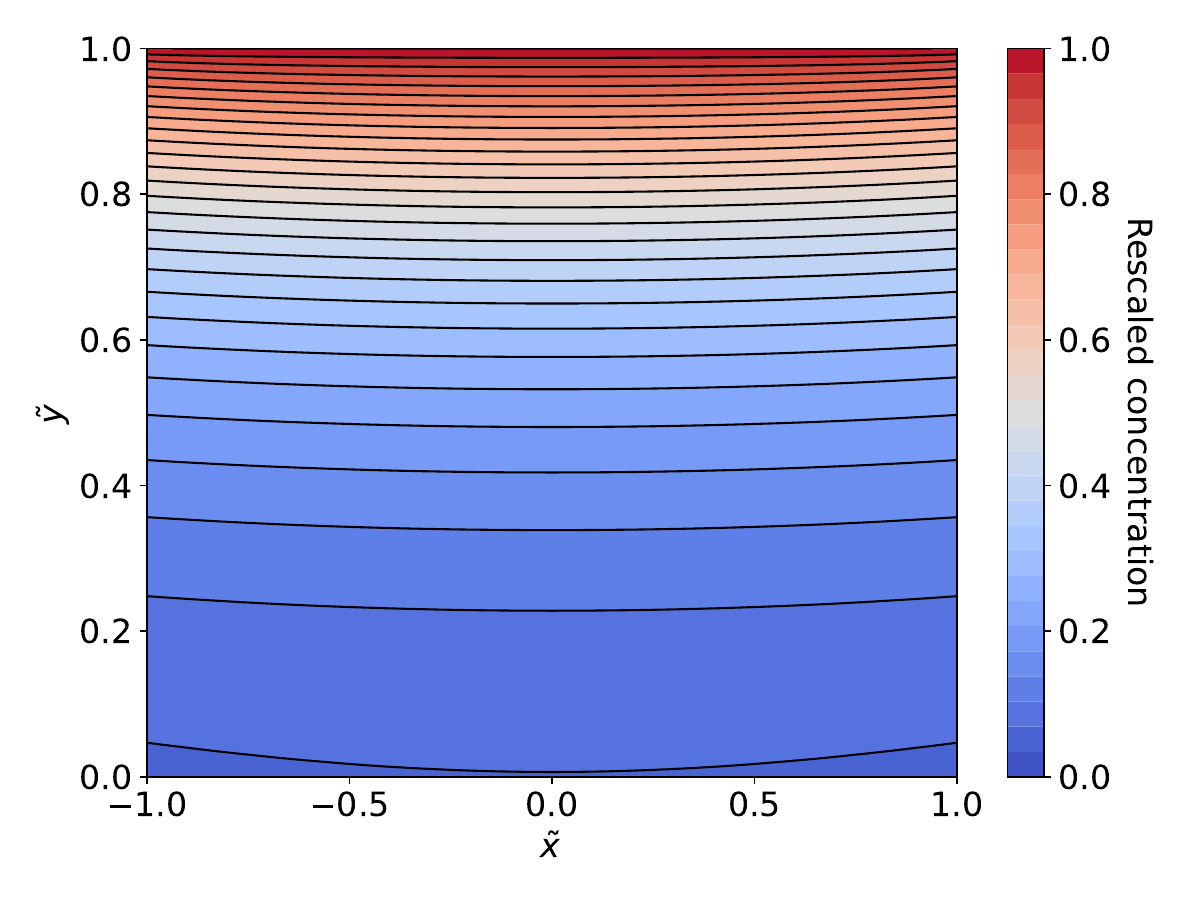}
  \caption{$\xi=0.1$ and $\Theta=1$}
\end{subfigure}
\hfill
\begin{subfigure}{.32\textwidth}
  \centering
  \includegraphics[width=.9\linewidth]{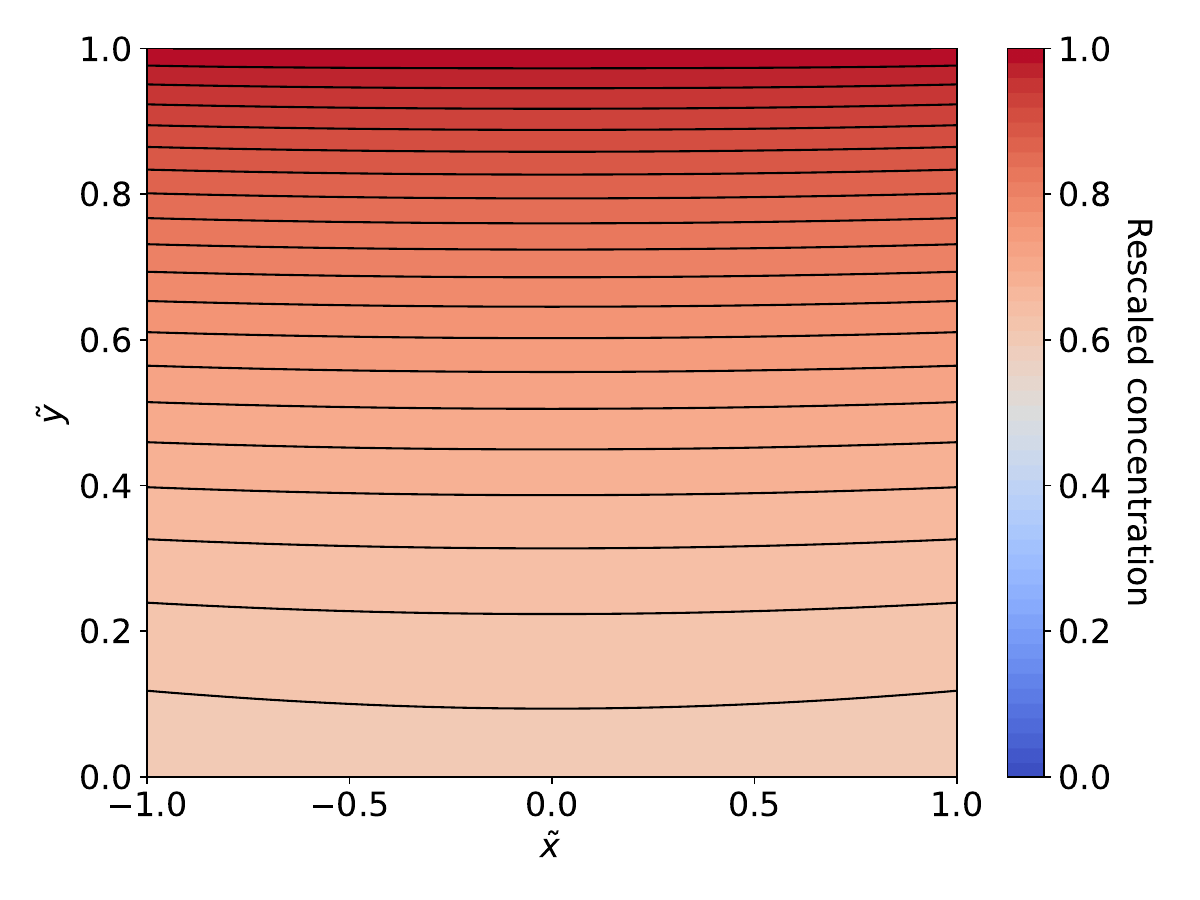}
  \caption{$\xi=0.1$ and $\Theta=10$}
\end{subfigure}
\begin{subfigure}{.32\textwidth}
  \centering
  \includegraphics[width=.9\linewidth]{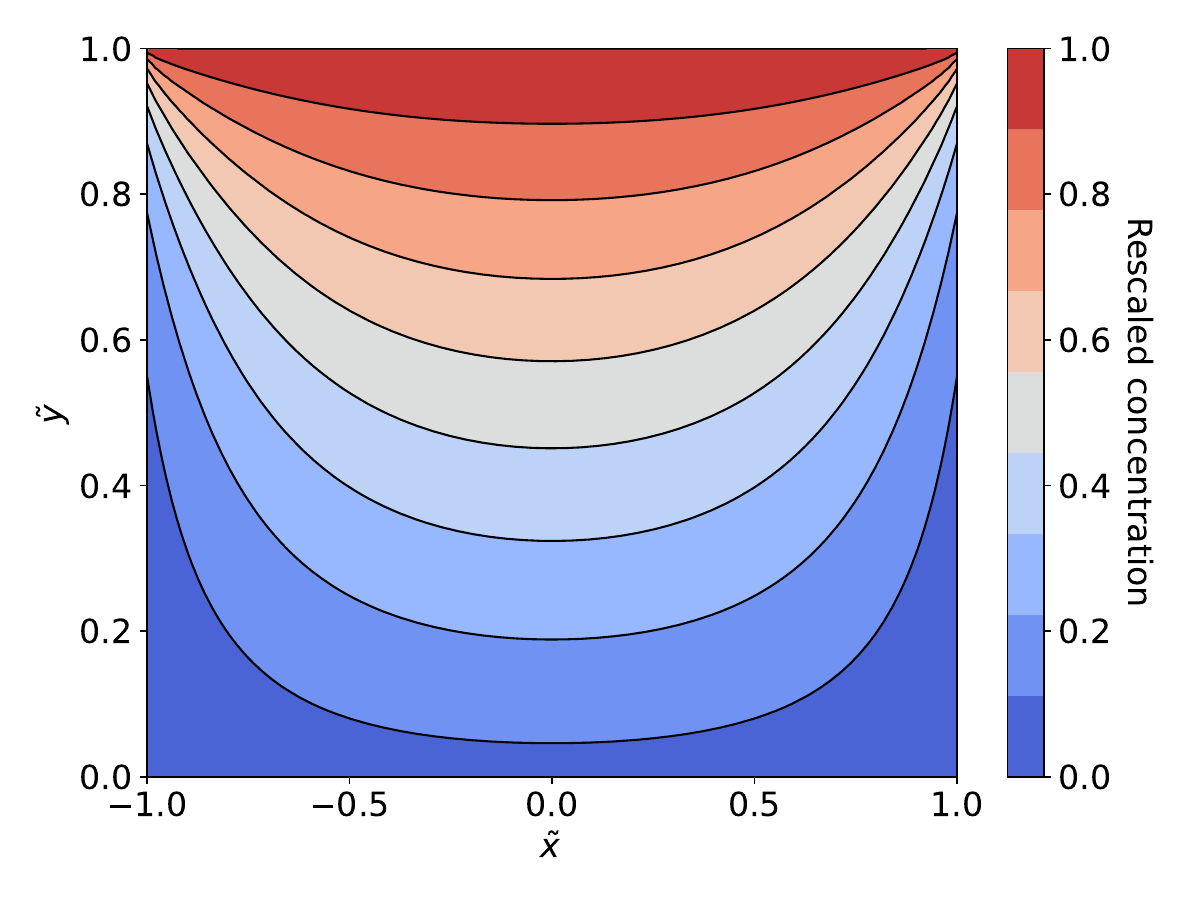}
  \caption{$\xi=1$ and $\Theta=0.1$}
\end{subfigure}%
\hfill
\begin{subfigure}{.32\textwidth}
  \centering
  \includegraphics[width=.9\linewidth]{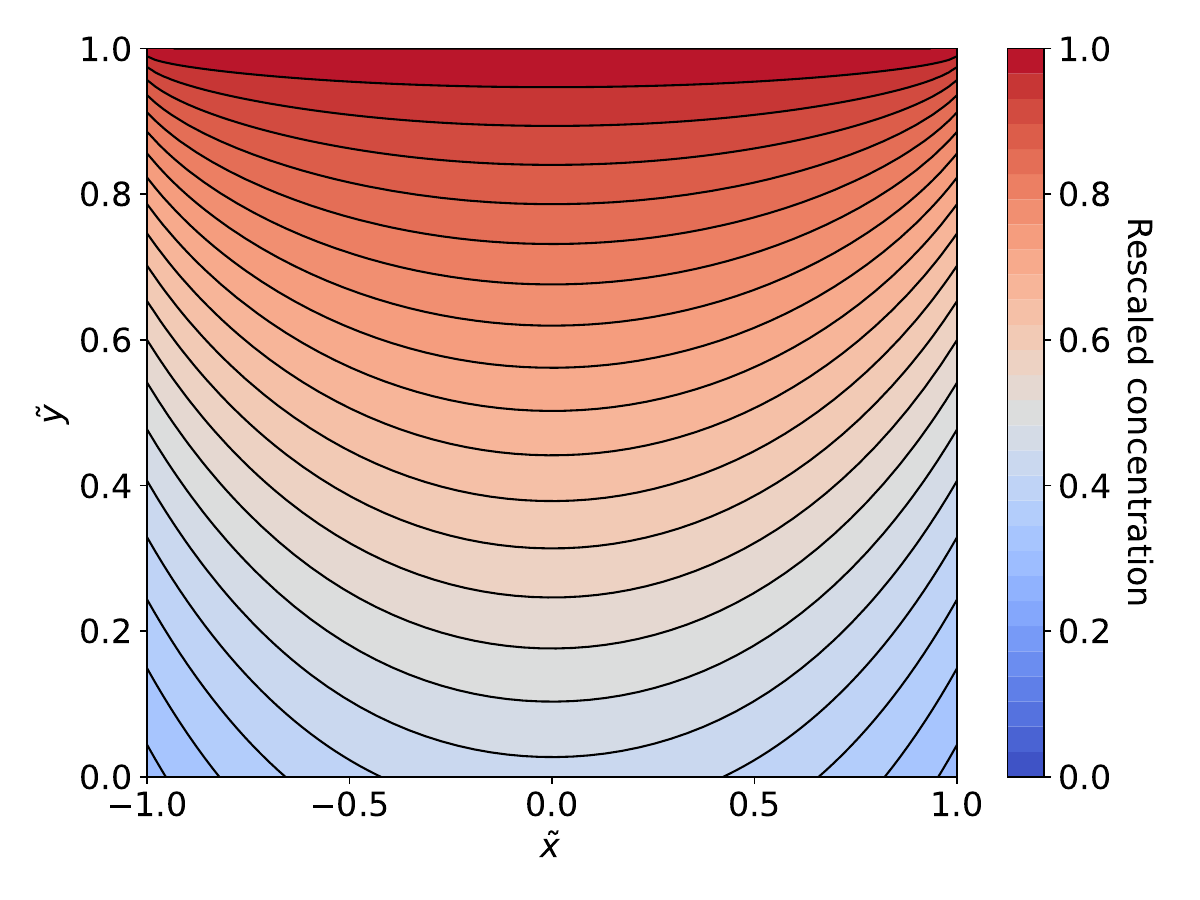}
  \caption{$\xi=1$ and $\Theta=1$}
\end{subfigure}
\hfill
\begin{subfigure}{.32\textwidth}
  \centering
  \includegraphics[width=.9\linewidth]{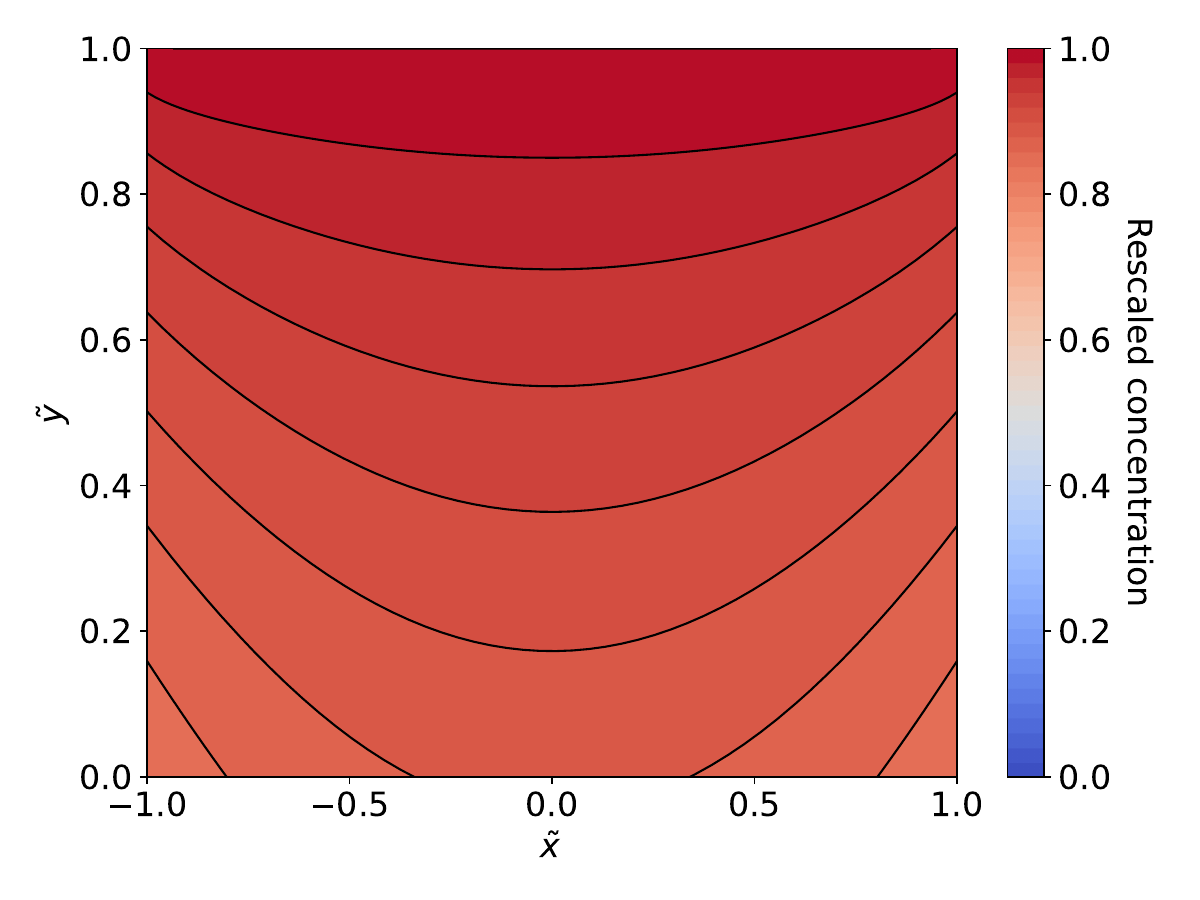}
  \caption{$\xi=1$ and $\Theta=10$}
\end{subfigure}
\begin{subfigure}{.32\textwidth}
  \centering
  \includegraphics[width=.9\linewidth]{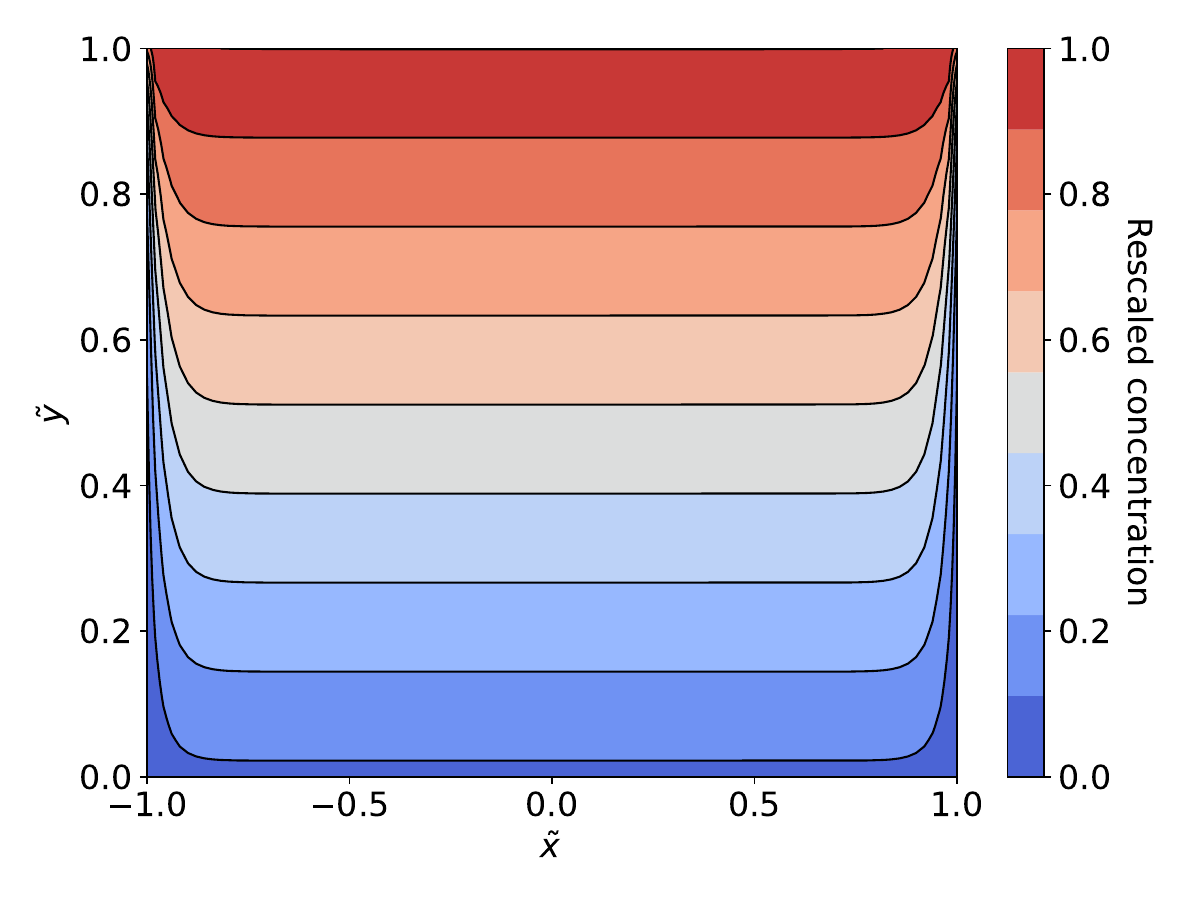}
  \caption{$\xi=10$ and $\Theta=0.1$}
\end{subfigure}%
\hfill
\begin{subfigure}{.32\textwidth}
  \centering
  \includegraphics[width=.9\linewidth]{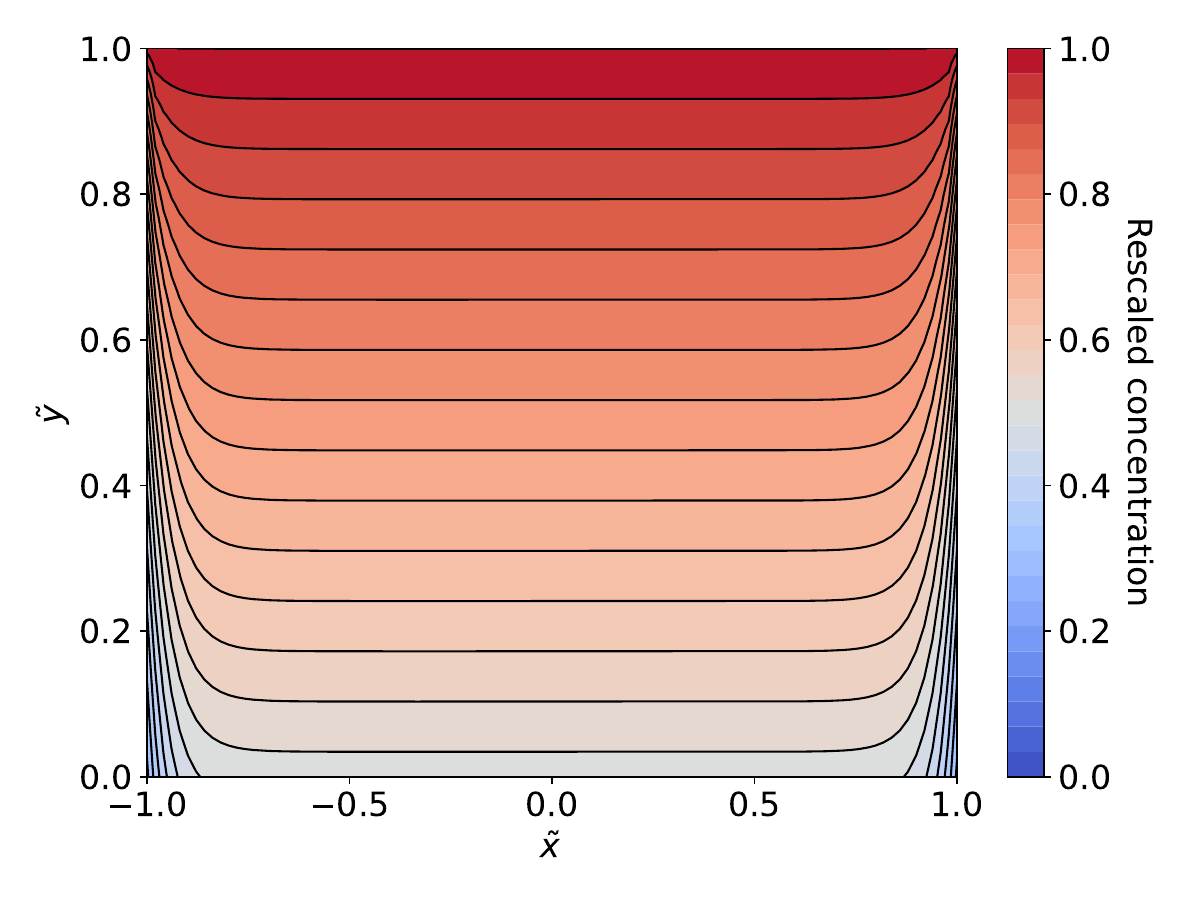}
  \caption{$\xi=10$ and $\Theta=1$}
\end{subfigure}
\hfill
\begin{subfigure}{.32\textwidth}
  \centering
  \includegraphics[width=.9\linewidth]{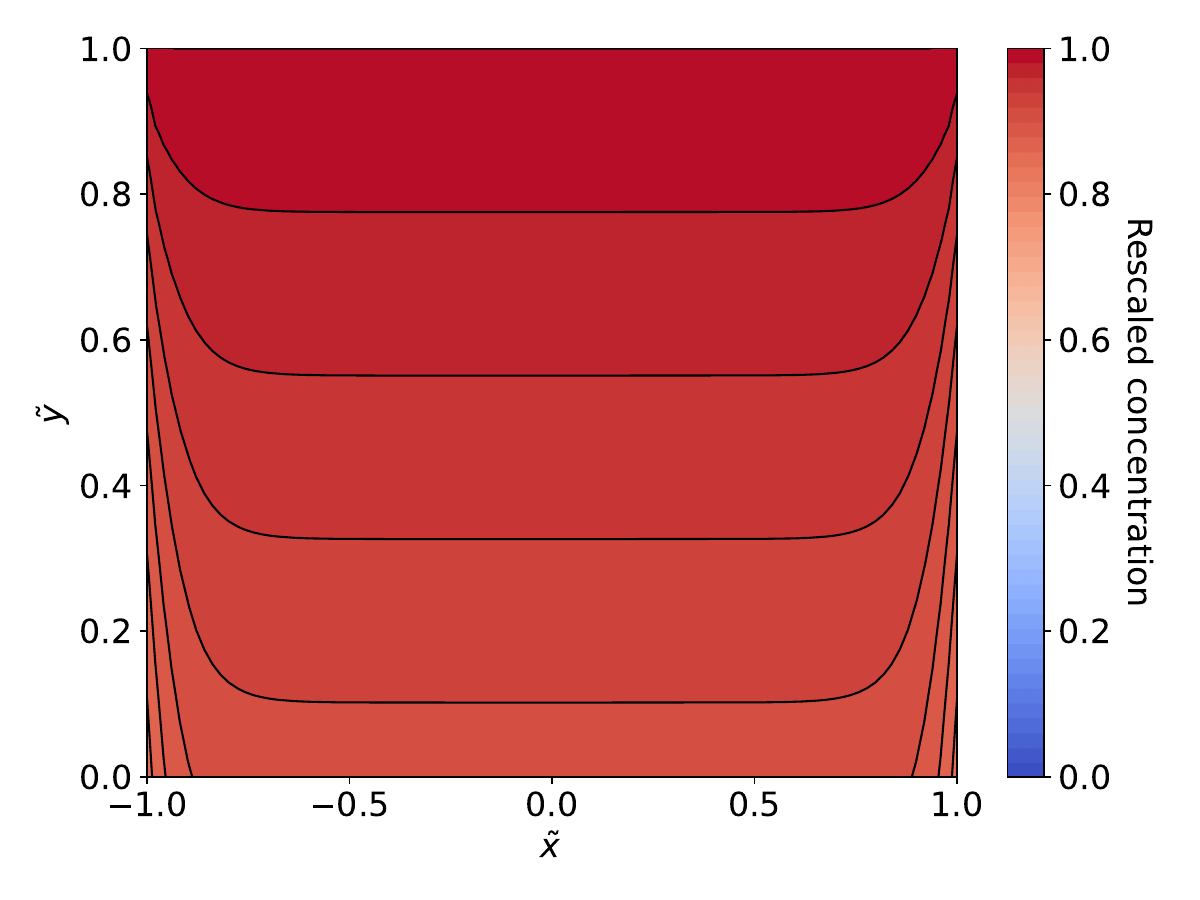}
  \caption{$\xi=10$ and $\Theta=10$}
\end{subfigure}

\caption{Contour plots of the concentration within an intervillous space, rescaled by the concentration in the lumen $c_0$, for the leaf-like villi geometry like in figure~\ref{fig:mainfig}, for various values of the rescaled intervillous space $\xi$ and the rescaled inverse surface absorptivity $\Theta$. The colors correspond to the rescaled concentration within this space. For each graph, the top is the lumen, the bottom the bottom of the intervilli space, and the right and left boundaries the villi surface. Left-right symmetry is a result of the symmetry of the system.}\label{figconcentrationleaf}
\end{figure}

This expression is plotted in figure~\ref{figconcentrationleaf}.

The expression of nutrient concentration flux integrated over one intervillous space is the following:

\begin{equation}
    J=-D\int \int_S\vec n \nabla C(x,y)d\vec S
\end{equation}

In dimensionless units this equation becomes:

\begin{equation}
    \tilde J=-(\xi\int_{-1}^1\frac{\partial c}{\partial \tilde y}\Bigr|_{\substack{\tilde y=0}}d\tilde x-\frac{1}{\xi}\int_0^1\frac{\partial c}{\partial \tilde x}\Bigr|_{\substack{\tilde x=1}}d\tilde y+\frac{1}{\xi}\int_0^1\frac{\partial c}{\partial \tilde x}\Bigr|_{\substack{\tilde x=-1}}d\tilde y)
\end{equation}
With $\tilde J=J/(Dc_0)$. The first term corresponds to the flux at the bottom of the intervillous space while the second and third terms correspond to the flux on each side of the intervillous space.
Integrating each term gives the following expression of the rescaled flux:

\begin{equation}
    \tilde J=\sum_{n=1}^\infty\frac{8\sin^2(\mu_n)\left(\Theta\frac{\mu_n}{\xi}\sinh\left(\frac{\mu_n}{\xi}\right)+\cosh\left(\frac{\mu_n}{\xi}\right)\right)}{\left(2\mu_n+\sin(2\mu_n)\right)\left(\Theta\frac{\mu_n}{\xi}\cosh\left(\frac{\mu_n}{\xi}\right)+\sinh\left(\frac{\mu_n}{\xi}\right)\right)}
\label{eq:flux_leaf}
\end{equation}

\begin{figure}[h]
\begin{subfigure}{.5\textwidth}
  \centering
  \includegraphics[width=.95\linewidth]{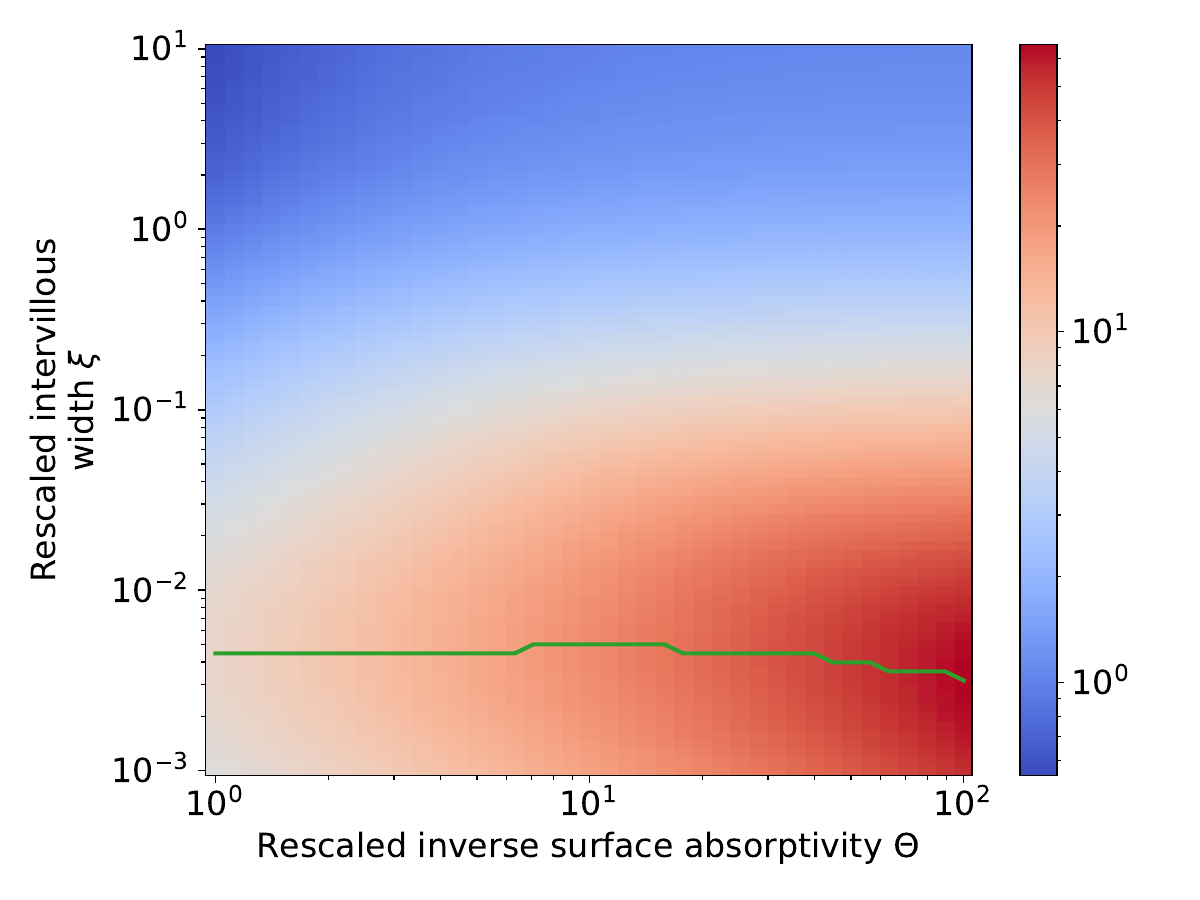}
  \caption{$\tilde e=0.01$}
\end{subfigure}%
\begin{subfigure}{.5\textwidth}
  \centering
  \includegraphics[width=.95\linewidth]{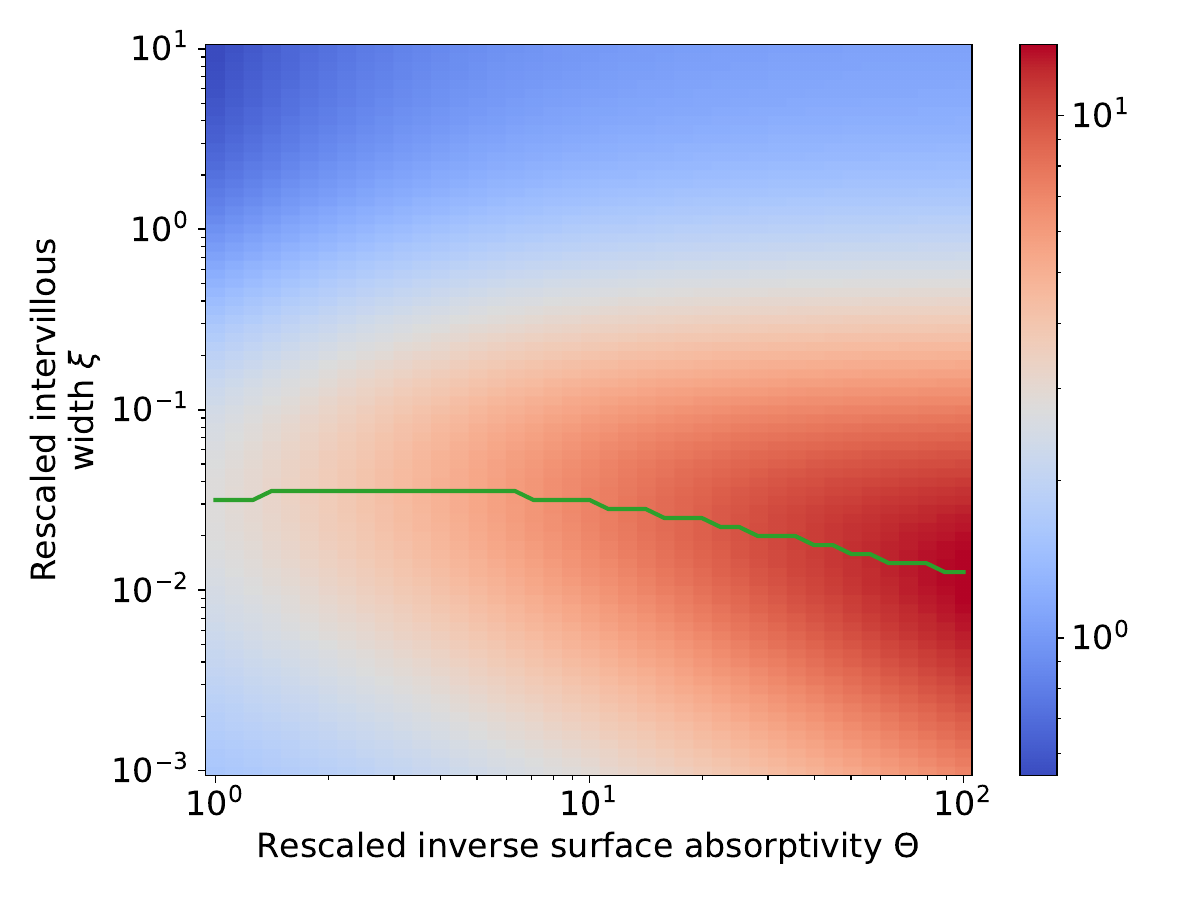}
  \caption{$\tilde e=0.1$}
\end{subfigure}
\begin{subfigure}{.5\textwidth}
  \centering
  \includegraphics[width=.95\linewidth]{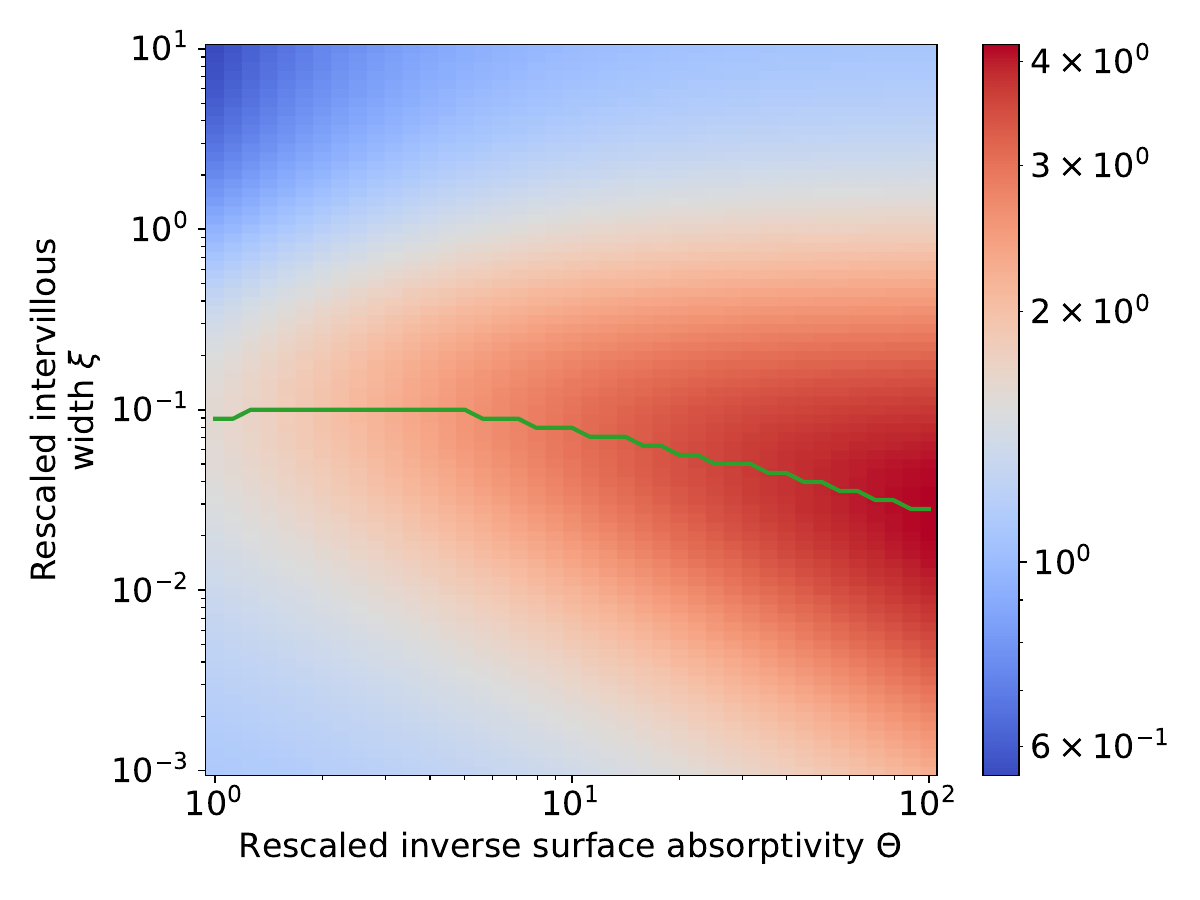}
  \caption{$\tilde e=0.5$}
\end{subfigure}%
\begin{subfigure}{.5\textwidth}
  \centering
  \includegraphics[width=.95\linewidth]{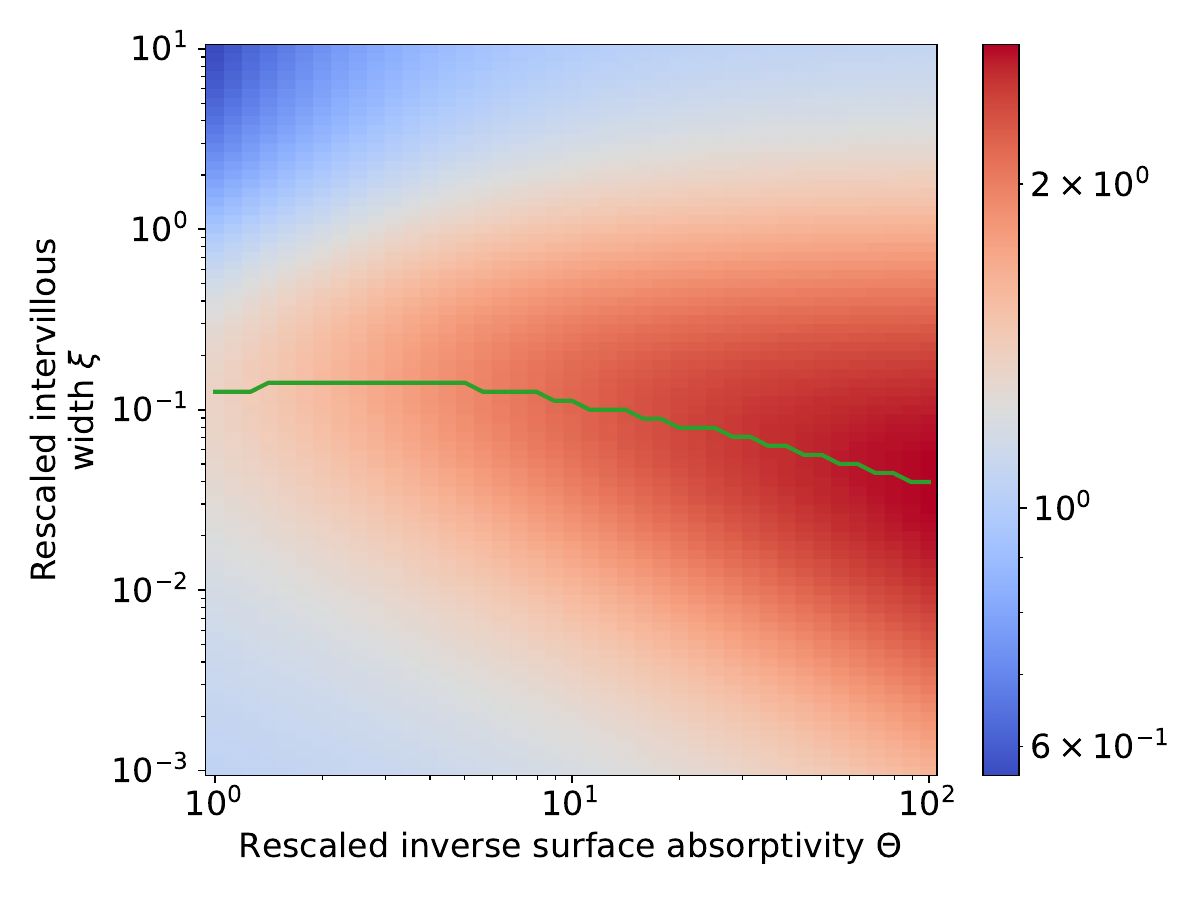}
  \caption{$\tilde e=1$}
\end{subfigure}

\caption{Rescaled log flux density per unit of gut length $\tilde j$ for the leaf-like geometry as a function of $\xi$ (the rescaled intervillous width) and $\Theta$ (the rescaled inverse surface absorptivity) for $\tilde e=0.01,0.1,0.5,1$ (the rescaled villi width). The green line represents the value of $\xi$ that maximizes absorption flux at a given value of $\Theta$.}\label{fig:rescaledfluxleaf}
\end{figure}

As explained in the main text, this expression is then used to calculate the flux density per unit of gut length, shown figure~\ref{fig:rescaledfluxleaf}.

\section{Finger-like geometry calculations}

\begin{figure}[h]
    \centering
    \includegraphics[width=0.5\linewidth]{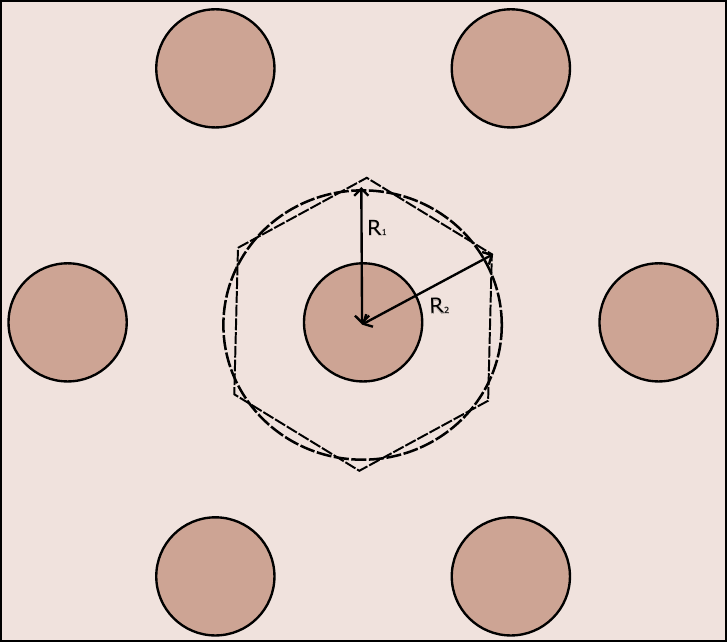}
    \caption{Sketch of how the hexagon is converted to a circle. We keep an equivalent surface such that $R_1=\sqrt{\frac{3\sqrt{3}}{2\pi}}R_2$.}
    \label{fig:hexagon_to_circle}
\end{figure}

Here we solve the Laplace equation in the finger-like villi case. The villi are modeled as evenly spaced cylinders of height $h$ on a triangular lattice such that the radius of a villus is $e/2$ and the shortest distance between the sides of two villi is $2R$ (consistent with the 2D model) as seen in figure~\ref{fig:mainfig}. There are repeating hexagons centered on the villi at the nodes of the triangular lattice of side length $R+e/2$. To solve this problem using cylindrical coordinates, the hexagons are approximated as circles of radius $R+e/2$. The cylinders are taken such that the surface area is conserved as seen in figure~\ref{fig:hexagon_to_circle}. The dimensionless units for the system are: $\tilde r=r/R$, $\tilde z=z/h$, $\xi=R/h$, $\tilde e=e/h$, $\Theta=\frac{D}{kh}$, and $c=C/c_0$. Using cylindrical coordinates, the system is reduced to an axisymmetric one, where equations~\ref{eq:Laplace}--\ref{eq:BC2} become:

\begin{equation}
    \frac{1}{\tilde r}\frac{\partial c}{\partial \tilde r}+\frac{\partial^2 c}{\partial \tilde r^2} + \xi^2\frac{\partial^2 c}{\partial \tilde z^2} = 0
    \label{eq:Laplace_finger}
\end{equation}

\begin{equation}
  c(\tilde r, 1)=1
  \label{eq:BC1_finger}
\end{equation}

\begin{equation}
  \Theta\left(\xi^{-1} \frac{\partial c}{\partial \tilde r} \vec e_r + \frac{\partial c}{\partial \tilde z}\vec e_z\right)\vec n + c = 0, (\tilde r, \tilde z) \in S,
  \label{eq:BC2_finger}
\end{equation}

where $S$ is either $\tilde r=\tilde e/2$ or $\tilde z=0$. The other boundary condition is the zero-flux on the outer circle:

\begin{equation}
    \vec n \cdot \vec \nabla c\left(1+\frac{\tilde e}{2\xi},\tilde z\right) = 0
    \label{eq:BC3_finger}
\end{equation}

The Laplace equation is solved using the method of separation of variables. The function \( c \) is expressed as the product of two functions, such that:

\begin{equation}
    c(\tilde r,\tilde z) = f(\tilde r)g(\tilde z)
    \label{eq:separation_finger}
\end{equation}

Substituting equation~\ref{eq:separation_finger} into equation~\ref{eq:Laplace_finger} yields:

\begin{equation}
    \frac{1}{\tilde r}f'g+f''g+fg''=0
\end{equation}

\begin{equation}
    \frac{1}{\tilde r}\frac{f'}{f}+\frac{f''}{f}=-\xi^2\frac{g''}{g}=-\lambda
\end{equation}

Where $f'(\tilde r)$ is the derivative of $f(\tilde r)$ with respect to $\tilde r$, $g'(\tilde z)$ is the derivative of $g(\tilde z)$ with respect to $\tilde z$ and $\lambda$ is chosen positive; otherwise, non-physical solutions are found, as in the previous case.

Let us first solve for $f(\tilde r)$:

\begin{equation}
    \frac{1}{\tilde r}f'(\tilde r)+ f''(\tilde r)+\lambda f(\tilde r) = 0
\end{equation}

$\mu$ is defined such that $\lambda=\mu^2$.

\begin{equation}
    f(\tilde r) = C J_0(\mu \tilde r)+D Y_0(\mu \tilde r)
\end{equation}

where $J_n$ and $Y_n$ are Bessel functions of the first and second kind of order $n$. Applying equation~\ref{eq:BC3_finger} yields:

\begin{equation}
    C= -D\frac{Y_1\left(\mu\left(1+\frac{\tilde e}{2\xi}\right)\right)}{J_1\left(\mu\left(1+\frac{\tilde e}{2\xi}\right)\right)}
\end{equation}

The boundary condition from equation~\ref{eq:BC2_finger} is then applied to determine $D$:

\begin{equation}
    \Theta \frac{\mu_n}{\xi} = -\frac{Y_1\left(\mu_n\left(1+\frac{\tilde e}{2\xi}\right)\right)J_0\left(\mu_n\frac{\tilde e}{2\xi}\right)-J_1\left(\mu_n\left(1+\frac{\tilde e}{2\xi}\right)\right)Y_0\left(\mu_n\frac{\tilde e}{2\xi}\right)}{Y_1\left(\mu_n\left(1+\frac{\tilde e}{2\xi}\right)\right)J_1\left(\mu_n\frac{\tilde e}{2\xi}\right)-J_1\left(\mu_n\left(1+\frac{\tilde e}{2\xi}\right)\right)Y_1\left(\mu_n\frac{\tilde e}{2\xi}\right)}
\label{eq:mu_finger}
\end{equation}

There exist an infinite number of solutions $\mu_n$ to equation~\ref{eq:mu_finger}, where $n$ is a positive integer. So, we define:

\begin{equation}
    f_n(\tilde r)=-\frac{Y_1\left(\mu_n\left(1+\frac{\tilde e}{2\xi}\right)\right)}{J_1\left(\mu_n\left(1+\frac{\tilde e}{2\xi}\right)\right)}J_0\left(\mu_n \tilde r\right)+Y_0\left(\mu_n \tilde r\right)
\end{equation}

Where $\mu_n$ is the nth solution to equation~\ref{eq:mu_finger}. Solving for $g_n(\tilde z)$ yields:

\begin{equation}
    \xi^2g_n''(\tilde z)-\lambda g_n(\tilde z) = 0
\end{equation}

\begin{equation}
    g_n(\tilde z) = B_n \cosh\left(\frac{\mu_n}{\xi} \tilde z\right)+A_n \sinh\left(\frac{\mu_n}{\xi} \tilde z\right)
\end{equation}

The boundary conditions are then applied to determine \( A_n \) and \( B_n \):

\begin{equation}
    -\Theta \frac{\mu_n}{\xi} A_n + B_n = 0
\end{equation}

\begin{equation}
    B_n = \Theta \frac{\mu_n}{\xi} A_n
\end{equation}

We then obtain $g_n(\tilde z)$ for a non-zero $A_n$:

\begin{equation}
    g_n(\tilde z) = \left(\Theta\frac{\mu_n}{\xi} \cosh\left(\frac{\mu_n}{\xi} \tilde z\right) + \sinh\left(\frac{\mu_n}{\xi} \tilde z\right)\right)A_n
\end{equation}

Then,

\begin{align}
  \begin{split}
  c_n(\tilde r, \tilde z)= &A_n \left(-\frac{Y_1\left(\mu_n\left(1+\frac{\tilde e}{2\xi}\right)\right)}{J_1\left(\mu_n\left(1+\frac{\tilde e}{2\xi}\right)\right)}J_0\left(\mu_n \tilde r\right)+Y_0\left(\mu_n \tilde r\right)\right)\\
  &\times\left(\Theta\frac{\mu_n}{\xi} \cosh\left(\frac{\mu_n}{\xi} \tilde z\right) + \sinh\left(\frac{\mu_n}{\xi} \tilde z\right)\right)
  \end{split}
\end{align}

And:

\begin{align}
  \begin{split}
    c(\tilde r,\tilde z) = \sum_{n=1}^{n=\infty}&A_n \left(-\frac{Y_1\left(\mu_n\left(1+\frac{\tilde e}{2\xi}\right)\right)}{J_1\left(\mu_n\left(1+\frac{\tilde e}{2\xi}\right)\right)}J_0\left(\mu_n \tilde r\right)+Y_0\left(\mu_n \tilde r\right)\right)\\
    &\times\left(\Theta\frac{\mu_n}{\xi} \cosh\left(\frac{\mu_n}{\xi} \tilde z\right) + \sinh\left(\frac{\mu_n}{\xi} \tilde z\right)\right)
  \end{split}
\end{align}

The coefficient \( A_n \) is determined by applying the final boundary condition \( c(\tilde r, 1) = 1 \) given in equation~\ref{eq:BC1_finger}:

\begin{align}
  \begin{split}
    c(\tilde r,1) = 1=\sum_{n=1}^{n=\infty}&A_n \left(-\frac{Y_1\left(\mu_n\left(1+\frac{\tilde e}{2\xi}\right)\right)}{J_1\left(\mu_n\left(1+\frac{\tilde e}{2\xi}\right)\right)}J_0\left(\mu_n \tilde r\right)+Y_0\left(\mu_n \tilde r\right)\right)\\
    &\times\left(\Theta\frac{\mu_n}{\xi} \cosh\left(\frac{\mu_n}{\xi}\right) + \sinh\left(\frac{\mu_n}{\xi}\right)\right)
  \end{split}
\end{align}

Substituting $\tilde A_n = A_n(\Theta \frac{\mu_n}{\xi}\cosh(\frac{\mu_n}{\xi}) + \sinh(\frac{\mu_n}{\xi}))$, gives the following equation:

\begin{equation}
    1 = \sum_{n=1}^{n=\infty}\tilde A_n \left(-\frac{Y_1\left(\mu_n\left(1+\frac{\tilde e}{2\xi}\right)\right)}{J_1\left(\mu_n\left(1+\frac{\tilde e}{2\xi}\right)\right)}J_0\left(\mu_n \tilde r\right)+Y_0\left(\mu_n \tilde r\right)\right)
\end{equation}

This expression corresponds to a generalized Fourier series. The coefficient \( \tilde{A}_n \) can be obtained as:

\begin{equation}
    \tilde A_n = \frac{\int_{\frac{\tilde e}{2\xi}}^{1+\frac{\tilde e}{2\xi}}rf_n(r)dr}{\int_{\frac{\tilde e}{2\xi}}^{1+\frac{\tilde e}{2\xi}}rf_n(r)^2dr}
\end{equation}

So the full expression of $c(\tilde r,\tilde z)$ is:

\begin{align}
  \begin{split}
    c(\tilde r,\tilde z)=&\sum_{n=1}^\infty \left(\int_{\frac{\tilde e}{2\xi}}^{1+\frac{\tilde e}{2\xi}}rf_n\left(r\right)dr\right)\left(-\frac{Y_1\left(\mu_n\left(1+\frac{\tilde e}{2\xi}\right)\right)}{J_1\left(\mu_n\left(1+\frac{\tilde e}{2\xi}\right)\right)}J_0\left(\mu_n \tilde r\right)+Y_0\left(\mu_n \tilde r\right)\right)\\
    & \times\frac{\left(\Theta\frac{\mu_n}{\xi} \cosh\left(\frac{\mu_n}{\xi} \tilde z\right) + \sinh\left(\frac{\mu_n}{\xi} \tilde z\right)\right)}{\left(\int_{\frac{\tilde e}{2\xi}}^{1+\frac{\tilde e}{2\xi}}rf_n\left(r\right)^2dr\right)\left(\Theta \frac{\mu_n}{\xi}\cosh\left(\frac{\mu_n}{\xi}\right) + \sinh\left(\frac{\mu_n}{\xi}\right)\right)}
  \end{split}
  \label{eq:concentration_finger}
\end{align}

\begin{figure}[h]
\begin{subfigure}{.32\textwidth}
  \centering
  \includegraphics[width=.9\linewidth]{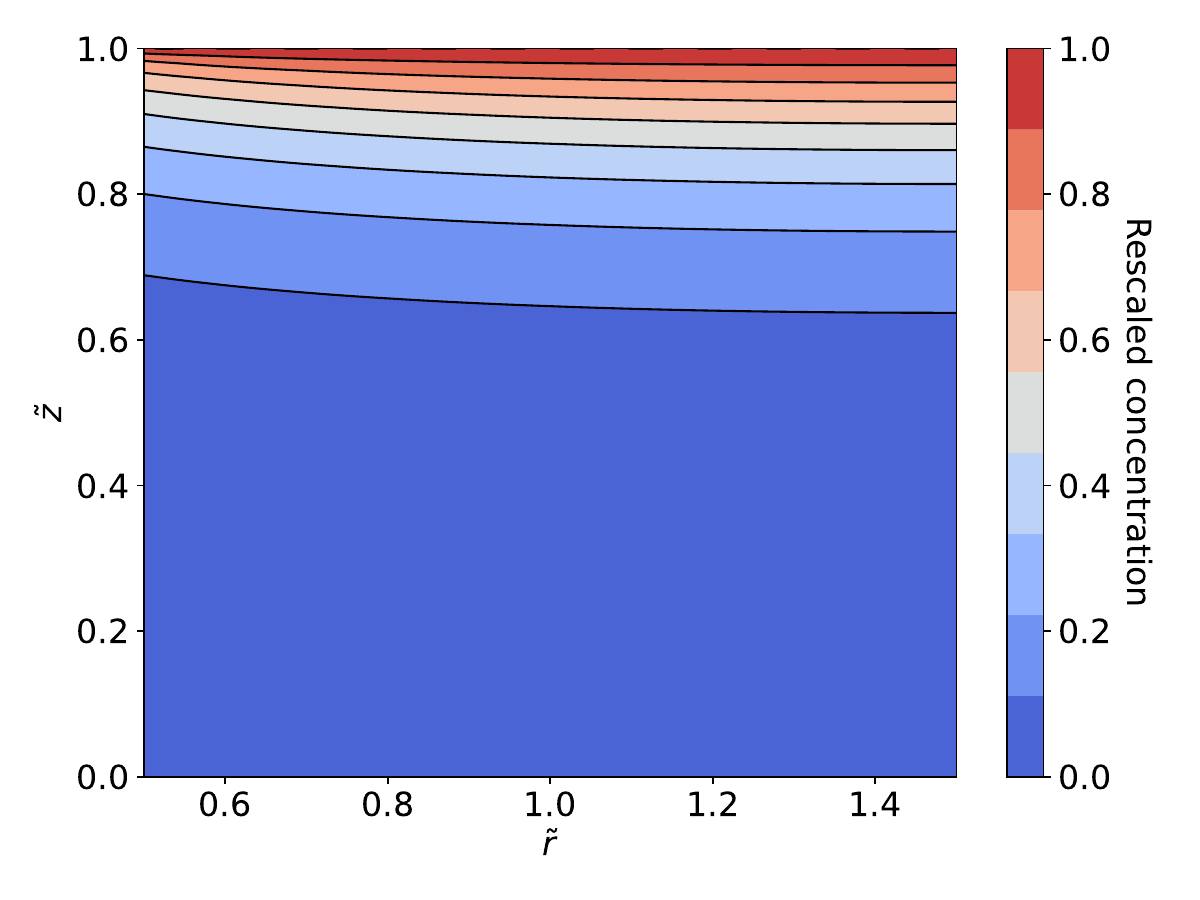}
  \caption{$\xi=0.1$ and $\Theta=0.1$}
\end{subfigure}%
\hfill
\begin{subfigure}{.32\textwidth}
  \centering
  \includegraphics[width=.9\linewidth]{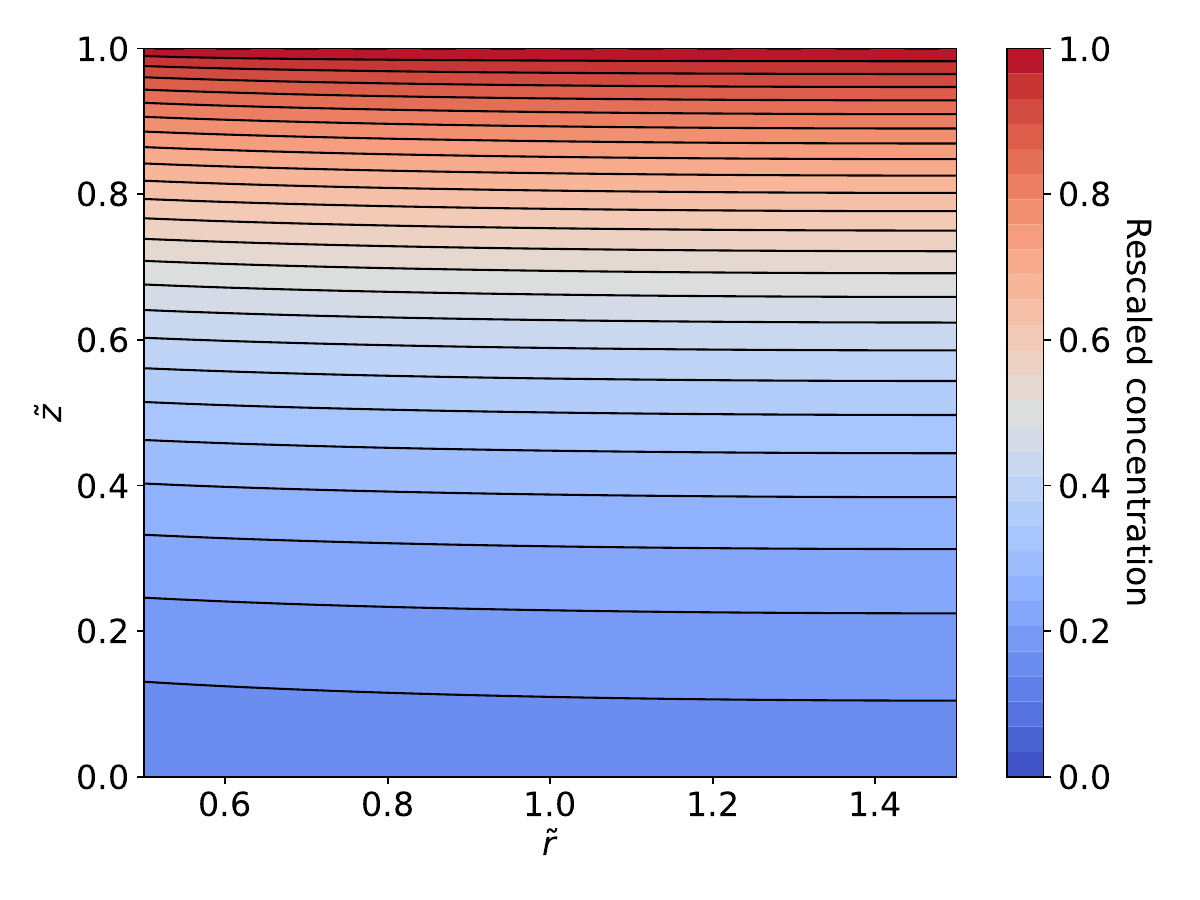}
  \caption{$\xi=0.1$ and $\Theta=1$}
\end{subfigure}
\hfill
\begin{subfigure}{.32\textwidth}
  \centering
  \includegraphics[width=.9\linewidth]{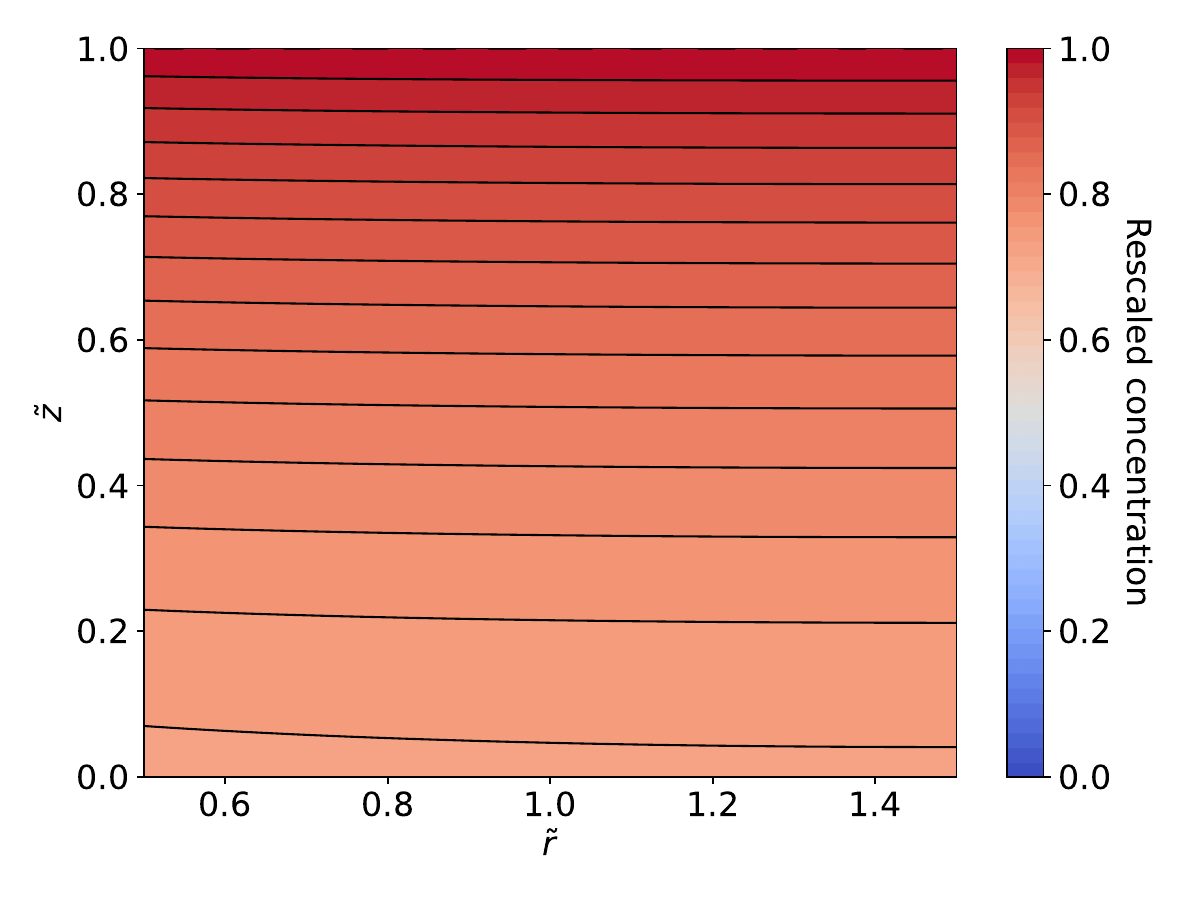}
  \caption{$\xi=0.1$ and $\Theta=10$}
\end{subfigure}
\begin{subfigure}{.32\textwidth}
  \centering
  \includegraphics[width=.9\linewidth]{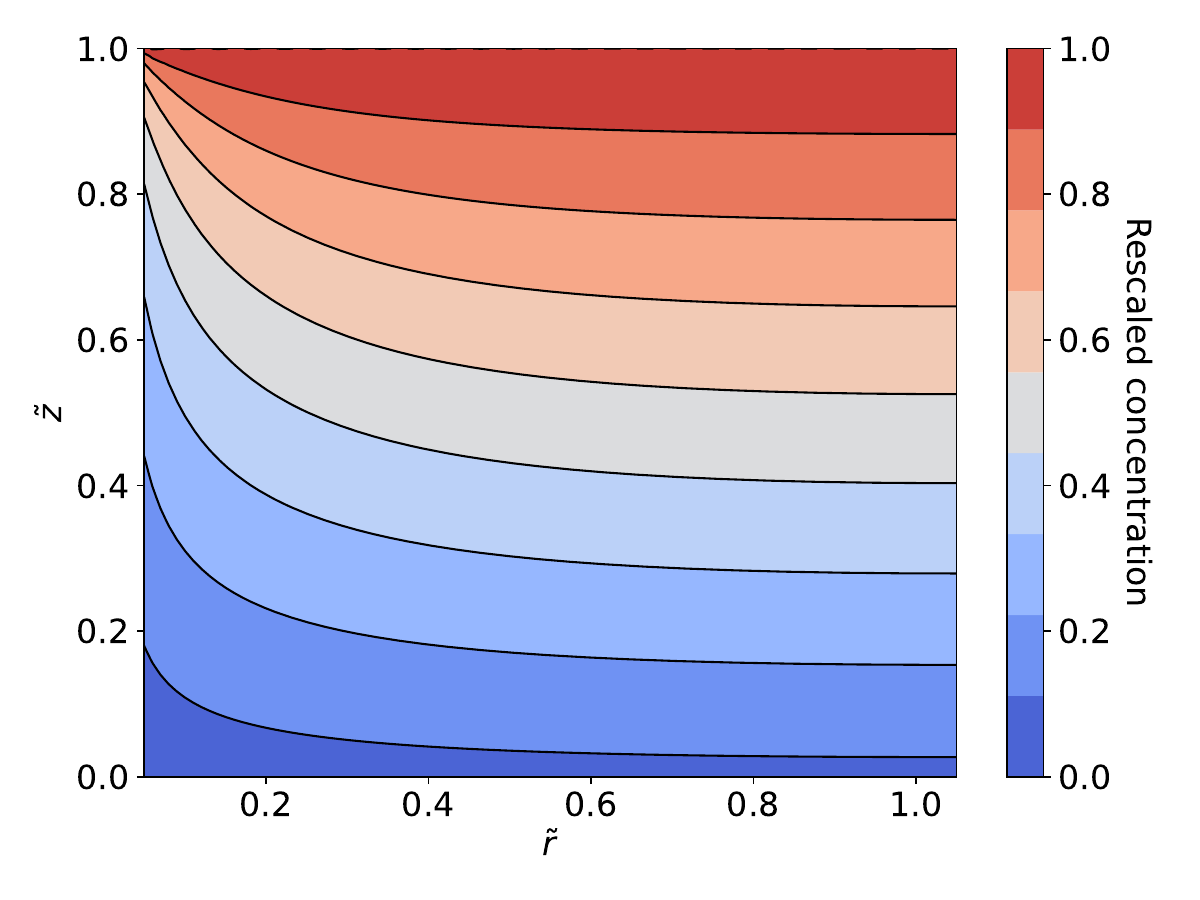}
  \caption{$\xi=1$ and $\Theta=0.1$}
\end{subfigure}%
\hfill
\begin{subfigure}{.32\textwidth}
  \centering
  \includegraphics[width=.9\linewidth]{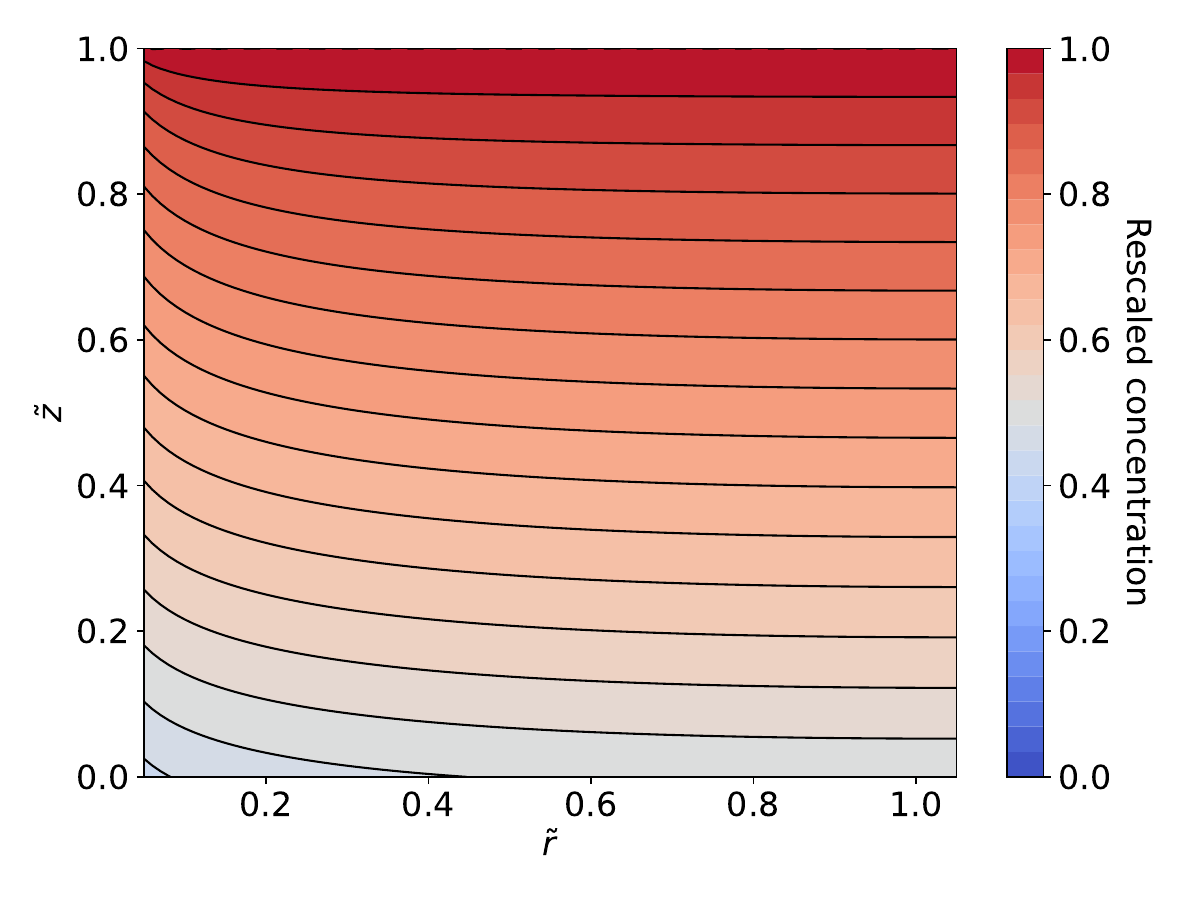}
  \caption{$\xi=1$ and $\Theta=1$}
\end{subfigure}
\hfill
\begin{subfigure}{.32\textwidth}
  \centering
  \includegraphics[width=.9\linewidth]{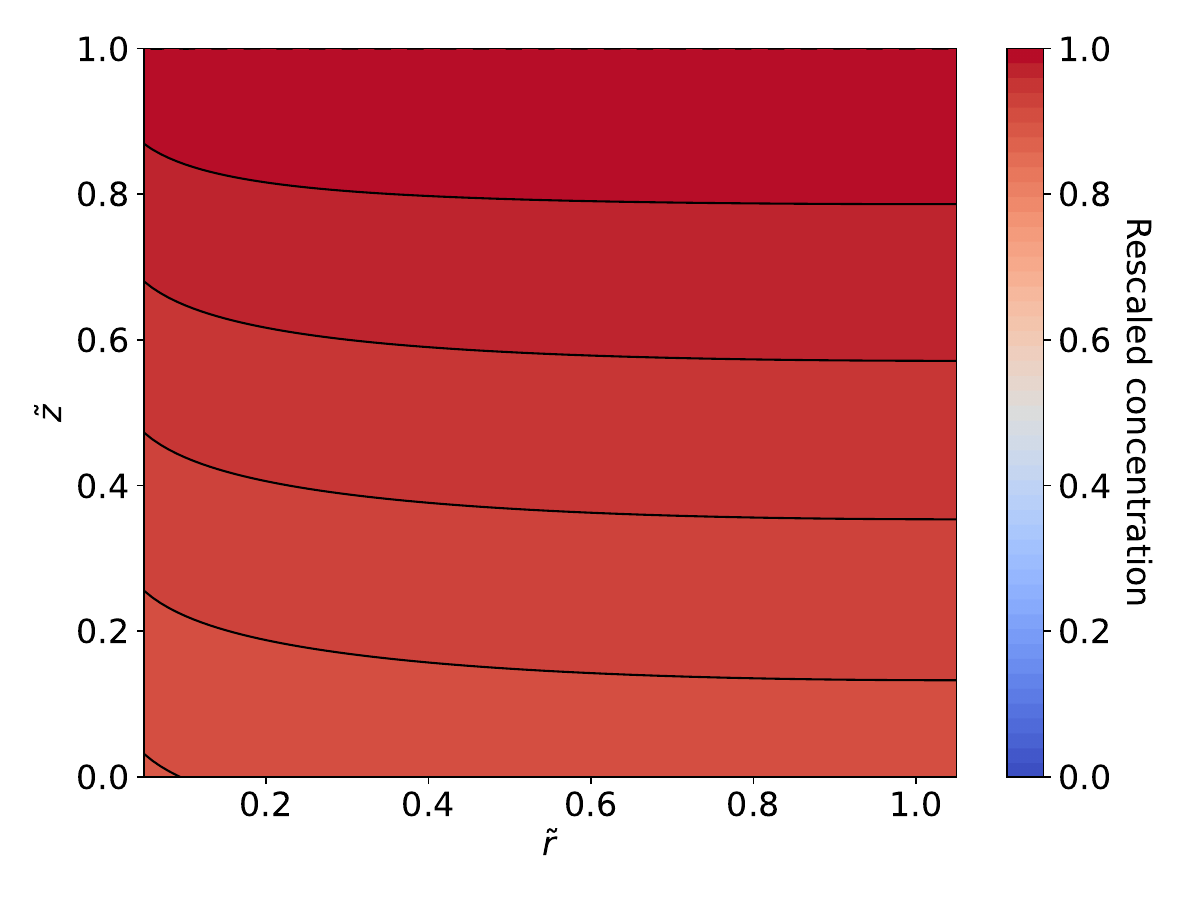}
  \caption{$\xi=1$ and $\Theta=10$}
\end{subfigure}
\begin{subfigure}{.32\textwidth}
  \centering
  \includegraphics[width=.9\linewidth]{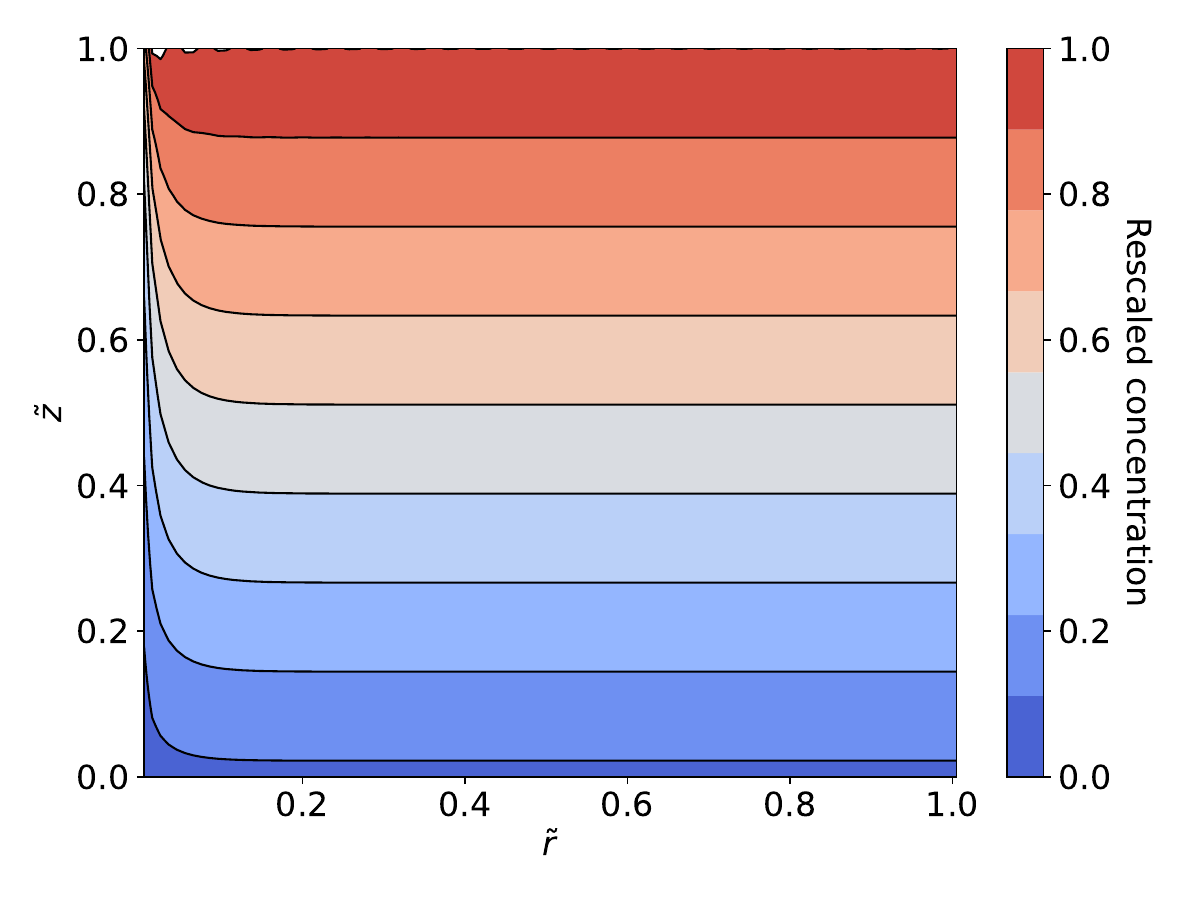}
  \caption{$\xi=10$ and $\Theta=0.1$}
\end{subfigure}%
\hfill
\begin{subfigure}{.32\textwidth}
  \centering
  \includegraphics[width=.9\linewidth]{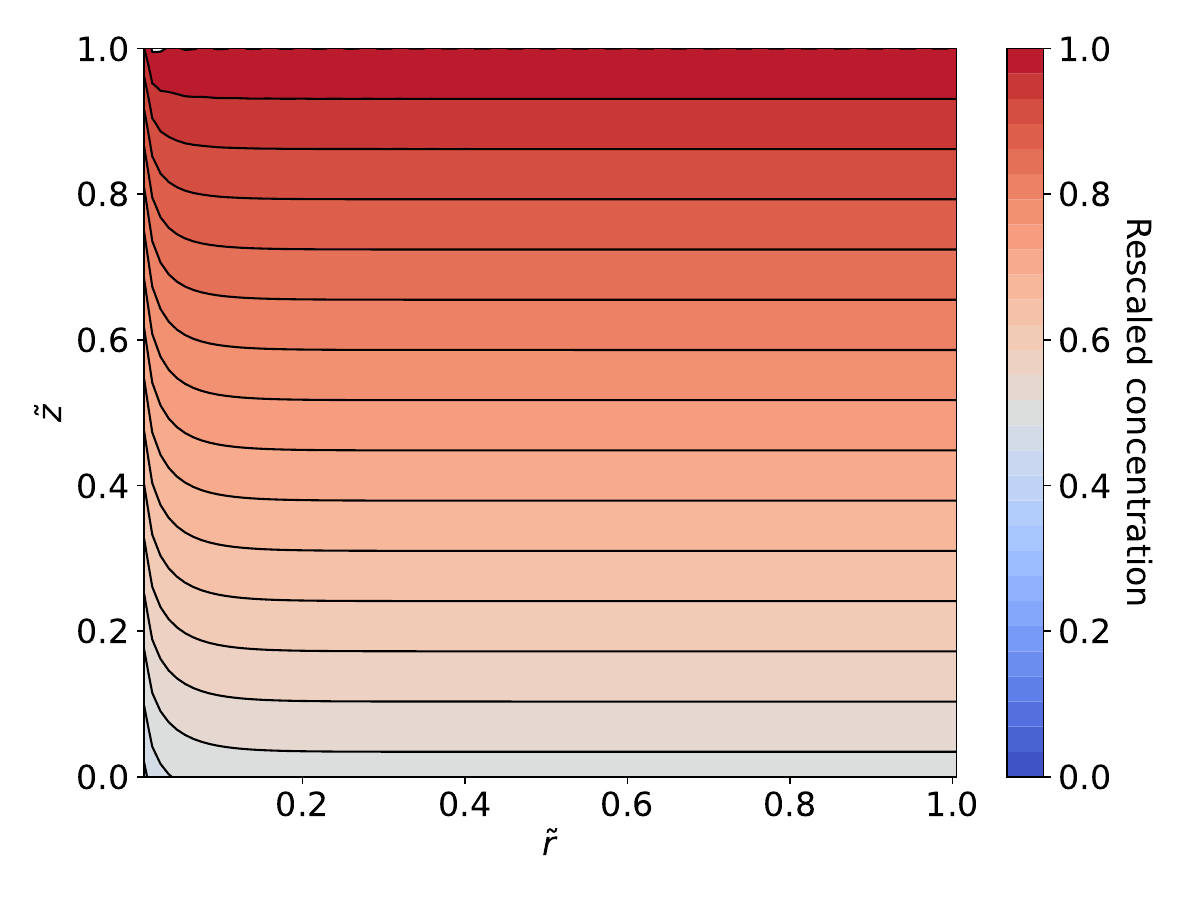}
  \caption{$\xi=10$ and $\Theta=1$}
\end{subfigure}
\hfill
\begin{subfigure}{.32\textwidth}
  \centering
  \includegraphics[width=.9\linewidth]{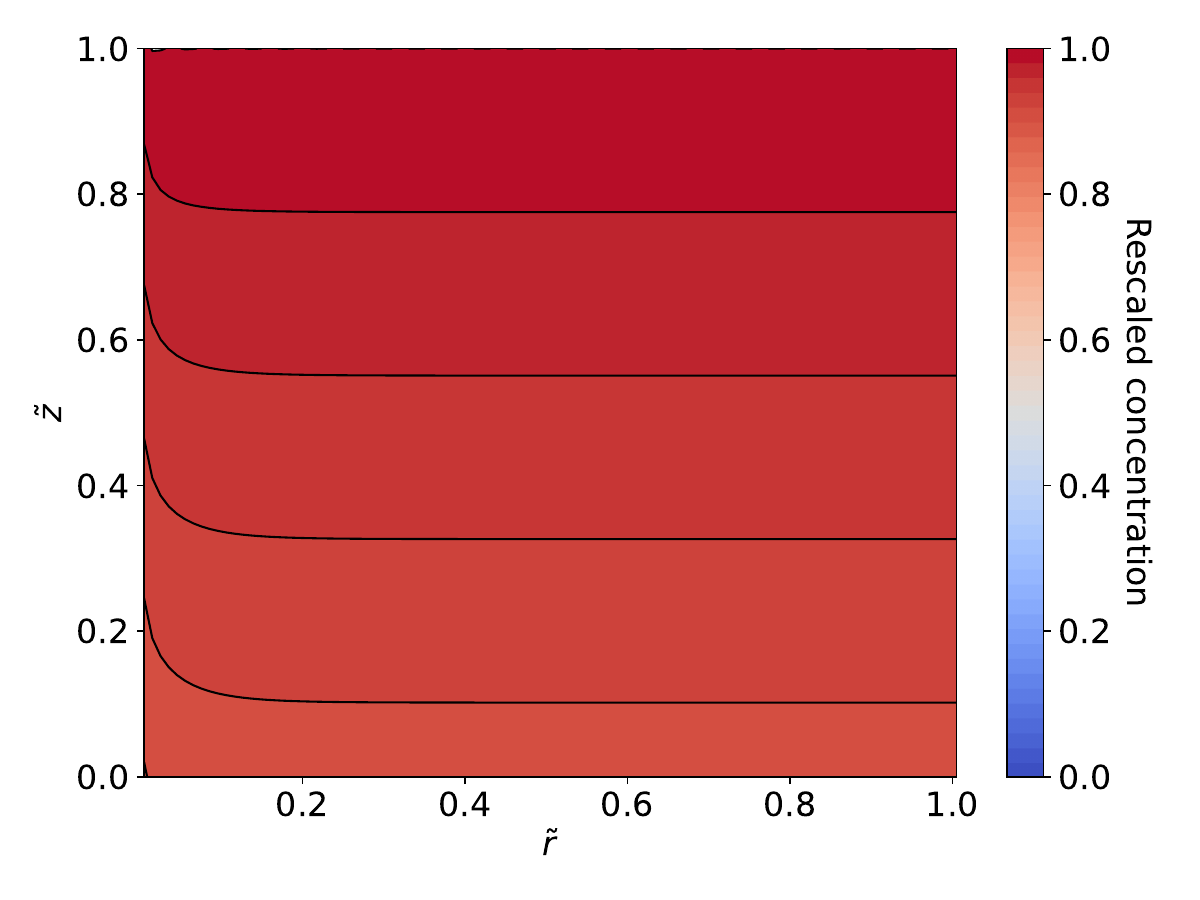}
  \caption{$\xi=10$ and $\Theta=10$}
\end{subfigure}
\caption{Contour plots of the rescaled concentration within an intervillous space for the finger-like villi geometry like in figure~\ref{fig:mainfig}. For each graph, the top corresponds to the lumen, the left boundary the villi surface towards the center of the cylindrical coordinates, and the right the edge of the cylindrical coordinates. Various values of the rescaled intervillous space $\xi$ and the inverse rescaled surface absorptivity $\Theta$. The exact solutions are infinite sums. Numerically, we went up to $n=100$, which was sufficient. The colors correspond to the rescaled concentration within this space (the lumen concentration is set to 1).}\label{fig:concentration_finger}
\end{figure}

It is plotted in figure~\ref{fig:concentration_finger}.

The expression of nutrient concentration flux integrated over one intervillous space is the following:

\begin{equation}
    J=-D\int \int_S\vec n \nabla C(r,z)d\vec S
\end{equation}

The flux is non-dimensionalized such that $\tilde J=J/(Dc_0h)$. In dimensionless units, this equation becomes:

\begin{equation}
	\tilde J=-2\pi\left(\xi^2\int_{\frac{\tilde e}{2\xi}}^{1+\frac{\tilde e}{2\xi}}r\frac{dc}{dz}\Bigr|_{\substack{ z=0}} dr + \frac{\tilde e}{2\xi} \int_0^1 \frac{dc}{dr}\Bigr|_{\substack{ r=\frac{\tilde e}{2\xi}}} dz \right),
\label{eq:flux_finger_integral}
\end{equation}

Integrating each term gives the following expression for the dimensionless flux:

\begin{align}
  \begin{split}
    \tilde J=&\sum_{n=1}^\infty \tilde e \pi\left(\int_{\frac{\tilde e}{2\xi}}^{1+\frac{\tilde e}{2\xi}}rf_n\left(r\right)dr\right)\left(-\frac{Y_1\left(\mu_n\left(1+\frac{\tilde e}{2\xi}\right)\right)}{J_1\left(\mu_n\left(1+\frac{\tilde e}{2\xi}\right)\right)}J_1\left(\mu_n \frac{\tilde e}{2\xi}\right)+Y_1\left(\mu_n \frac{\tilde e}{2\xi}\right)\right)\\
    &\times\frac{\left(\Theta\frac{\mu_n}{\xi} \sinh\left(\frac{\mu_n}{\xi}\right) + \cosh\left(\frac{\mu_n}{\xi}\right)\right)}{\left(\int_{\frac{\tilde e}{2\xi}}^{1+\frac{\tilde e}{2\xi}}rf_n\left(r\right)^2dr\right)\left(\Theta \frac{\mu_n}{\xi}\cosh\left(\frac{\mu_n}{\xi}\right) + \sinh\left(\frac{\mu_n}{\xi}\right)\right)}
  \end{split}
  \label{eq:flux_finger}
\end{align}

\begin{figure}[h]
    \centering
    \includegraphics[width=1\linewidth]{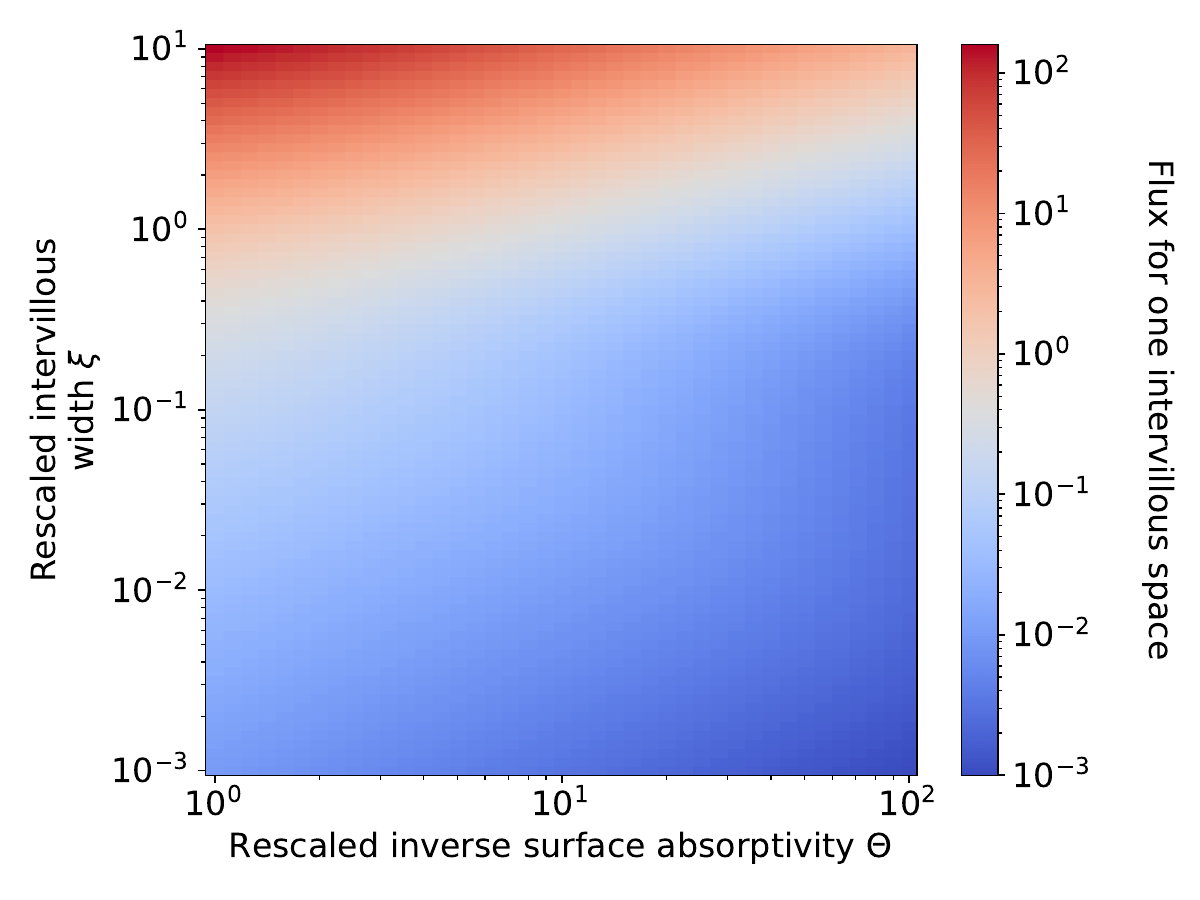}
    \caption{Non-dimensionalized flux for one intervillous space for the finger-like villi geometry for a wide range of $\xi$ and $\Theta$ in log-scale.}\label{figfluxfinger}
\end{figure}

\begin{figure}[h]
\begin{subfigure}{.5\textwidth}
  \centering
  \includegraphics[width=.95\linewidth]{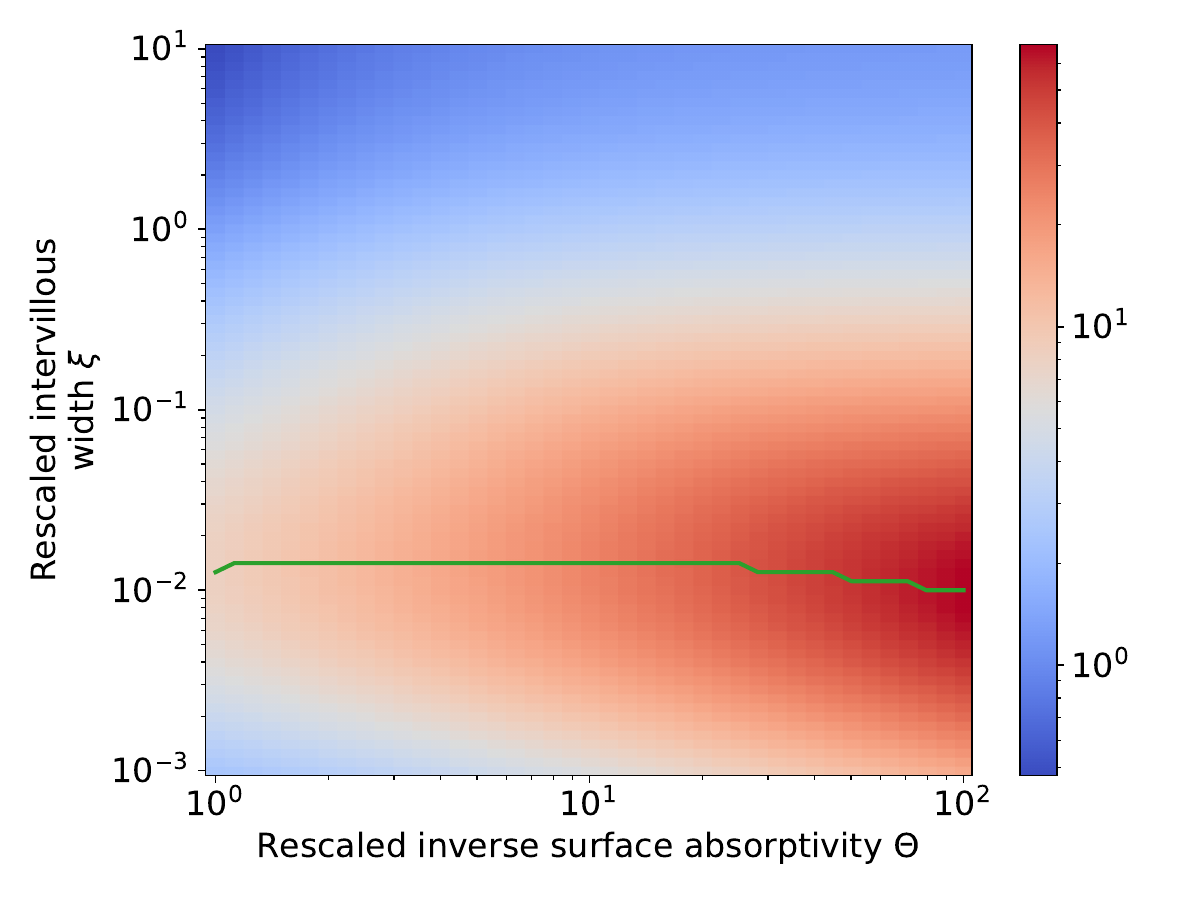}
  \caption{$\tilde e=0.01$}
\end{subfigure}%
\begin{subfigure}{.5\textwidth}
  \centering
  \includegraphics[width=.95\linewidth]{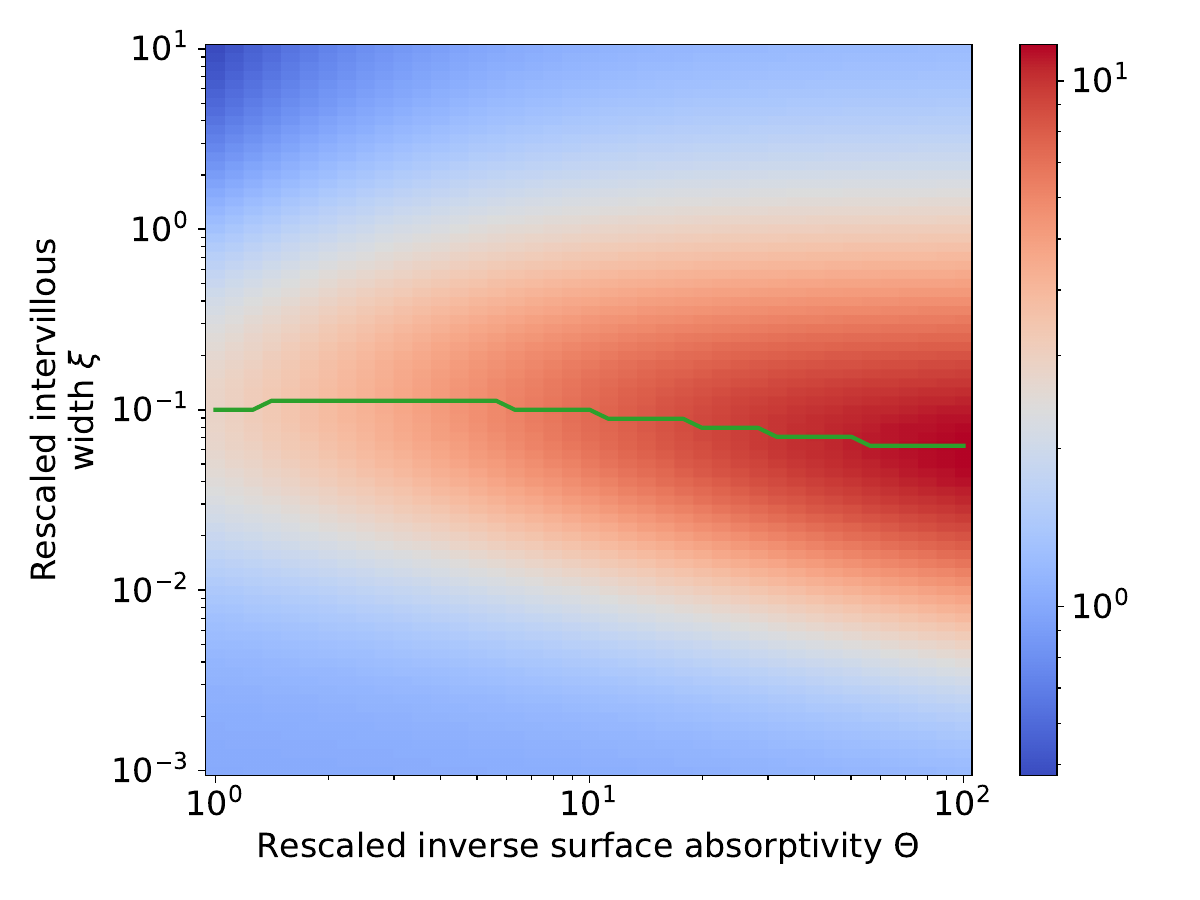}
  \caption{$\tilde e=0.1$}
\end{subfigure}
\begin{subfigure}{.5\textwidth}
  \centering
  \includegraphics[width=.95\linewidth]{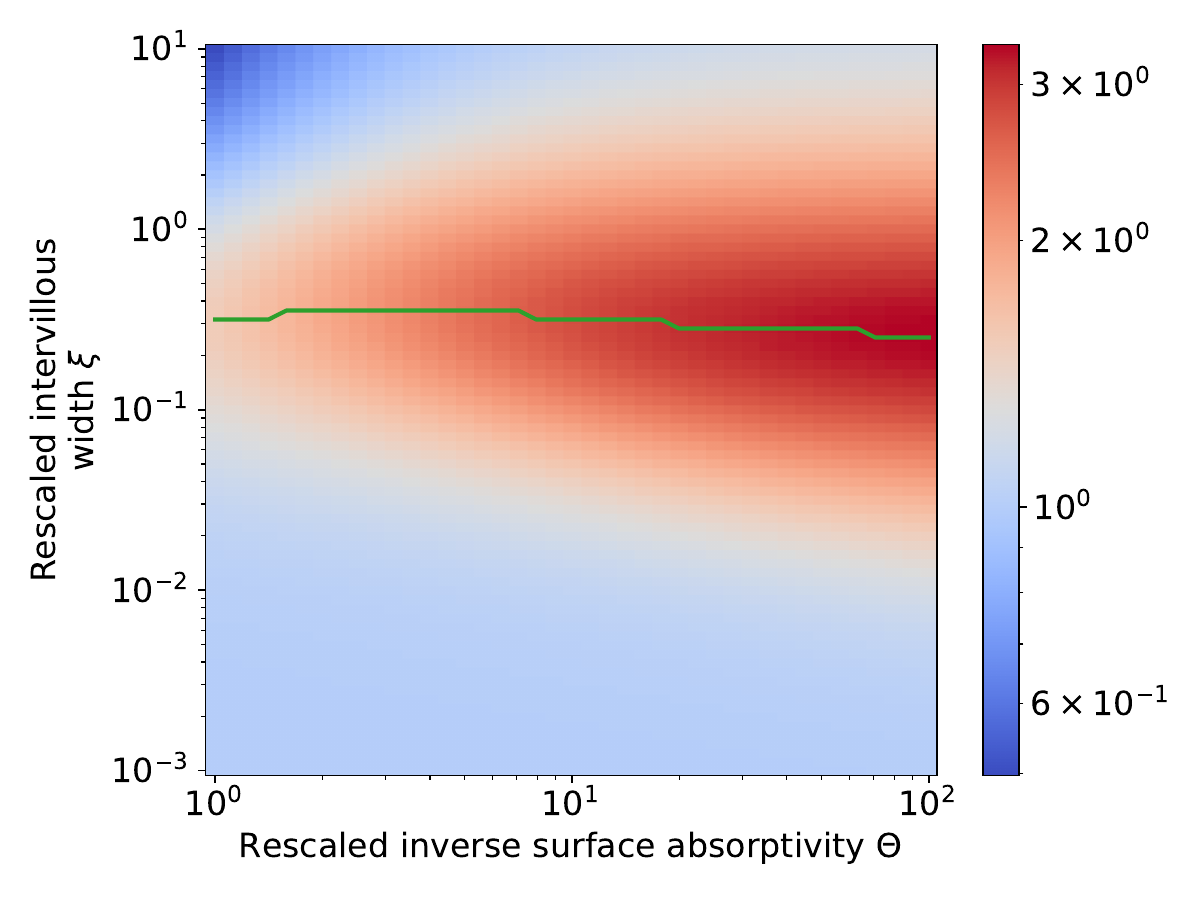}
  \caption{$\tilde e=0.5$}
\end{subfigure}%
\begin{subfigure}{.5\textwidth}
  \centering
  \includegraphics[width=.95\linewidth]{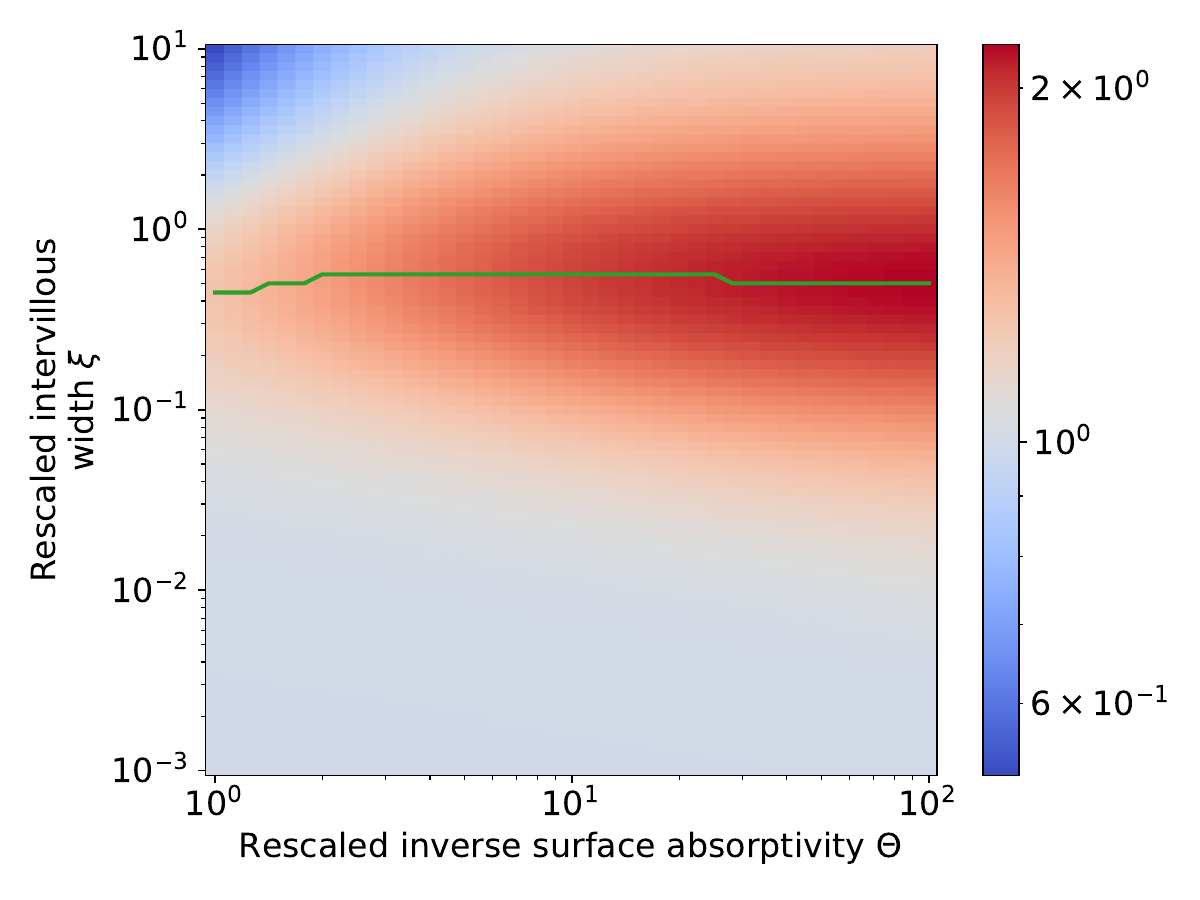}
  \caption{$\tilde e=1$}
\end{subfigure}

\caption{Rescaled log flux density per unit of gut length $\tilde j$ for the finger-like geometry as a function of $\xi$ (the rescaled intervillous width) and $\Theta$ (the rescaled inverse surface absorptivity) for $\tilde e=0.01,0.1,0.5,1$ (the rescaled villi width). The green line represents the value of $\xi$ that maximizes absorption flux at a given value of $\Theta$.}\label{figrescaledfluxfinger}
\end{figure}

It is plotted in figure~\ref{figfluxfinger}. As explained in the main text, this result is used in the calculation of the flux per unit of gut length, plotted in figure~\ref{figrescaledfluxfinger}.

\clearpage

\section{Crypt geometry calculations}

Here we solve the Laplace equation in the colonic crypt case. The crypts are on a similar hexagonal lattice as the finger-like structures in this model. The dimensionless units for the system are: $\tilde r=r/R$, $\tilde z=z/h$, $\xi=R/h$, $\tilde e=e/h$, $\Theta=\frac{D}{kh}$, and $c=C/c_0$. The differential equation to solve is:

\begin{equation}
  \frac{1}{\tilde r}\frac{\partial c}{\partial \tilde r} + \frac{\partial^2 c}{\partial \tilde r^2}+\xi^2\frac{\partial^2 c}{\partial \tilde z^2} = 0
  \label{eq:Laplace_crypt}
\end{equation}

With the following boundary conditions:

\begin{equation}
  c\left(\tilde r, 1\right)=1
  \label{eq:BC1_crypt}
\end{equation}

\begin{equation}
  \Theta\left(\xi^{-1} \frac{\partial c}{\partial \tilde r} \vec e_r + \frac{\partial c}{\partial \tilde z}\vec e_z\right)\vec n + c = 0, \left(\tilde r, \tilde z\right) \in S
  \label{eq:BC2_crypt}
\end{equation}

where $S$ is either $\tilde r= 1$ or $\tilde z=0$. The cylindrical Laplace equation is solved by assuming that \( c \) can be written as the product of two functions, such that:

\begin{equation}
    c\left(\tilde r,\tilde z\right) = f\left(\tilde r\right)g\left(\tilde z\right)
    \label{eq:separation_crypt}
\end{equation}

Substituting equation~\ref{eq:separation_crypt} into equation~\ref{eq:Laplace_crypt} yields:

\begin{equation}
  \frac{1}{\tilde r} f'\left(\tilde r\right)g\left(\tilde z\right)+ f''\left(\tilde r\right)g\left(\tilde z\right) + \xi^2f\left(\tilde r\right)g''\left(\tilde z\right) = 0
\end{equation}

\begin{equation}
  \frac{1}{\tilde r} \frac{f'\left(\tilde r\right)}{f\left(\tilde r\right)}+ \frac{f''\left(\tilde r\right)}{f\left(\tilde r\right)} = -\xi^2\frac{g''\left(\tilde z\right)}{g\left(\tilde z\right)} = -\lambda
\end{equation}

Where $f'\left(\tilde r\right)$ is the derivative of $f\left(\tilde r\right)$ with respect to $\tilde r$, $g'\left(\tilde z\right)$ is the derivative of $g\left(\tilde z\right)$ with respect to $\tilde z$ and $\lambda$ is chosen positive otherwise non-physical solutions are found. We define $\lambda=\mu^2$.

Let us first solve for $f\left(\tilde r\right)$:

\begin{equation}
    \frac{1}{\tilde r}f'(\tilde r) + f''(\tilde r)+\lambda f(\tilde r) = 0
\end{equation}

The resulting expression corresponds to a Bessel differential equation:

\begin{equation}
    f\left(\tilde r\right) = C J_0\left(\mu \tilde r\right) + D Y_0\left(\mu \tilde r\right)
\end{equation}

Finiteness of the solution implies that $D=0$. Using the imperfect absorption boundary condition:

\begin{equation}
    \Theta \xi^{-1} \frac{\partial c}{\partial \tilde r}\vert_{\tilde r=1} + c\left(1,z\right)=0
\end{equation}

Assuming \( C \neq 0 \), the result is:

\begin{equation}
    \Theta\frac{\mu}{\xi} = \frac{J_0\left(\mu\right)}{J_1\left(\mu\right)}
\label{eq:mu_crypt}
\end{equation}

There exists an infinite number of solutions $\mu_n$, where n is a positive integer. So, for a non-zero $C$ we have $f_n\left(\tilde r\right)$:

\begin{equation}
    f_n\left(\tilde r\right) = J_0\left(\mu_n\tilde r\right)
\end{equation}

Where $\mu_n$ is the nth solution to equation~\ref{eq:mu_crypt}. Solving for $g_n\left(\tilde z\right)$ yields:

\begin{equation}
    g_n''\left(\tilde z\right)-\xi^2\lambda g_n\left(\tilde z\right) = 0
\end{equation}

\begin{equation}
    g_n\left(\tilde z\right) = B_n \cosh\left(\frac{\mu_n}{\xi} \tilde z\right)+A_n \sinh\left(\frac{\mu_n}{\xi} \tilde z\right)
\end{equation}

The boundary conditions are then applied to determine $A_n$ and $B_n$:

\begin{equation}
    -\Theta \frac{\mu_n}{\xi} A_n + B_n = 0
\end{equation}

\begin{equation}
    B_n = \Theta \frac{\mu_n}{\xi} A_n
\end{equation}

For a non-zero \( A_n \), the function \( g_n(\tilde{z}) \) is obtained as:

\begin{equation}
    g_n\left(\tilde z\right) = A_n\left(\Theta\frac{\mu_n}{\xi} \cosh\left(\frac{\mu_n}{\xi} \tilde z\right) + \sinh\left(\frac{\mu_n}{\xi} \tilde z\right)\right)
\end{equation}

Then,

\begin{equation}
  c_n\left(\tilde r,\tilde z\right) = A_n J_0\left(\mu_n\tilde r\right)\left(\Theta\frac{\mu_n}{\xi} \cosh\left(\frac{\mu_n}{\xi} \tilde z\right) + \sinh\left(\frac{\mu_n}{\xi}\tilde z\right)\right)
\end{equation}

And:

\begin{equation}
  c\left(\tilde r,\tilde z\right) = \sum_{n=1}^{n=\infty}A_n J_0\left(\mu_n\tilde r\right)\left(\Theta\frac{\mu_n}{\xi} \cosh\left(\frac{\mu_n}{\xi} \tilde z\right) + \sinh\left(\frac{\mu_n}{\xi} \tilde z\right)\right)
\end{equation}

The coefficient \( A_n \) is determined by applying the final boundary condition \( c(\tilde{r}, 1) = 1 \). This leads to the equation:

\begin{equation}
  1 = \sum_{n=1}^{n=\infty}\tilde A_n J_0\left(\mu_n\tilde r\right)
\end{equation}

Where $\tilde A_n = A_n\left(\Theta\frac{\mu_n}{\xi} \cosh\left(\frac{\mu_n}{\xi}\right) + \sinh\left(\frac{\mu_n}{\xi}\right)\right)$. It is possible to find $\tilde A_n$:

\begin{equation}
  \int_0^1rJ_0\left(\mu_n r\right)dr = \sum_{n=1}^{n=\infty}\tilde A_n \int_0^1rJ_0\left(\mu_n r\right)J_0\left(\mu_n r\right)dr
\end{equation}

\begin{equation}
  \tilde A_n=\frac{2J_1\left(\mu_n\right)}{\mu_n\left(J_0\left(\mu_n\right)^2+J_1\left(\mu_n\right)^2\right)}
\end{equation}

So, the complete expression of $c\left(\tilde r,\tilde z\right)$ is:

\begin{equation}
  c\left(\tilde r,\tilde z\right) = \sum_{n=1}^{n=\infty} \frac{2 J_1\left(\mu_n\right)J_0\left(\mu_n\tilde r\right)\left(\Theta\frac{\mu_n}{\xi} \cosh\left(\frac{\mu_n}{\xi} \tilde z\right)+\sinh\left(\frac{\mu_n}{\xi} \tilde z\right)\right)}{\mu_n\left(J_0\left(\mu_n\right)^2+ J_1\left(\mu_n\right)^2\right)\left(\Theta\frac{\mu_n}{\xi} \cosh\left(\frac{\mu_n}{\xi}\right) + \sinh\left(\frac{\mu_n}{\xi}\right)\right)}
\label{eq:concentration_crypt}
\end{equation}

\begin{figure}[h]
\begin{subfigure}{.32\textwidth}
  \centering
  \includegraphics[width=.9\linewidth]{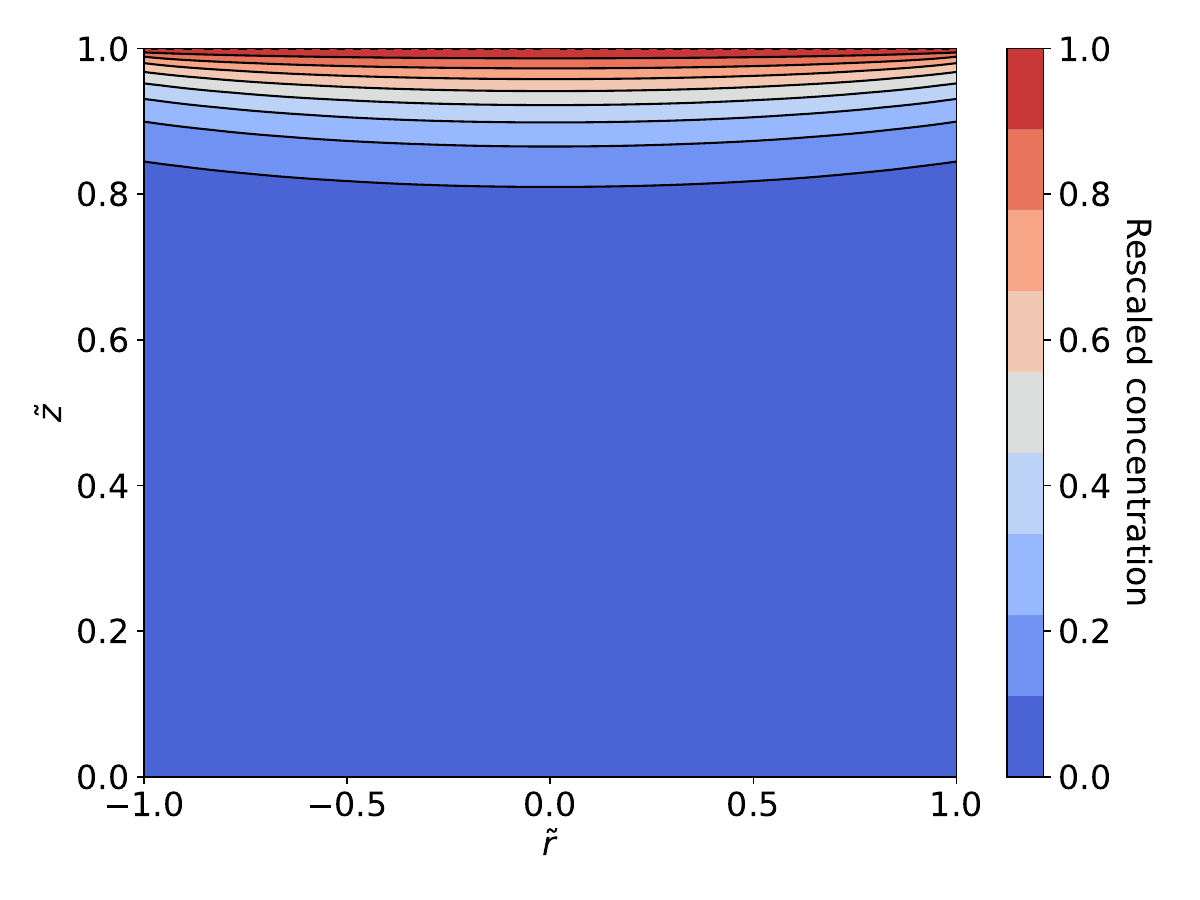}
  \caption{$\xi=0.1$ and $\Theta=0.1$}
\end{subfigure}%
\hfill
\begin{subfigure}{.32\textwidth}
  \centering
  \includegraphics[width=.9\linewidth]{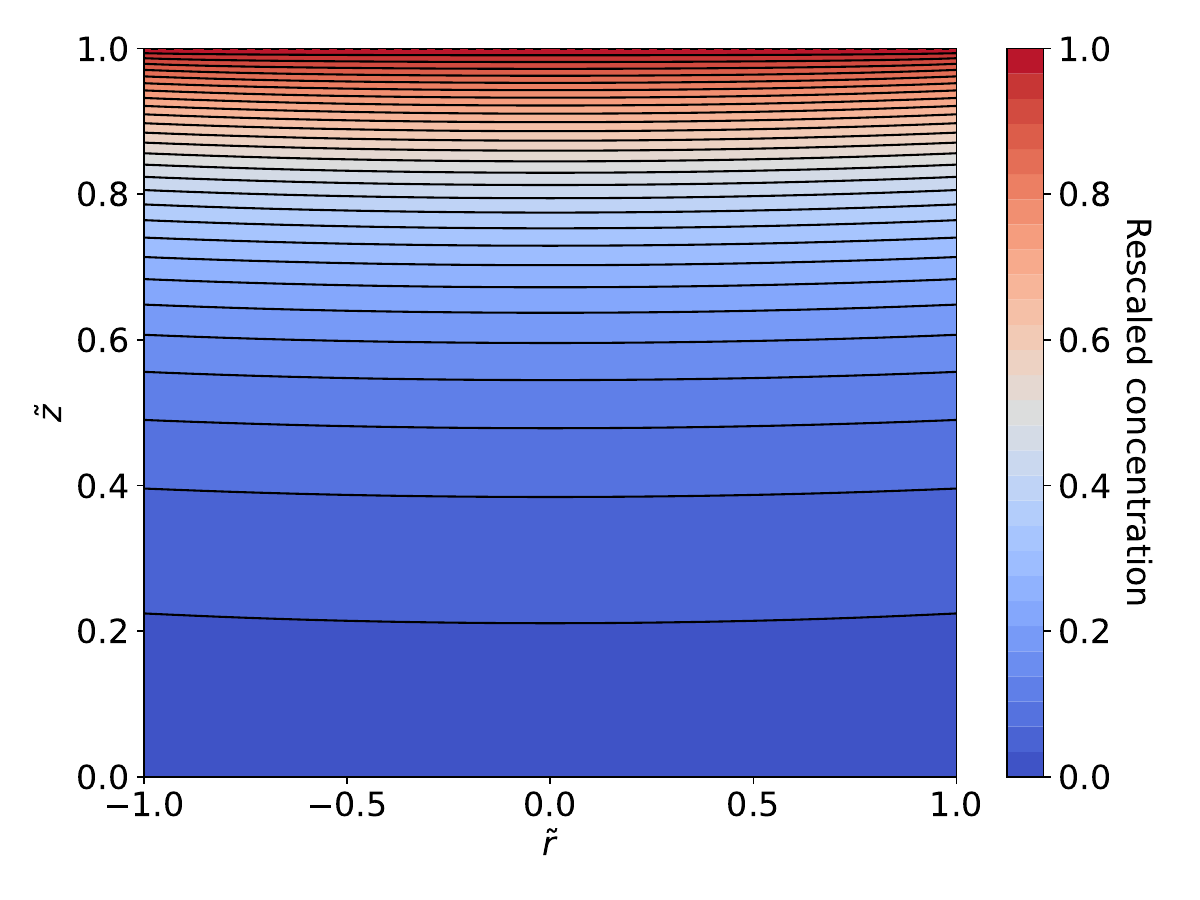}
  \caption{$\xi=0.1$ and $\Theta=1$}
\end{subfigure}
\hfill
\begin{subfigure}{.32\textwidth}
  \centering
  \includegraphics[width=.9\linewidth]{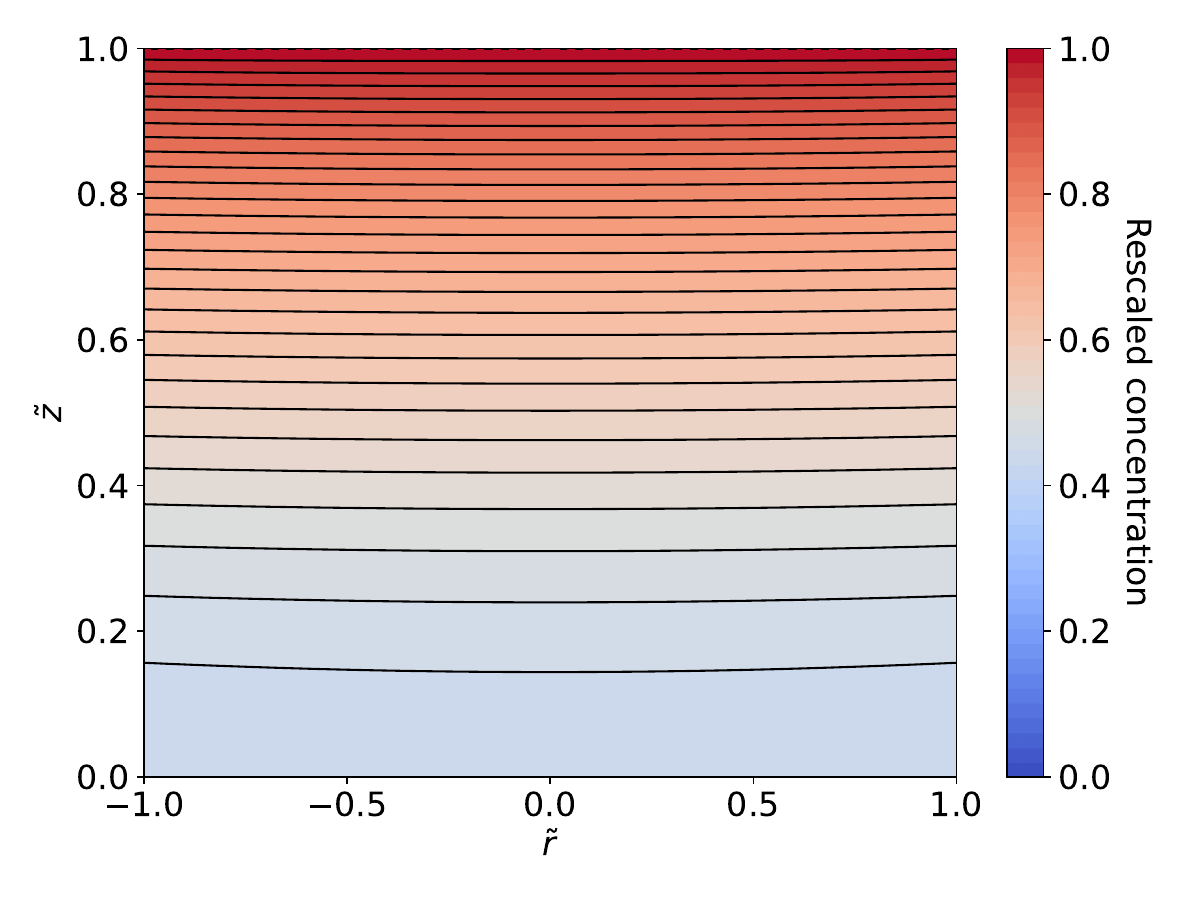}
  \caption{$\xi=0.1$ and $\Theta=10$}
\end{subfigure}
\begin{subfigure}{.32\textwidth}
  \centering
  \includegraphics[width=.9\linewidth]{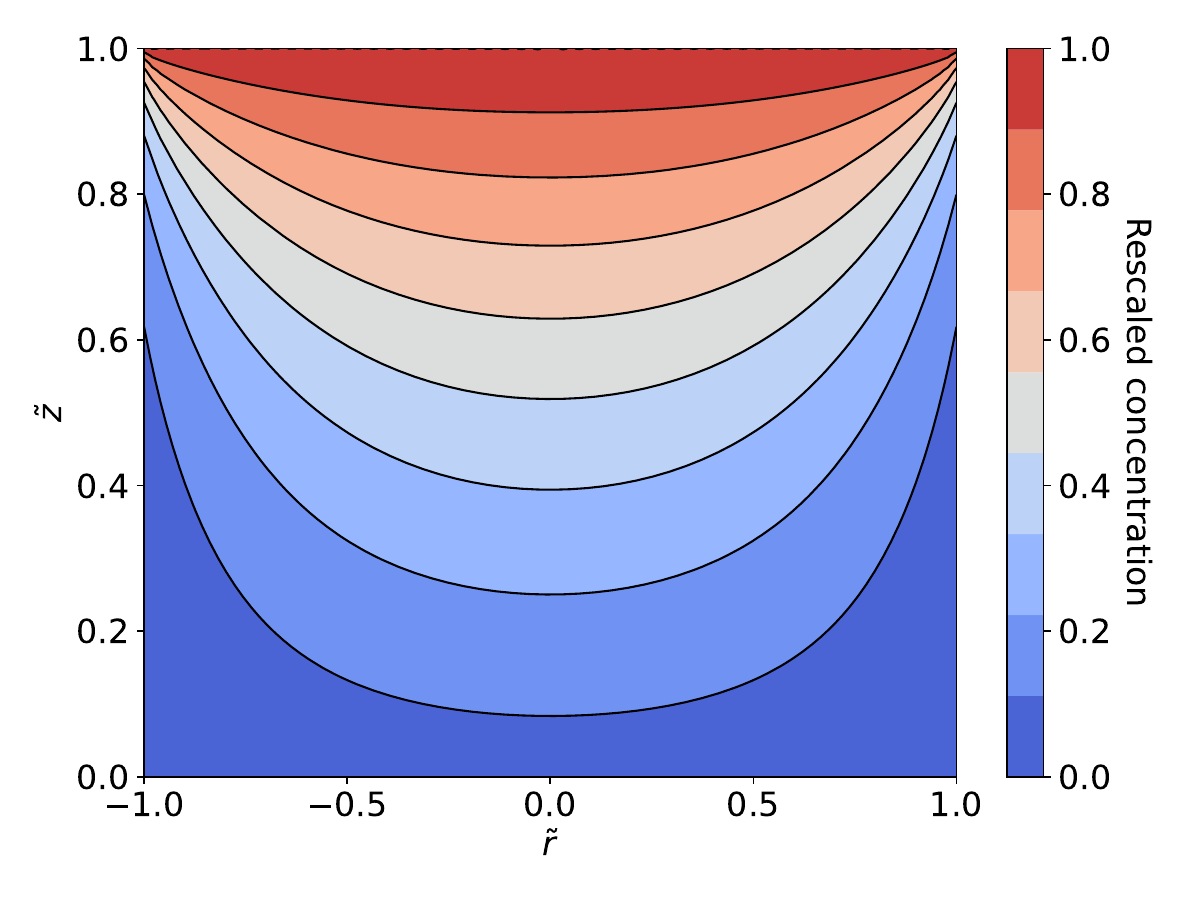}
  \caption{$\xi=1$ and $\Theta=0.1$}
\end{subfigure}%
\hfill
\begin{subfigure}{.32\textwidth}
  \centering
  \includegraphics[width=.9\linewidth]{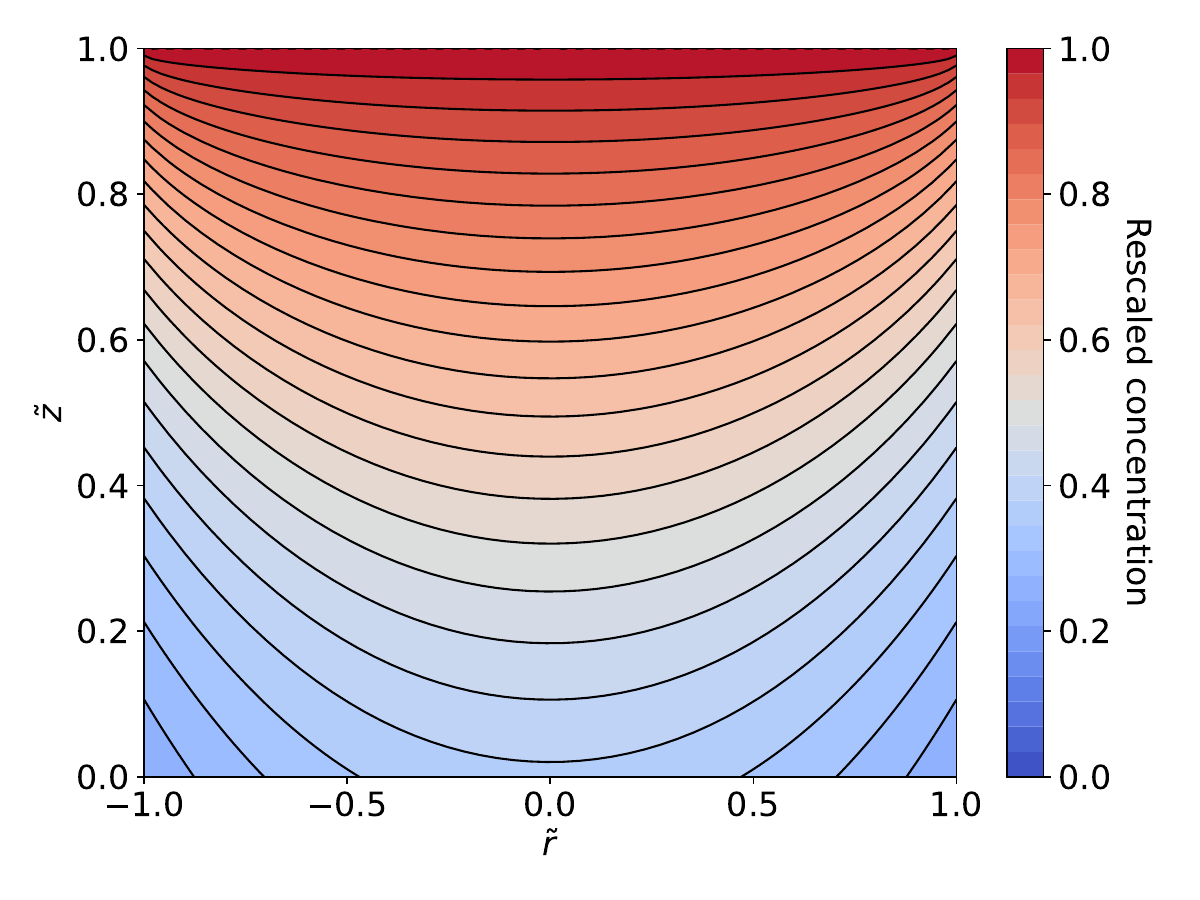}
  \caption{$\xi=1$ and $\Theta=1$}
\end{subfigure}
\hfill
\begin{subfigure}{.32\textwidth}
  \centering
  \includegraphics[width=.9\linewidth]{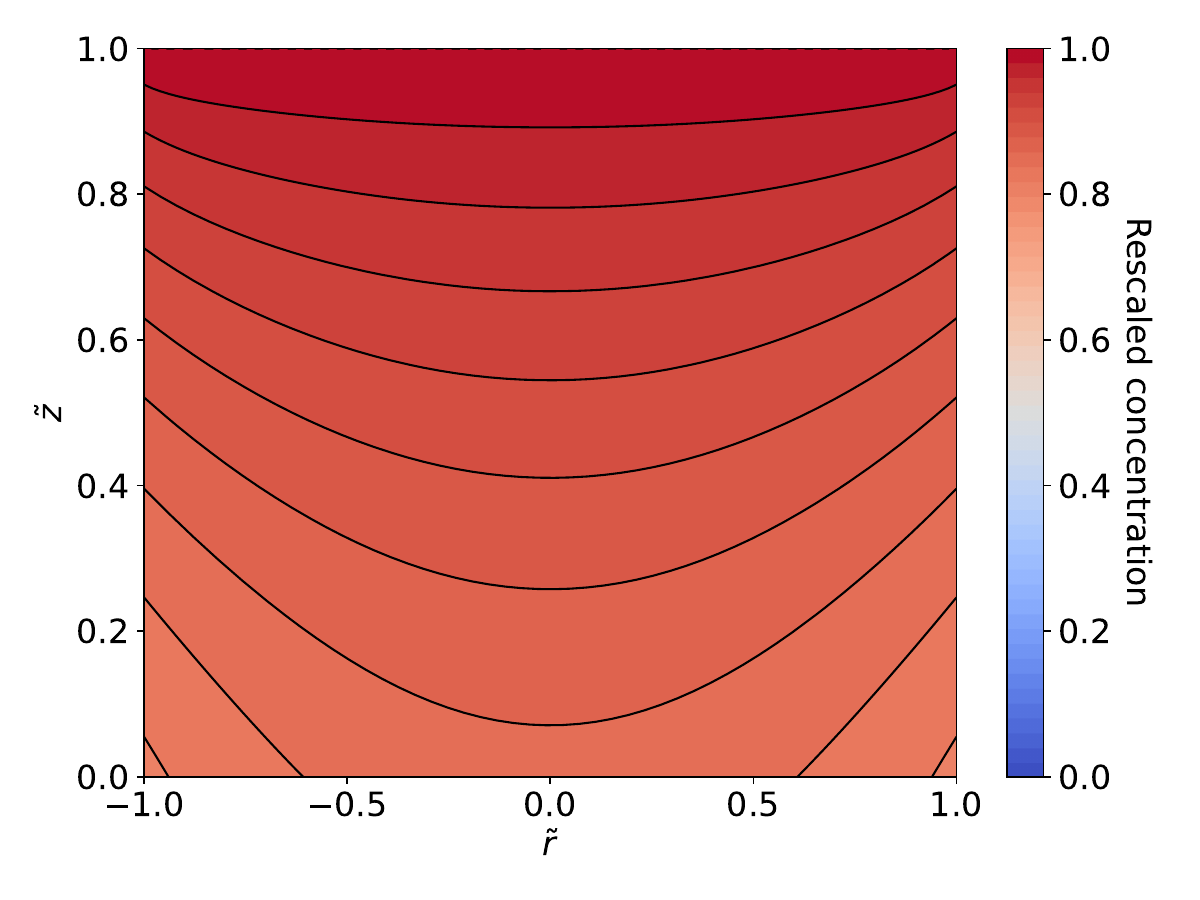}
  \caption{$\xi=1$ and $\Theta=10$}
\end{subfigure}
\begin{subfigure}{.32\textwidth}
  \centering
  \includegraphics[width=.9\linewidth]{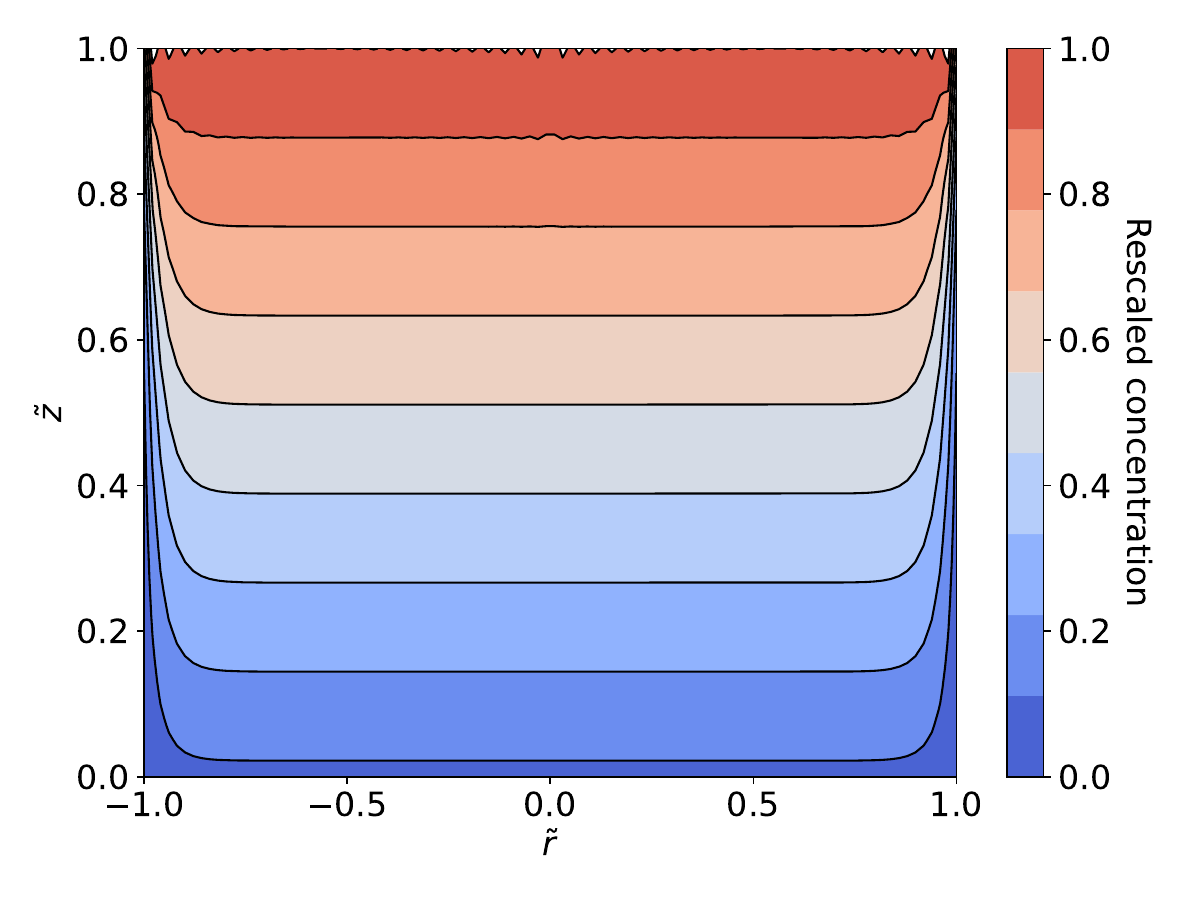}
  \caption{$\xi=10$ and $\Theta=0.1$}
\end{subfigure}%
\hfill
\begin{subfigure}{.32\textwidth}
  \centering
  \includegraphics[width=.9\linewidth]{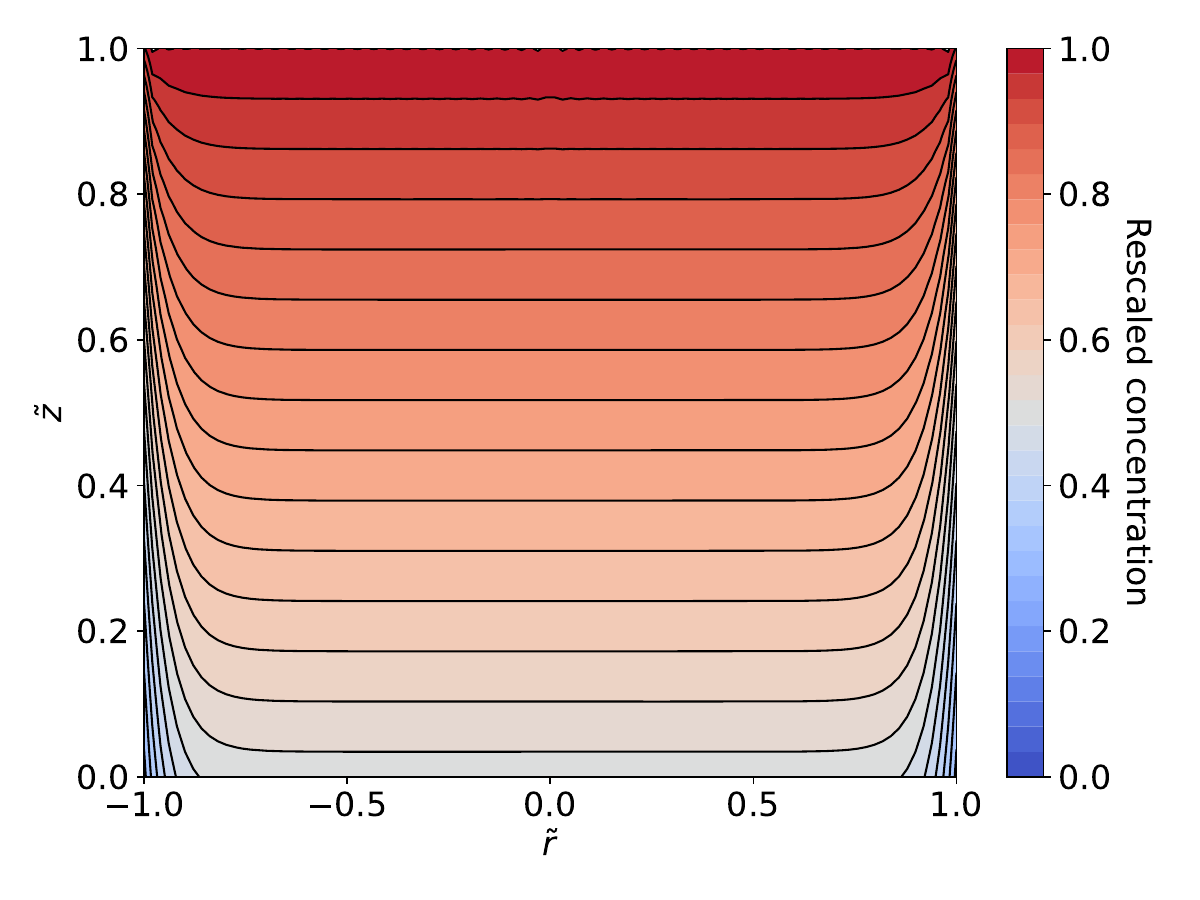}
  \caption{$\xi=10$ and $\Theta=1$}
\end{subfigure}
\hfill
\begin{subfigure}{.32\textwidth}
  \centering
  \includegraphics[width=.9\linewidth]{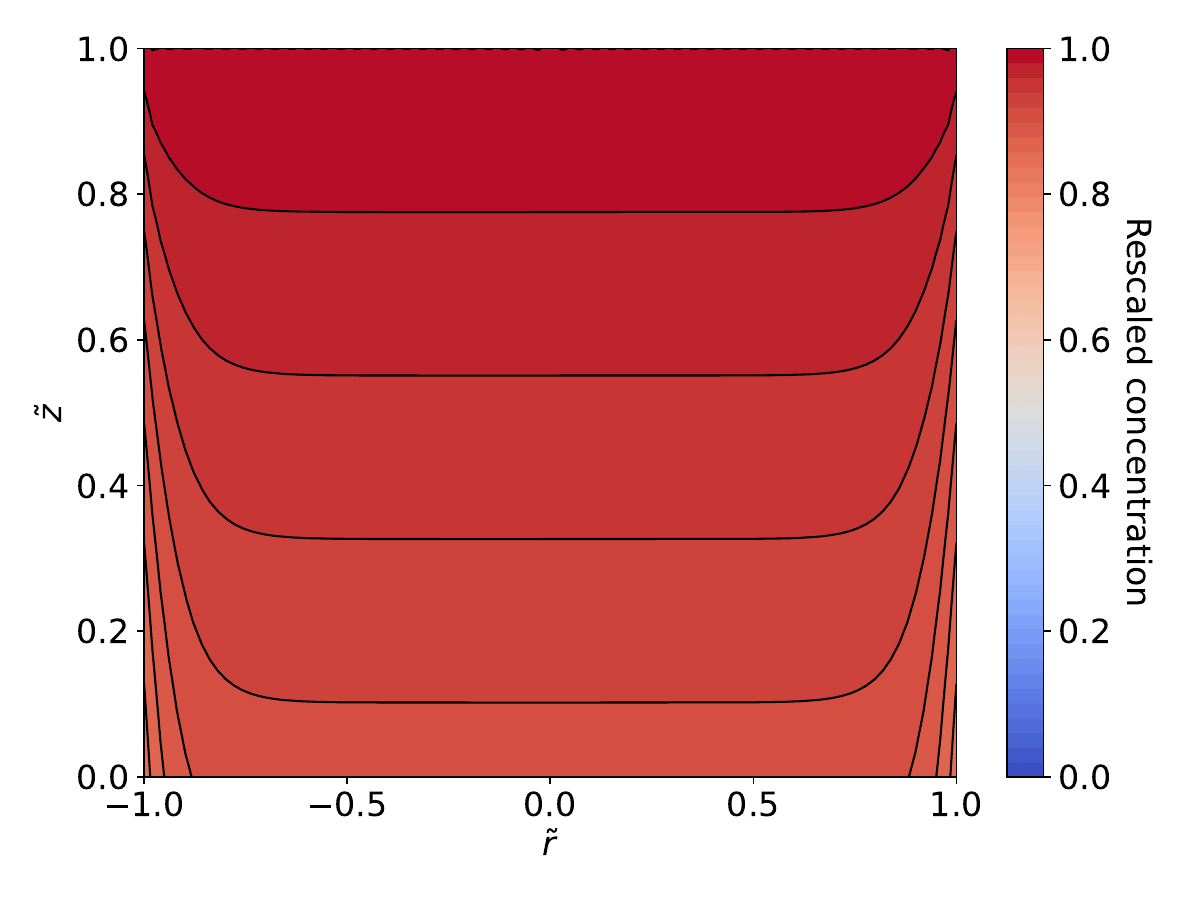}
  \caption{$\xi=10$ and $\Theta=10$}
\end{subfigure}
\caption{Contour plots of the rescaled concentration within a crypt like in figure~\ref{fig:mainfig}. The top corresponds to the lumen while the bottom corresponds to the bottom of the crypt. The left and right boundaries represent the sides of the crypt in cylindrical geometry. Various values of the rescaled crypt radius $\xi$ and the inverse rescaled surface absorptivity $\Theta$ are used. The exact solutions are infinite sums. Numerically, we went up to $n=100$, which was sufficient. The colors correspond to the rescaled concentration within this space (the lumen concentration is set to 1).}\label{fig:concentration_crypt}
\end{figure}

This expression is plotted in figure~\ref{fig:concentration_crypt}.

The expression of nutrient concentration flux integrated over one intervillous space is the following:

\begin{equation}
    J=-D\int \int_S\vec n \nabla C(r,z)d\vec S
\end{equation}

In dimensionless units this equation becomes:

\begin{equation}
  \tilde J=-2\pi(\int_0^1\frac{\partial c}{\partial \tilde r}\Bigr|_{\substack{\tilde r=1}}d\tilde z - \xi^2 \int_0^1 \tilde r\frac{\partial c}{\partial \tilde z}\Bigr|_{\substack{\tilde z=0}}d\tilde r)
\label{eq:flux_crypt_integral}
\end{equation}

With $\tilde J=J/(Dc_0h)$. The first term is the absorption on the walls and the second term is the absorption at the bottom of the crypt. Integrating each term gives the following expression of the rescaled flux:

\begin{equation}
    \tilde J = \sum_{n=1}^\infty \frac{4\pi J_1^2\left(\mu_n\right)\left(\Theta \frac{\mu_n}{\xi} \sinh\left(\frac{\mu_n}{\xi}\right)+\cosh\left(\frac{\mu_n}{\xi}\right)\right)}{\mu_n\left(J_0^2\left(\mu_n\right)+J_1^2\left(\mu_n\right)\right)\left(\Theta \frac{\mu_n}{\xi} \cosh\left(\frac{\mu_n}{\xi}\right)+\sinh\left(\frac{\mu_n}{\xi}\right)\right)}
\label{eq:flux_crypt}
\end{equation}

\begin{figure}[h]
    \centering
    \includegraphics[width=1\linewidth]{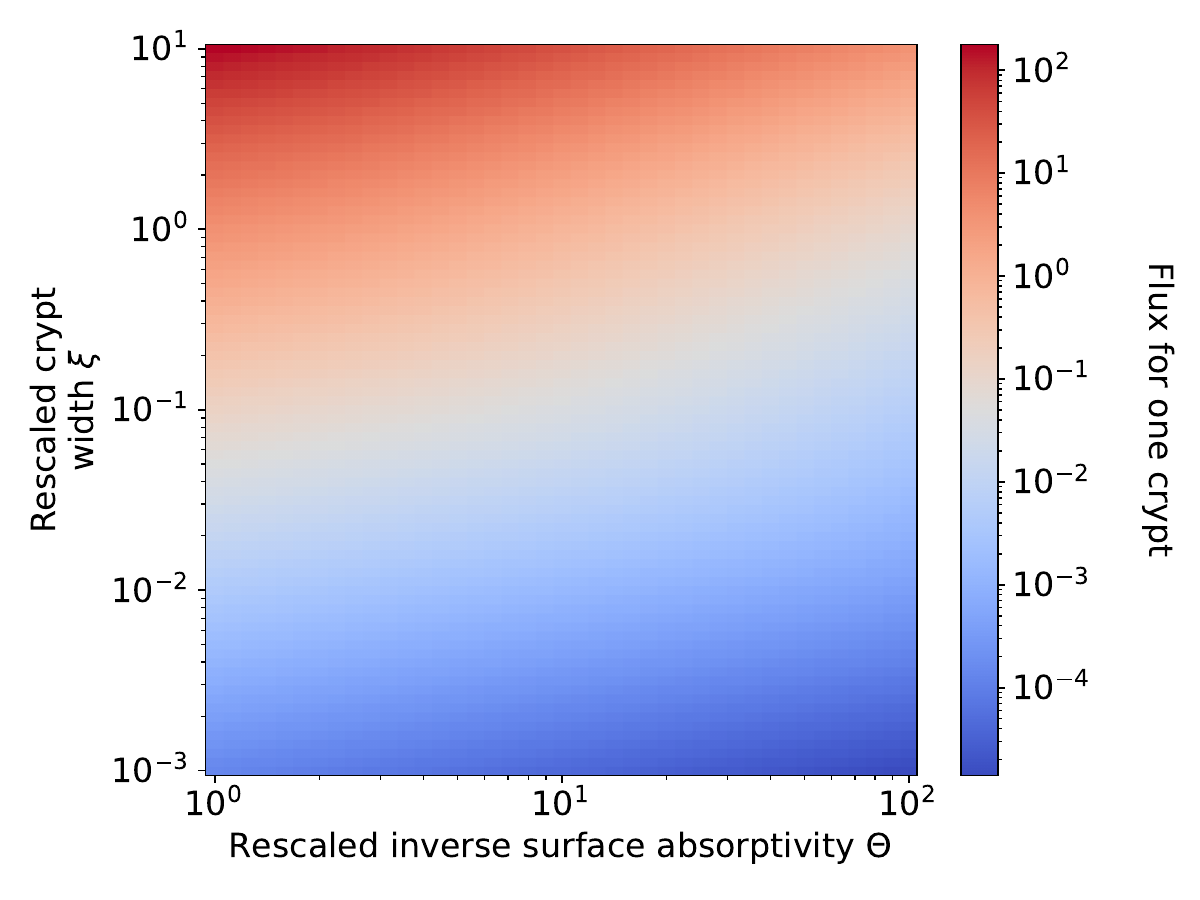}
    \caption{Non-dimensionalized flux for one crypt for the crypt-like geometry for a wide range of $\xi$ and $\Theta$ in log-scale.}\label{figfluxcrypt}
\end{figure}

\begin{figure}[h]
\begin{subfigure}{.5\textwidth}
  \centering
  \includegraphics[width=.95\linewidth]{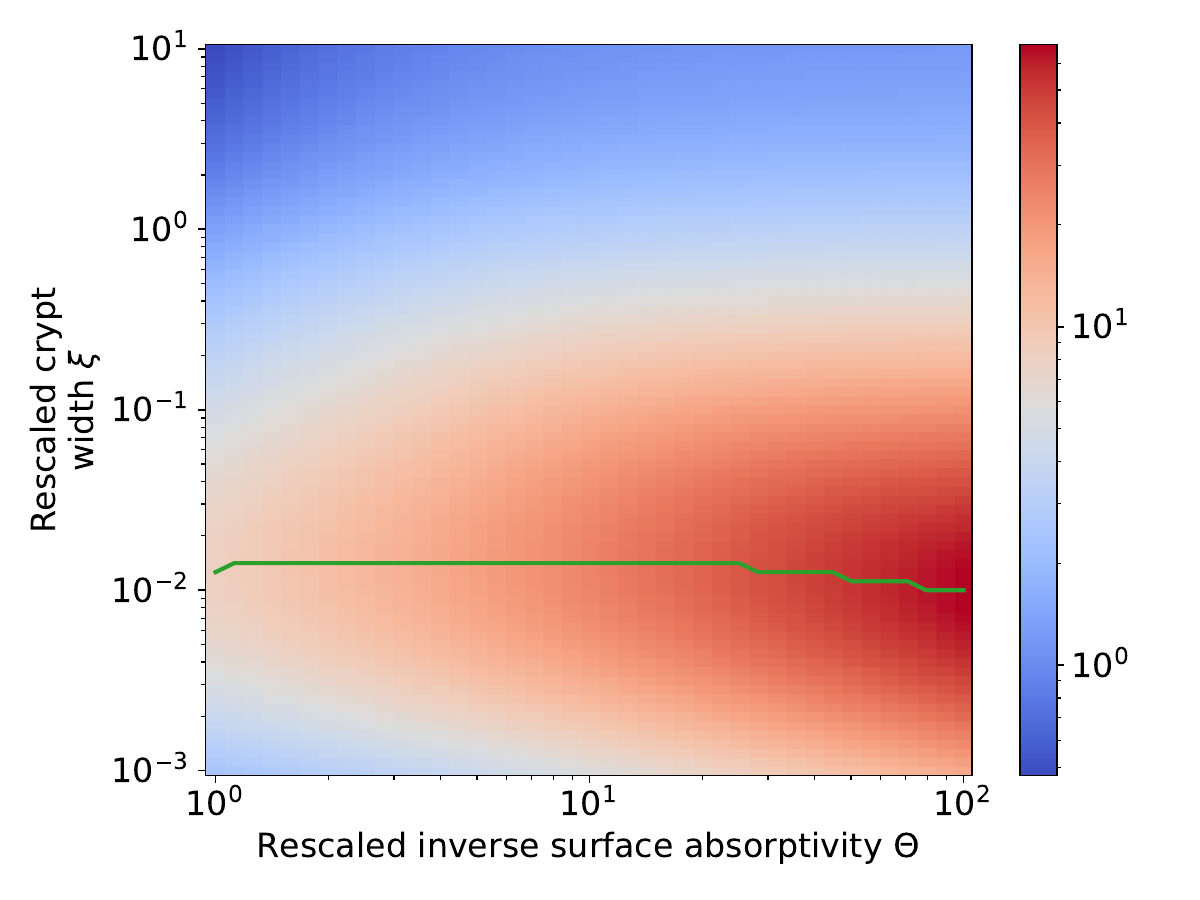}
  \caption{$\tilde e=0.01$}
\end{subfigure}%
\begin{subfigure}{.5\textwidth}
  \centering
  \includegraphics[width=.95\linewidth]{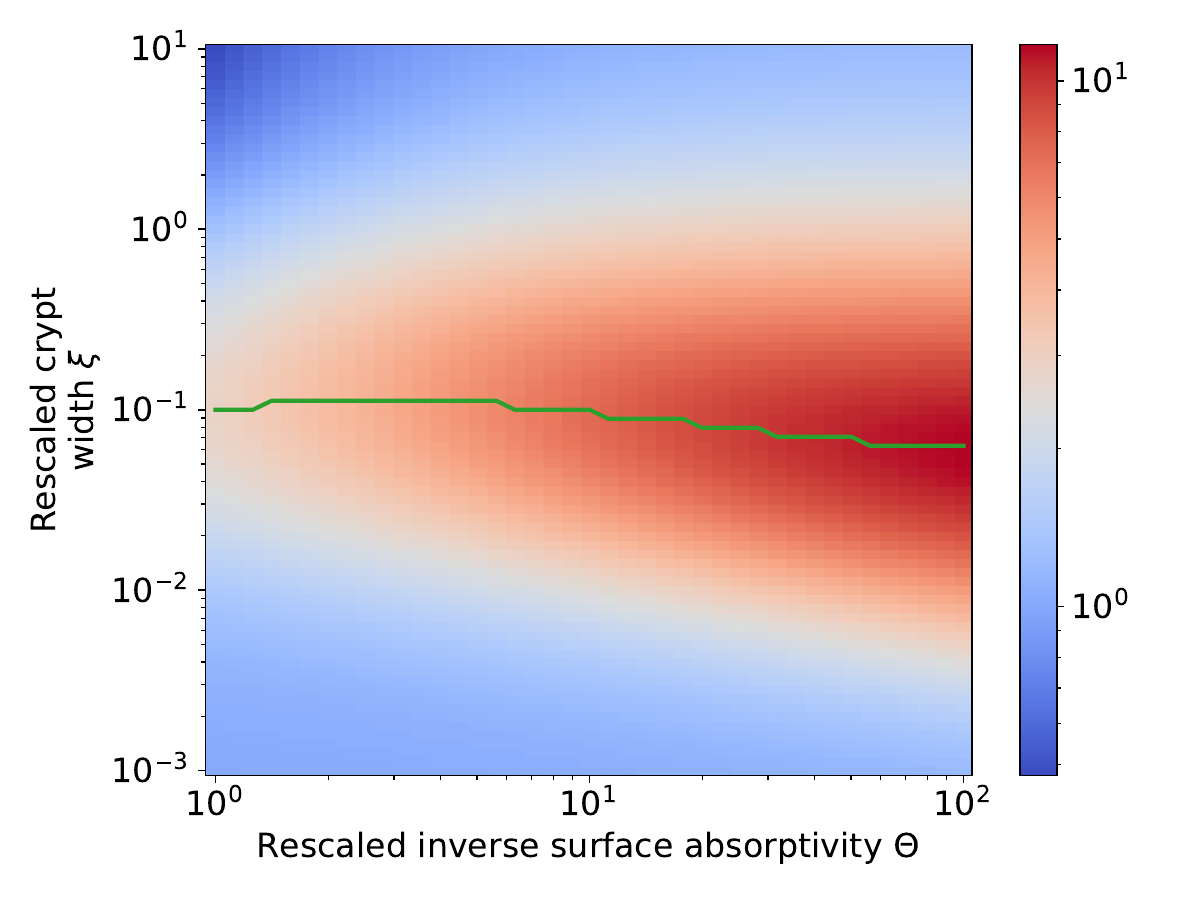}
  \caption{$\tilde e=0.1$}
\end{subfigure}
\begin{subfigure}{.5\textwidth}
  \centering
  \includegraphics[width=.95\linewidth]{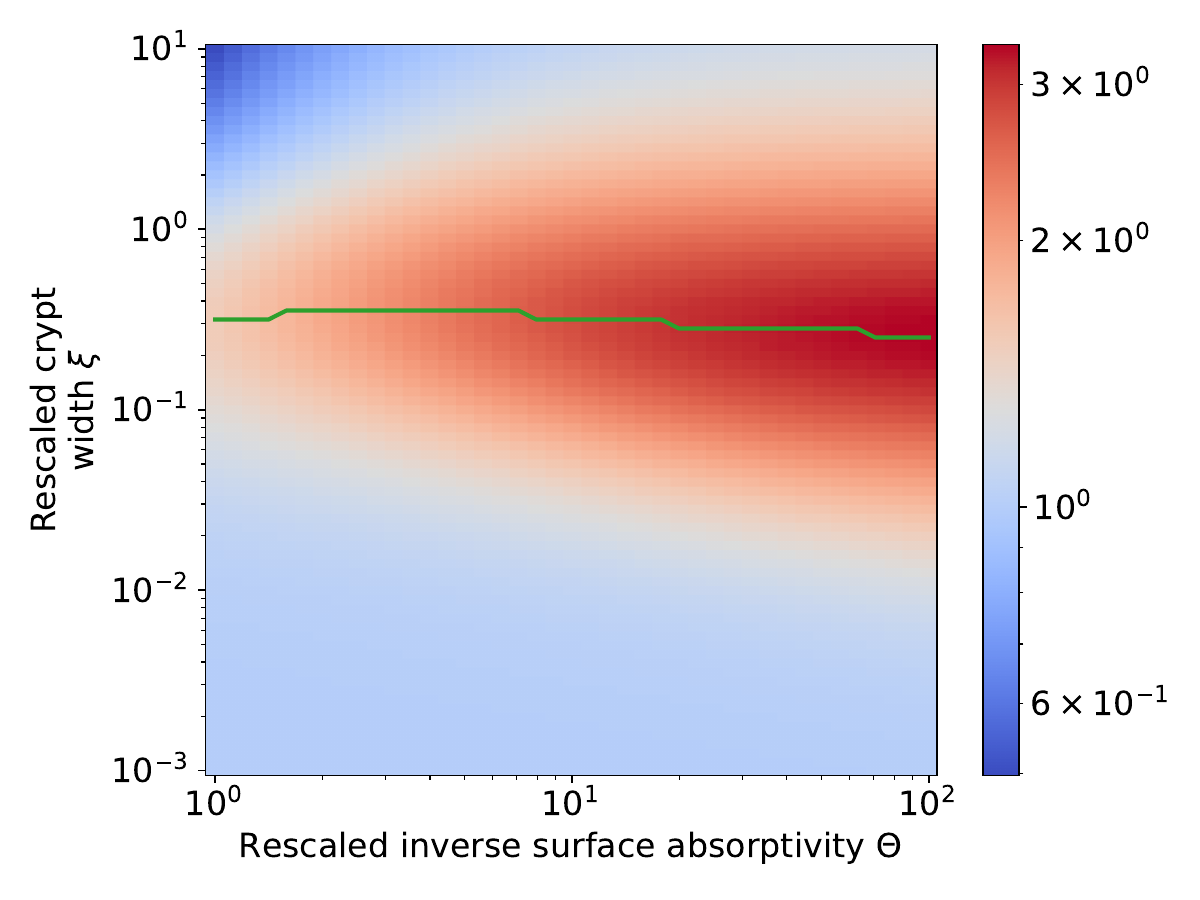}
  \caption{$\tilde e=0.5$}
\end{subfigure}%
\begin{subfigure}{.5\textwidth}
  \centering
  \includegraphics[width=.95\linewidth]{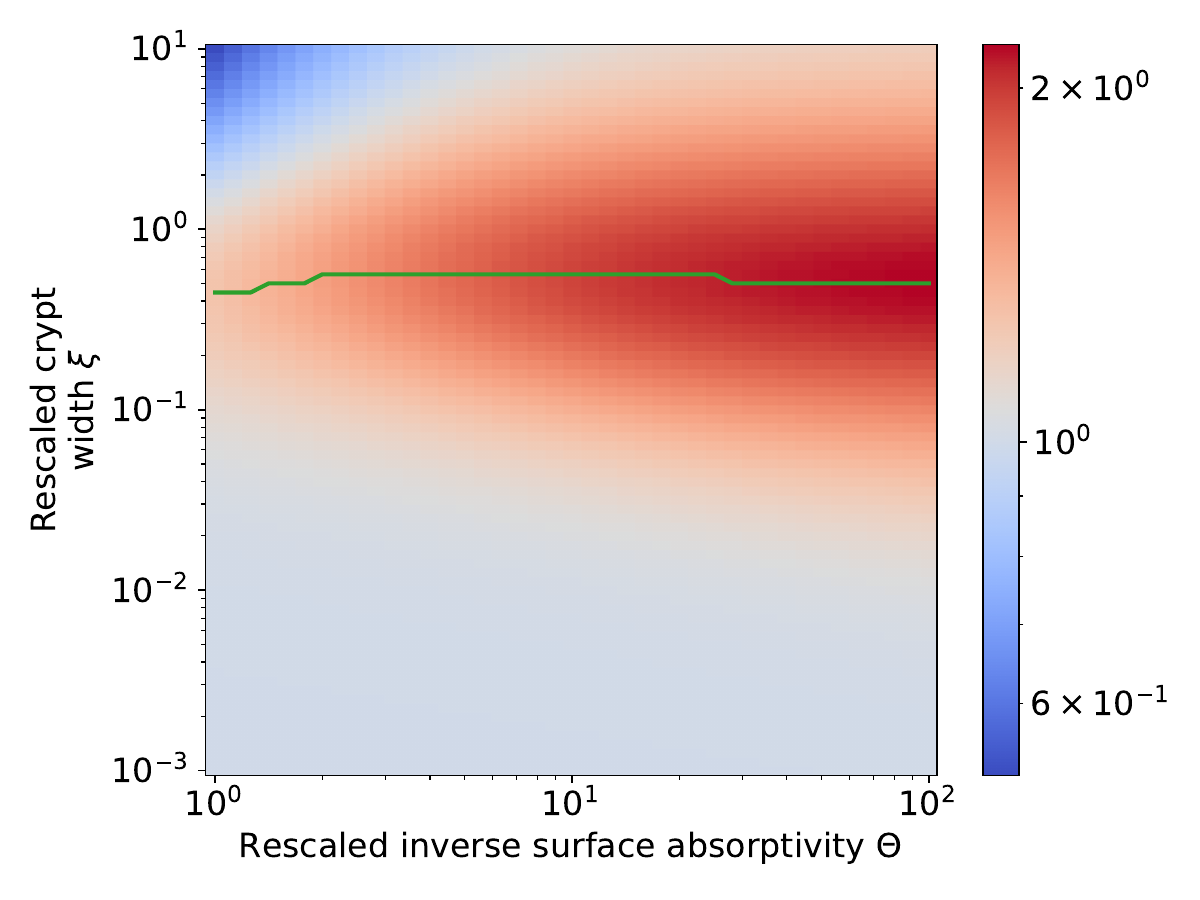}
  \caption{$\tilde e=1$}
\end{subfigure}

\caption{Rescaled log flux density per unit of gut length $\tilde j$ for the crypt geometry as a function of $\xi$ (the rescaled distance between two crypts) and $\Theta$ (the rescaled inverse surface absorptivity) for $\tilde e=0.01,0.1,0.5,1$ (the rescaled crypt radius). The green line represents the value of $\xi$ that maximizes absorption flux at a given value of $\Theta$.}\label{figfluxresccrypt}
\end{figure}

It is plotted in figure~\ref{figfluxcrypt}. As explained in the main text, this result is then used to calculate the flux per unit of gut length, plotted in figure~\ref{figfluxresccrypt}.

\clearpage

\section{Methods for collecting physiological data}

\begin{table}[t!]
\centering
\begin{tabular}{ |c |c | c|c| }
 \hline
 Species & Section & Source & Direct measures or images\\
 &&  &within the source\\
 \hline
 Human    & villi J & \cite{marsh_study_1969}  & Figures 3 and 7 \\
 Rat      & villi D &  \cite{seyyedin_histomorphometric_2017}& Table 1 and Figure 1 \\
 Rat      & villi D & \cite{fujimiya_serotonin-containing_1991}  & Figure 1  \\
 Mouse    & villi J &   \cite{abbas_internal_1989} & Figure 1 and results section text \\
 Piglet   & villi D   & \cite{skrzypek_mechanisms_2018} & Figures 1 and 2 \\
 Pheasant & villi J & \cite{goodarzi_histology_2021}  & Figures 3 and 5 \\
 Horse (au)& villi J&\cite{roberts_mucosa_1974}  & Figures 5 and 7 \\
 Chicken  & villi D & \cite{samanya_histological_2002} & Figures 1 and 2 \\
 Human    & crypts&\cite{qi2008automated} & Figures 3 and 5 \\
 Mouse    &crypts &\cite{farkas2015cryosectioning} & Figure 2 \\
 \hline
\end{tabular}
    \caption{Summary of sources and corresponding figures or tables used for extracting geometrical measurements of villi and crypts. D stands for duodenum, J for jejunum}\label{tablesup}
\end{table}

Here we present briefly the methods used to collect data for Table~\ref{tab:physiological_data} of the main text. We use published physiological data. When the quantities we are interested in are already measured in the source article, we directly use the data. In other cases, we exploit the microscopy images, measuring in ImageJ~\cite{schneider2012nih}, and converting the pixels into real distances using the scale bars. In table~\ref{tablesup}, we give the list of sources, with for each source where we took the relevant information from.  

For finger-like villi there is a conversion to make. We measure the villi diameter $e$ and the distance between two villi $2R_2$, villi arrangement is approximately hexagonal. As shown in figure~\ref{fig:hexagon_to_circle}, we approximate it by a cylindrical geometry, with $R_1=\sqrt{\frac{3\sqrt{3}}{2\pi}}R_2$ giving the same surface, and $R_1 = e/2 + R$. As $e$ is directly measured, this expression gives $R$ which is reported in the table from the main text.

For the horse data, there was no scale on the images but we could get the measurments of the different lengths on the same images. So the ratio are correct. The length in pixels were $h=775\pm 20$, $e=400\pm 5$, and $2R=80\pm 10$.


\section{Absorption by unit length}

\begin{figure}[h]
\begin{subfigure}{.32\textwidth}
  \centering
  \includegraphics[width=.9\linewidth]{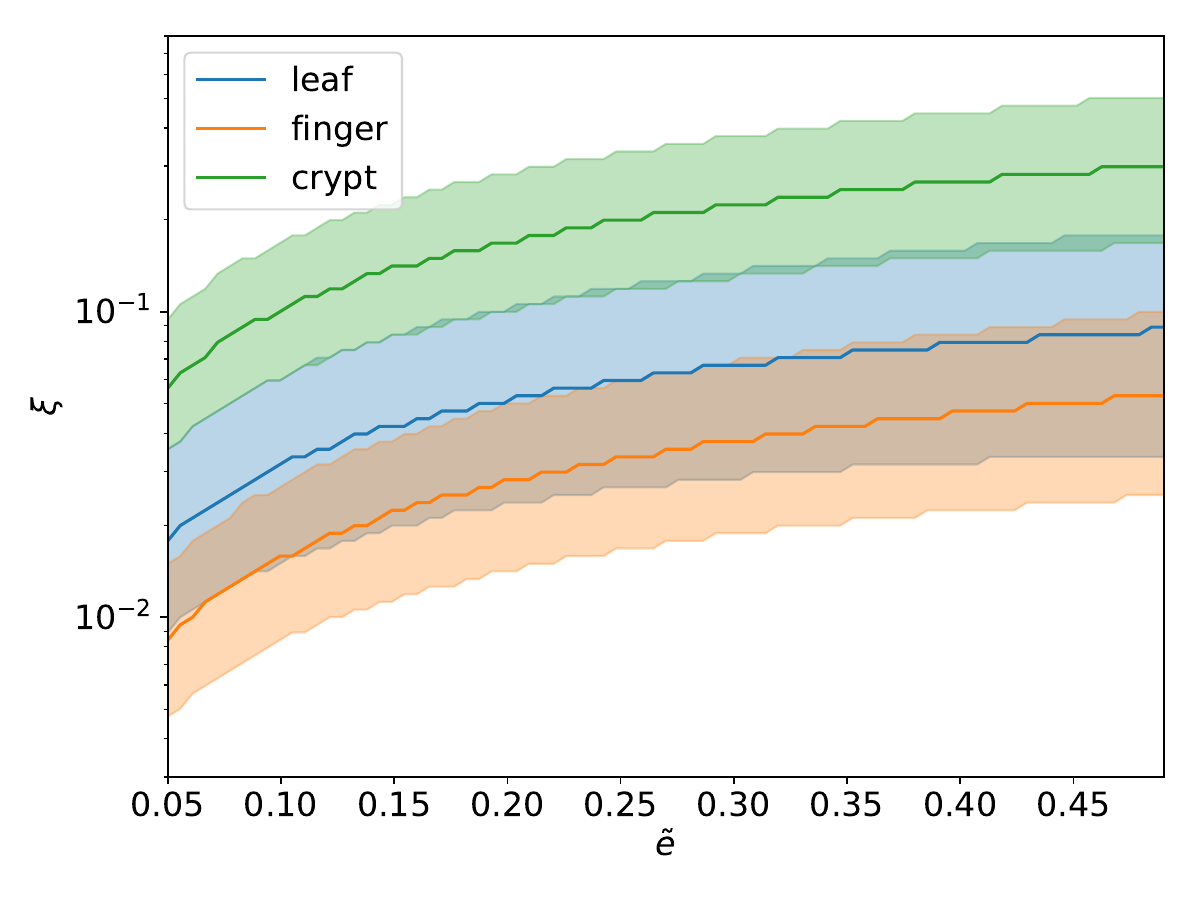}
  \caption{$\Theta=1$}
\end{subfigure}%
\begin{subfigure}{.32\textwidth}
  \centering
  \includegraphics[width=.9\linewidth]{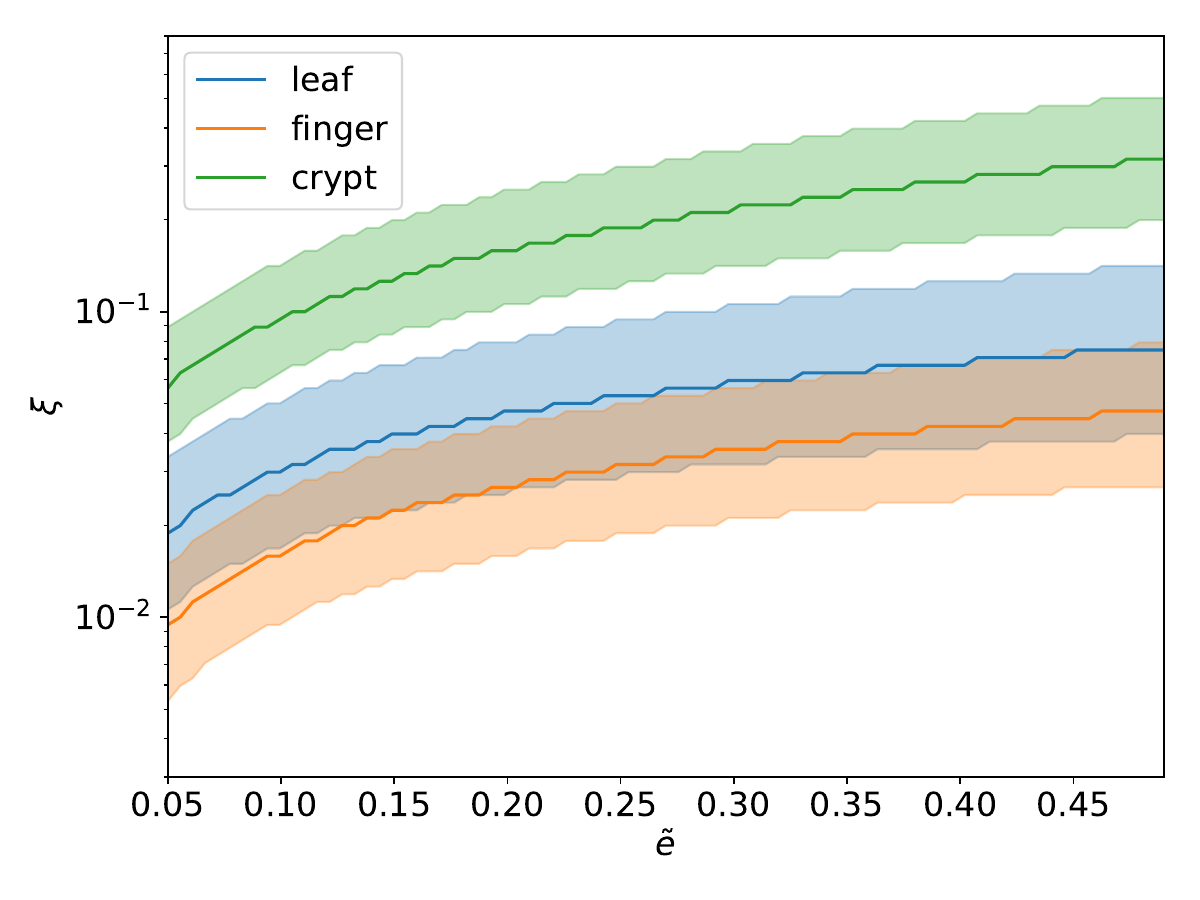}
  \caption{$\Theta=10$}
\end{subfigure}%
\begin{subfigure}{.32\textwidth}
  \centering
  \includegraphics[width=.9\linewidth]{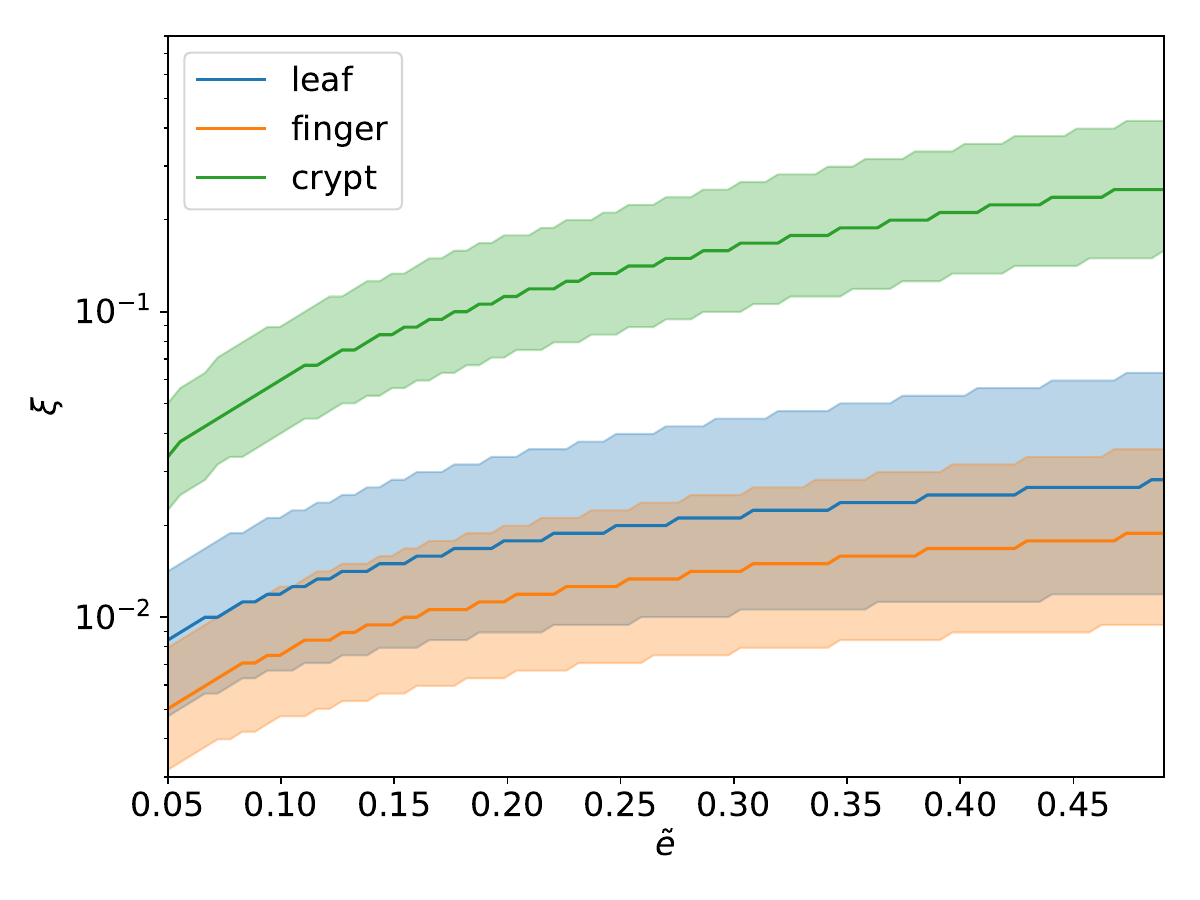}
  \caption{$\Theta=100$}
\end{subfigure}

\caption{Plot of the rescaled intervillous distance or crypt radius maximizing absorption per unit length as a function of the rescaled villi width or intercrypt distance for the three geometries for several values of $\Theta$, the rescaled inverse surface absorptivity. The shaded regions correspond to 95\% of the optimum for each geometry.}\label{rescaledabs}
\end{figure}

Figure~\ref{rescaledabs} shows the optimal geometry for different $\Theta$.

\begin{figure}[h]
\centering
\includegraphics[width=1\linewidth]{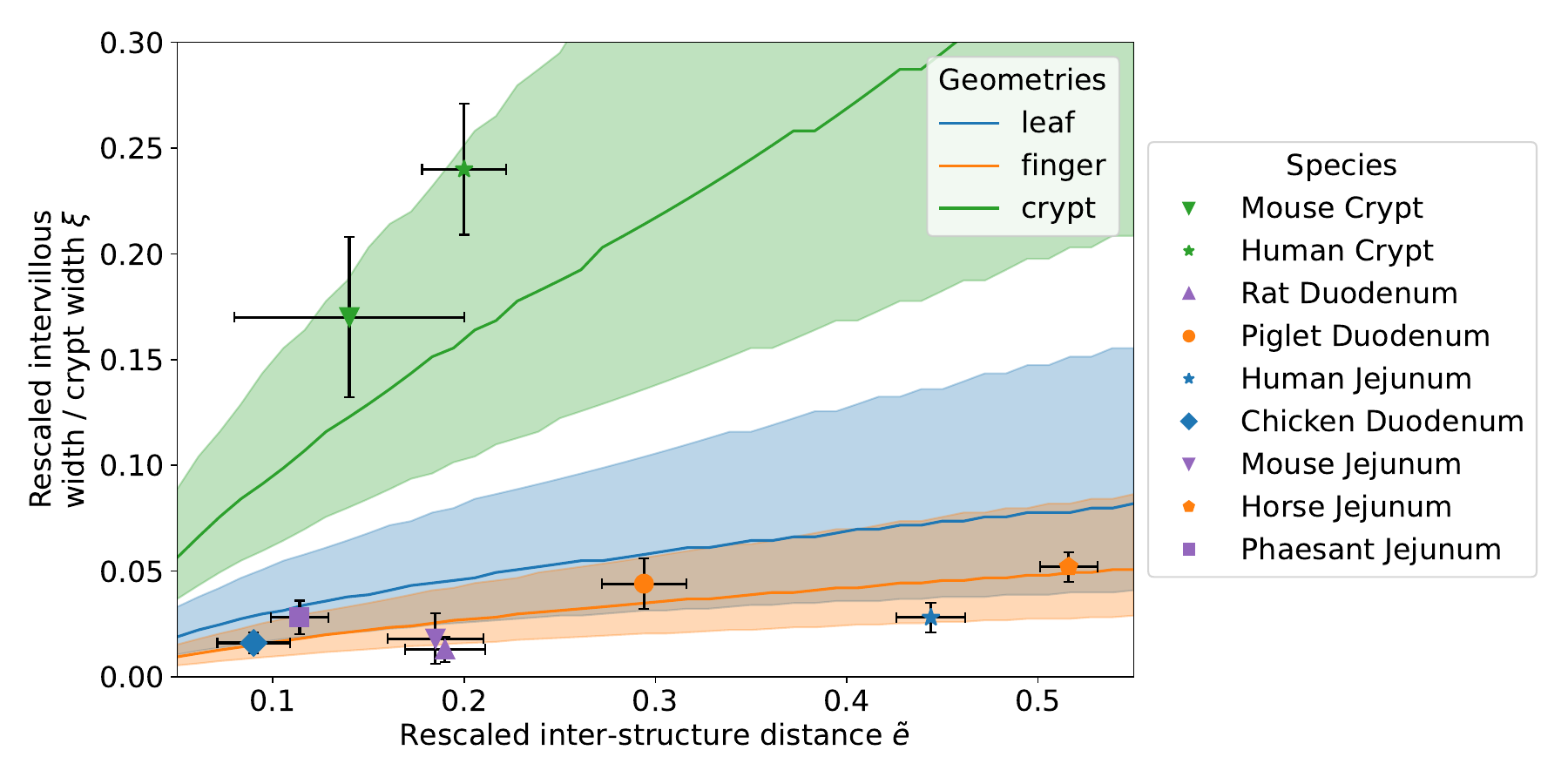}
\caption{Linear-scale version of figure~\ref{fig:CompareAllAnalVC_Log}: rescaled intervillous width / crypt radius $\xi$ maximizing absorption per unit of gut length as a function of the rescaled inter-structure distance $\tilde e$, at $\Theta=10$, for the three geometries. The shaded regions correspond to $95\%$ of the maximal absorption for each geometry. Data points show physiological data from various species (Table~\ref{tab:physiological_data}), with the same colour coding as in figure~\ref{fig:CompareAllAnalVC_Log}.}\label{fig:CompareAllAnalVC_LinLin}
\end{figure}

Figure~\ref{fig:CompareAllAnalVC_LinLin} reproduces main-text figure~\ref{fig:CompareAllAnalVC_Log} on a linear vertical axis instead of a logarithmic one. The agreement between the physiological measurements and the predicted $95\%$-optimal regions is unchanged; the linear axis makes the absolute spread of the optimal $\xi$ across geometries more directly readable.

\begin{figure}[h]
\centering
\includegraphics[width=1\linewidth]{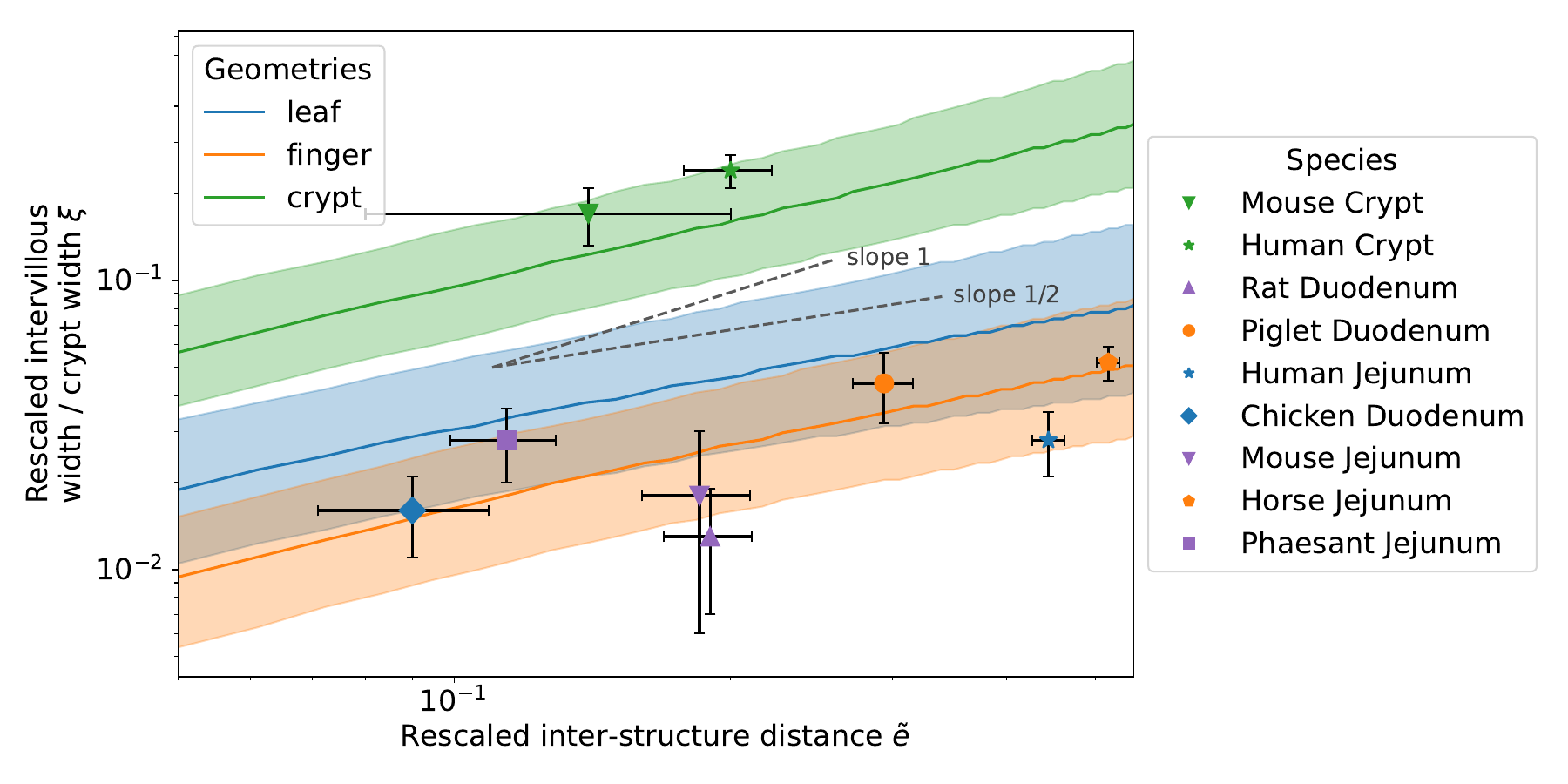}
\caption{Log-log version of figure~\ref{fig:CompareAllAnalVC_Log}: rescaled intervillous width / crypt radius $\xi$ maximizing absorption per unit of gut length as a function of the rescaled inter-structure distance $\tilde e$, at $\Theta=10$, for the three geometries. The shaded regions correspond to $95\%$ of the maximal absorption for each geometry. Data points show physiological data from various species (Table~\ref{tab:physiological_data}), with the same colour coding as in figure~\ref{fig:CompareAllAnalVC_Log}.}\label{fig:CompareAllAnalVC_LogLog}
\end{figure}

Figure~\ref{fig:CompareAllAnalVC_LogLog} shows the same data on log-log axes. Over the physiological range of $\tilde e$ the optimal $\xi$ grows sublinearly with $\tilde e$ for the three geometries, but the range is too narrow to extract a robust power law.

\section{Experimental images}\label{sec:experimental_images}

To illustrate the different geometries of the villi, we took images of the inner structures of the rat small intestine.

An adult Male Wistar rat (350g) was euthanized via intraperitoneal injection of pentobarbital (180 mg/kg). A midline abdominal incision was made to access the small intestine. The first 7 cm distal to the stomach (corresponding to the duodenum and jejunum) and the final 4 cm proximal to the cecum (corresponding to the ileum) were dissected and immersed in Tyrode buffer at room temperature (pH=7.4). The 7 cm segment was then divided into two portions representing the duodenum and jejunum. All three segments were flushed with the Tyrode buffer to remove intestinal contents. They were then cut along the mesenteric border to expose the lumen. Tissues were pinned in a Petri dish previously coated with PDMS (polydimethylsiloxane), and images were captured using a Basler acA4112-20uc camera equipped with 0.75x lenses from Opto Engineering. The pictures were edited to enhance contrast using Adobe Lightroom Classic CC.

In the ileum, the villi are more leaf-like, whereas in the duodenum, they have a more intermediate shape.

\begin{figure}
    \centering
    \includegraphics[width=\linewidth]{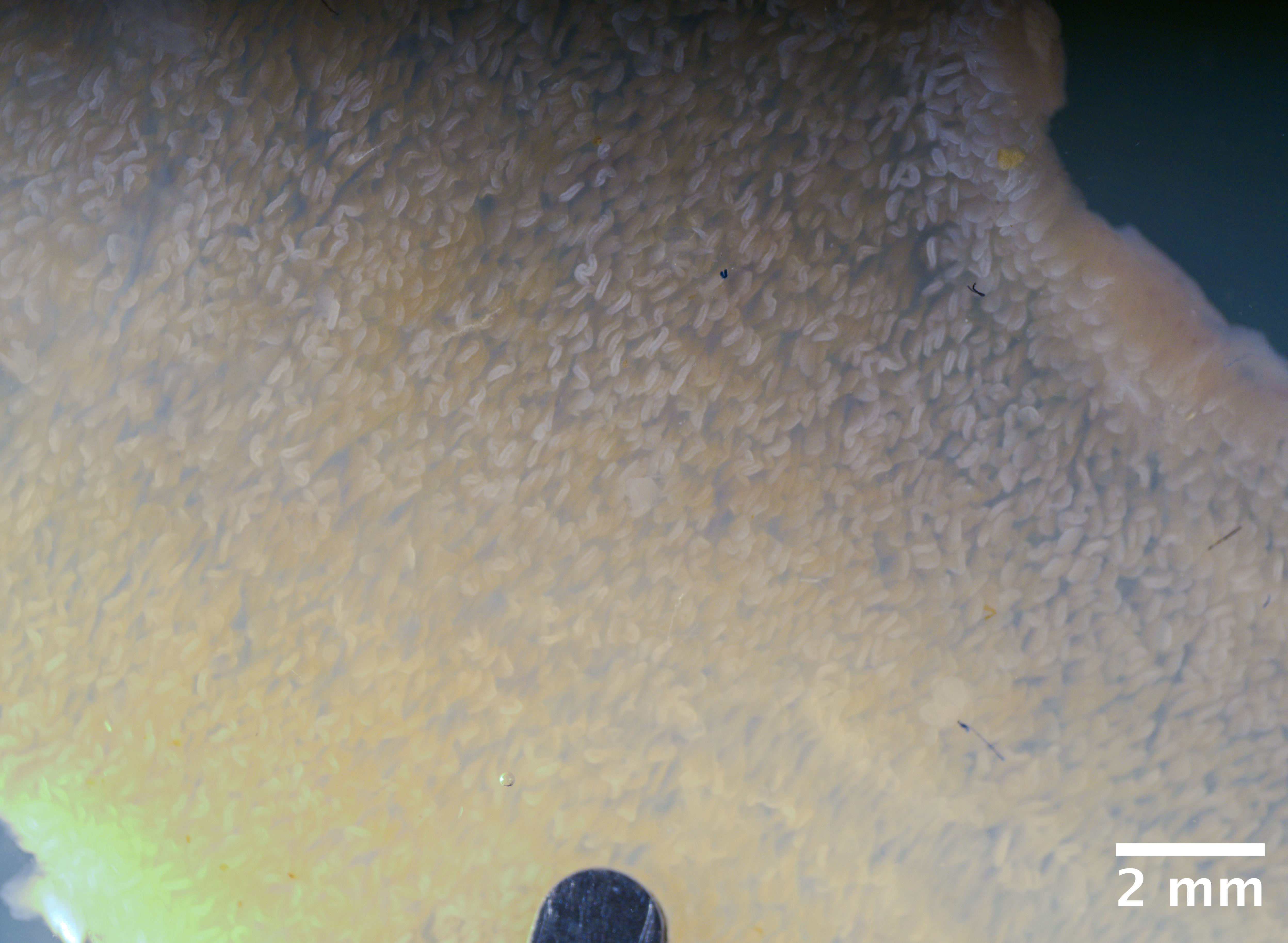}
    \caption{An image of the inner surface of a rat duodenum}\label{duo2}
\end{figure}

\begin{figure}
    \centering
    \includegraphics[width=\linewidth]{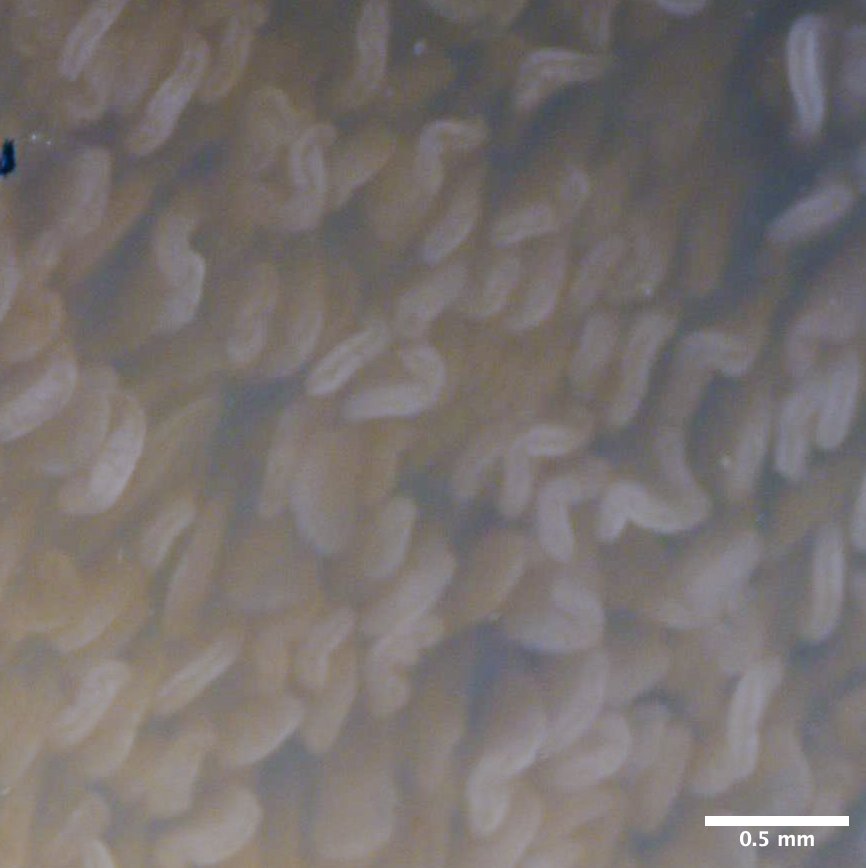}
    \caption{Zoom of an image of the inner surface of a rat duodenum}\label{duoz}
\end{figure}

\begin{figure}
    \centering
    \includegraphics[width=\linewidth]{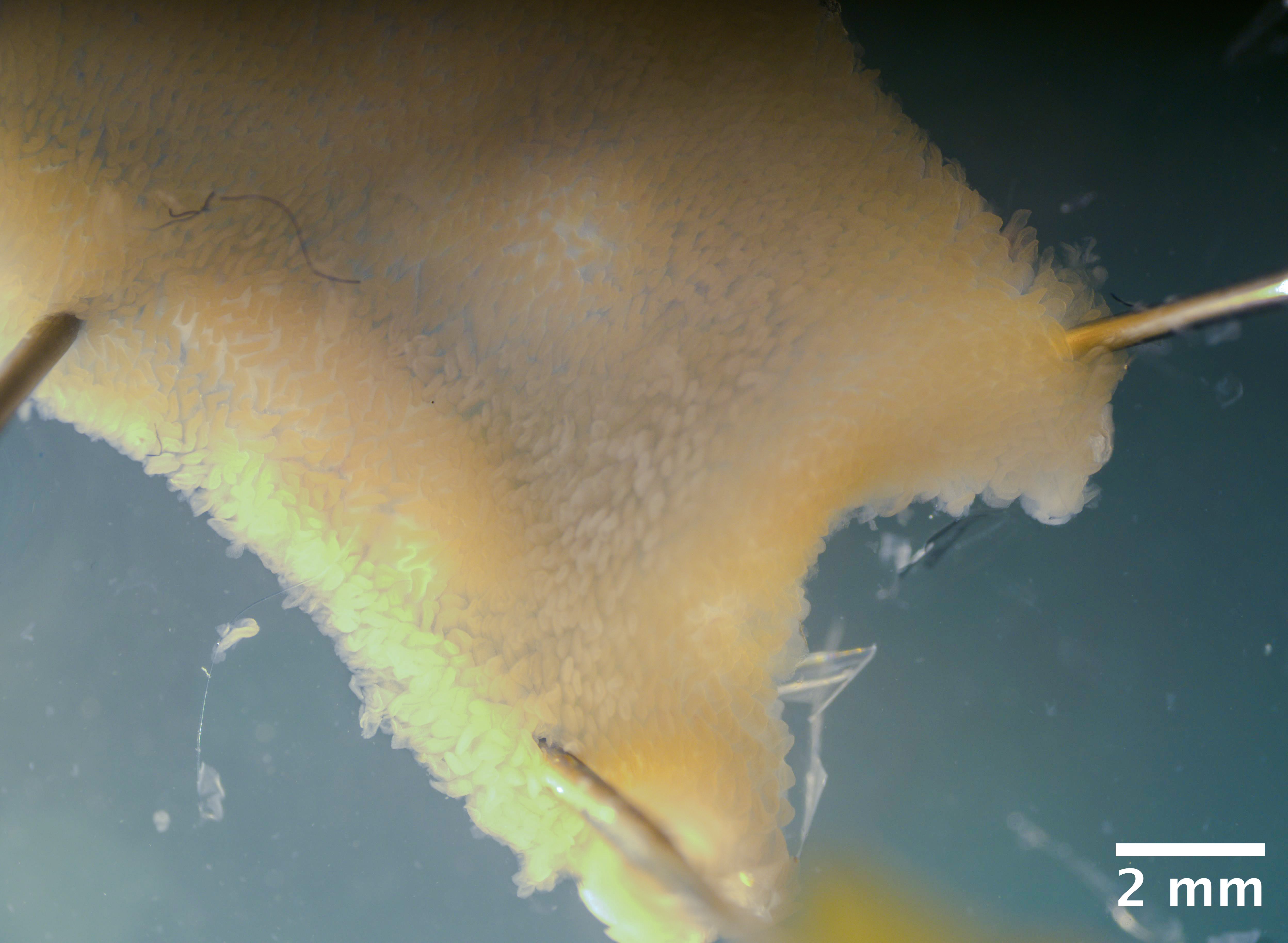}
    \caption{An image of the inner surface of a rat jejunum}\label{jej5}
\end{figure}

\begin{figure}
    \centering
    \includegraphics[width=\linewidth]{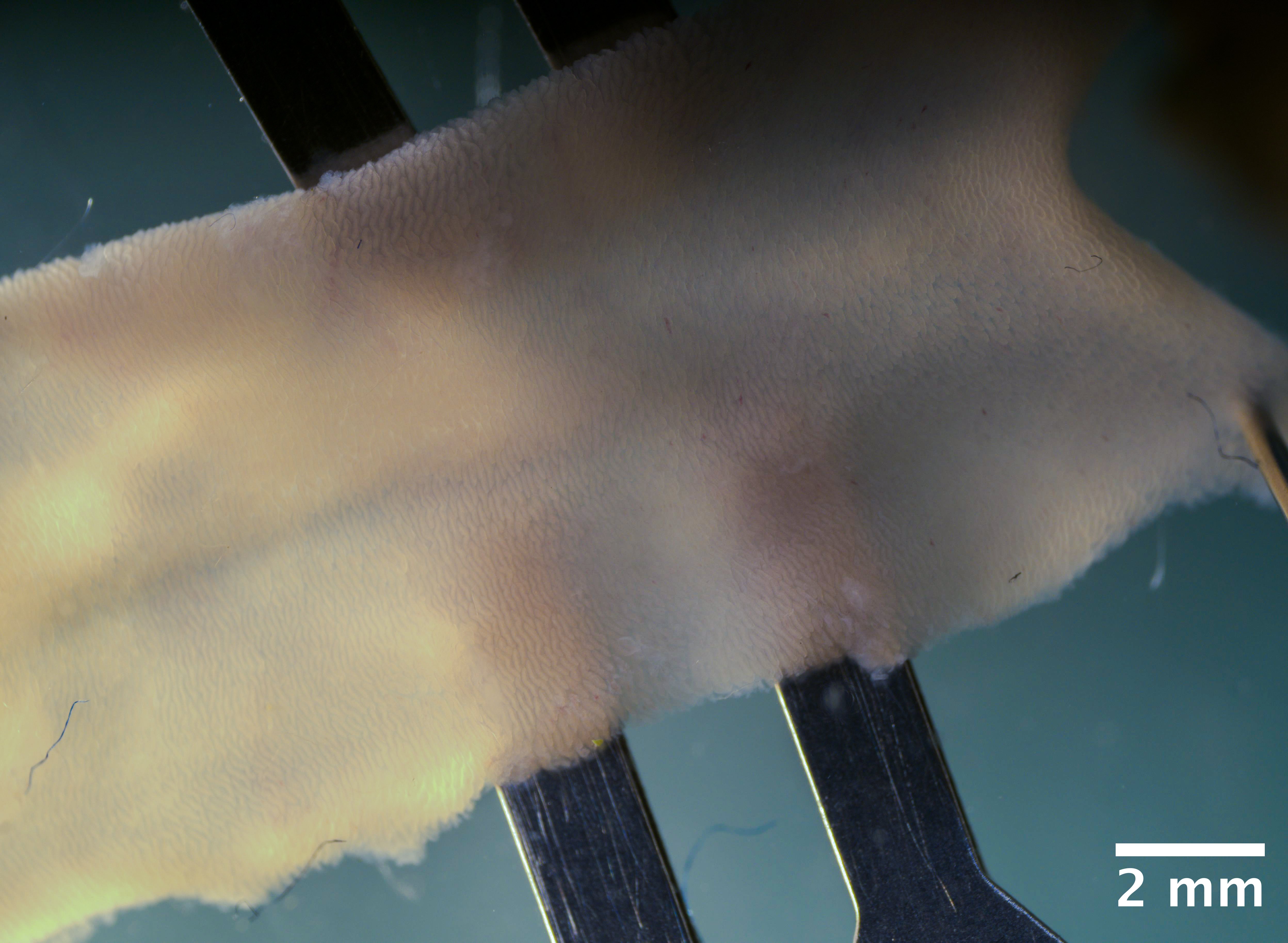}
    \caption{An image of the inner surface of a rat ileum, used in figure~\ref{fig:mainfig} of the main text}\label{iel}
\end{figure}

\begin{figure}
    \centering
    \includegraphics[width=\linewidth]{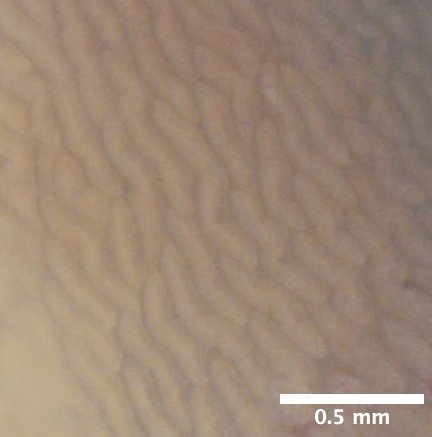}
    \caption{Zoom of an image of the inner surface of a rat ileum, used in figure~\ref{fig:mainfig} of the main text}\label{ielz}
\end{figure}

\clearpage

\clearpage

\printbibliography

@article{reed_review_2009,
	series = {Gastrointestinal {Malignancies}},
	title = {Review of the {Gastrointestinal} {Tract}: {From} {Macro} to {Micro}},
	volume = {25},
	issn = {0749-2081},
	shorttitle = {Review of the {Gastrointestinal} {Tract}},
	url = {https://www.sciencedirect.com/science/article/pii/S0749208108000739},
	doi = {10.1016/j.soncn.2008.10.002},
	number = {1},
	urldate = {2024-09-02},
	journal = {Seminars in Oncology Nursing},
	author = {Reed, Kathleen K. and Wickham, Rita},
	month = feb,
	year = {2009},
	pages = {3--14},
}

@article{fiori2022pharmacokinetics,
  title={Pharmacokinetics of oral mannitol for bowel preparation for colonoscopy},
  author={Fiori, Giancarla and Spada, Cristiano and Soru, Pietro and Tontini, Gian Eugenio and Bravi, Ivana and Cesana, Bruno Mario and Cesaro, Paola and Manes, Gianpiero and Orsatti, Anna and Prada, Alberto and others},
  journal={Clinical and Translational Science},
  volume={15},
  number={10},
  pages={2448--2457},
  year={2022},
  publisher={Wiley Online Library}
}

@article{schneider2012nih,
  title={NIH Image to ImageJ: 25 years of image analysis},
  author={Schneider, Caroline A and Rasband, Wayne S and Eliceiri, Kevin W},
  journal={Nature methods},
  volume={9},
  number={7},
  pages={671--675},
  year={2012},
  publisher={Nature Publishing Group US New York}
}

@article{henderson_mechanism_1928,
	title = {The mechanism of intestinal peristalsis},
	volume = {86},
	issn = {0002-9513},
	url = {https://journals.physiology.org/doi/abs/10.1152/ajplegacy.1928.86.1.82},
	doi = {10.1152/ajplegacy.1928.86.1.82},
	number = {1},
	urldate = {2024-09-02},
	journal = {American Journal of Physiology-Legacy Content},
	author = {Henderson, V. E.},
	month = aug,
	year = {1928},
	note = {Publisher: American Physiological Society},
	pages = {82--98},
}

@article{lentle_review_2015,
	title = {A review of mixing and propulsion of chyme in the small intestine: fresh insights from new methods},
	volume = {185},
	issn = {1432-136X},
	shorttitle = {A review of mixing and propulsion of chyme in the small intestine},
	url = {https://doi.org/10.1007/s00360-015-0889-5},
	doi = {10.1007/s00360-015-0889-5},
	language = {en},
	number = {4},
	urldate = {2024-09-02},
	journal = {J Comp Physiol B},
	author = {Lentle, R. G. and de Loubens, C.},
	month = may,
	year = {2015},
	pages = {369--387},
}

@article{eltantaway_effect_2015,
	title = {Effect of experimentally induced chronic pancreatitis on the structure of hepatocytes and the duodenal villi in adult male albino rats},
	volume = {38},
	issn = {2090-2417},
	url = {https://journals.lww.com/ejhistology/abstract/2015/03000/effect_of_experimentally_induced_chronic.1.aspx},
	doi = {10.1097/01.EHX.0000457300.50130.58},
	language = {en-US},
	number = {1},
	urldate = {2024-09-02},
	journal = {Egyptian Journal of Histology},
	author = {Eltantaway, Eman H. and Khalaf, Gehan and Hamam, Ghada G.},
	month = mar,
	year = {2015},
	pages = {1},
}

@article{walton_blueprint_2018,
	title = {Blueprint for an intestinal villus: {Species}-specific assembly required},
	volume = {7},
	copyright = {© 2018 Wiley Periodicals, Inc.},
	issn = {1759-7692},
	shorttitle = {Blueprint for an intestinal villus},
	url = {https://onlinelibrary.wiley.com/doi/abs/10.1002/wdev.317},
	doi = {10.1002/wdev.317},
	language = {en},
	number = {4},
	urldate = {2024-09-02},
	journal = {WIREs Developmental Biology},
	author = {Walton, Katherine D. and Mishkind, Darcy and Riddle, Misty R. and Tabin, Clifford J. and Gumucio, Deborah L.},
	year = {2018},
	note = {\_eprint: https://onlinelibrary.wiley.com/doi/pdf/10.1002/wdev.317},
	pages = {e317},
}

@article{n_i_contribution_1984,
	title = {The contribution of the large intestine to energy supplies in man},
	volume = {39},
	issn = {0002-9165},
	url = {https://www.sciencedirect.com/science/article/pii/S0002916523245992},
	doi = {10.1093/ajcn/39.2.338},
	number = {2},
	urldate = {2024-09-02},
	journal = {The American Journal of Clinical Nutrition},
	author = {N i, McNeil},
	month = feb,
	year = {1984},
	pages = {338--342},
}

@article{prieto2019differences,
  title={Differences in colonic crypt morphology of spontaneous and colitis-associated murine models via second harmonic generation imaging to quantify colon cancer development},
  author={Prieto, Sandra P and Reed, Cassandra L and James, Haley M and Quinn, Kyle P and Muldoon, Timothy J},
  journal={BMC cancer},
  volume={19},
  pages={1--11},
  year={2019},
  publisher={Springer}
}

@article{helander1973morphological,
  title={Morphological studies on the development of the rat colonic mucosa},
  author={Helander, Herbert F},
  journal={Cells Tissues Organs},
  volume={85},
  number={2},
  pages={153--176},
  year={1973},
  publisher={Karger Publishers}
}

@article{walton_generation_2016,
	title = {Generation of intestinal surface: an absorbing tale},
	volume = {143},
	issn = {0950-1991},
	shorttitle = {Generation of intestinal surface},
	url = {https://doi.org/10.1242/dev.135400},
	doi = {10.1242/dev.135400},
	number = {13},
	urldate = {2024-09-02},
	journal = {Development},
	author = {Walton, Katherine D. and Freddo, Andrew M. and Wang, Sha and Gumucio, Deborah L.},
	month = jul,
	year = {2016},
	pages = {2261--2272},
}

@article{lim2015flow,
  title={Flow and mixing by small intestine villi},
  author={Lim, YF and de Loubens, Cl{\'e}ment and Love, RJ and Lentle, RG and Janssen, PWM},
  journal={Food \& function},
  volume={6},
  number={6},
  pages={1787--1795},
  year={2015},
  publisher={Royal Society of Chemistry}
}

@article{helander_surface_2014,
	title = {Surface area of the digestive tract – revisited},
	volume = {49},
	issn = {0036-5521},
	url = {https://doi.org/10.3109/00365521.2014.898326},
	doi = {10.3109/00365521.2014.898326},
	number = {6},
	urldate = {2024-09-02},
	journal = {Scandinavian Journal of Gastroenterology},
	author = {Helander, Herbert F and Fändriks, Lars},
	month = jun,
	year = {2014},
	pmid = {24694282},
	note = {Publisher: Taylor \& Francis
\_eprint: https://doi.org/10.3109/00365521.2014.898326},
	pages = {681--689},
}

@article{strocchi_role_1993,
	title = {Role of villous surface area in absorption science versus religion},
	volume = {38},
	issn = {1573-2568},
	url = {https://doi.org/10.1007/BF01316488},
	doi = {10.1007/BF01316488},
	language = {en},
	number = {3},
	urldate = {2024-09-02},
	journal = {Digest Dis Sci},
	author = {Strocchi, Alessandra and Levitt, Michael D.},
	month = mar,
	year = {1993},
	pages = {385--387},
}

@article{wilson_intestinal_1974,
	title = {The intestinal unstirred layer: {Its} surface area and effect on active transport kinetics},
	volume = {363},
	issn = {0005-2736},
	shorttitle = {The intestinal unstirred layer},
	url = {https://www.sciencedirect.com/science/article/pii/0005273674900108},
	doi = {10.1016/0005-2736(74)90010-8},
	number = {1},
	urldate = {2024-09-02},
	journal = {Biochimica et Biophysica Acta (BBA) - Biomembranes},
	author = {Wilson, Frederick A. and Dietschy, John M.},
	month = aug,
	year = {1974},
	pages = {112--126},
}

@article{smithson_intestinal_1981,
	title = {Intestinal {Diffusion} {Barrier}: {Unstirred} {Water} {Layer} or {Membrane} {Surface} {Mucous} {Coat}?},
	volume = {214},
	shorttitle = {Intestinal {Diffusion} {Barrier}},
	url = {https://www.science.org/doi/10.1126/science.7302593},
	doi = {10.1126/science.7302593},
	number = {4526},
	urldate = {2024-09-02},
	journal = {Science},
	author = {Smithson, Kenneth W. and Millar, David B. and Jacobs, Lucien R. and Gray, Gary M.},
	month = dec,
	year = {1981},
	note = {Publisher: American Association for the Advancement of Science},
	pages = {1241--1244},
}

@incollection{lentle_flow_2011,
	address = {New York, NY},
	title = {Flow, {Mixing} and {Absorption} at the {Mucosa}},
	isbn = {978-1-4419-9449-3},
	url = {https://doi.org/10.1007/978-1-4419-9449-3_10},
	language = {en},
	urldate = {2024-09-02},
	booktitle = {The {Physical} {Processes} of {Digestion}},
	publisher = {Springer},
	author = {Lentle, Roger G. and Janssen, Patrick W. M.},
	editor = {Lentle, Roger G. and Janssen, Patrick W.M.},
	year = {2011},
	doi = {10.1007/978-1-4419-9449-3_10},
	pages = {221--274},
}

@article{fagerholm_experimental_1995,
	title = {Experimental estimation of the effective unstirred water layer thickness in the human jejunum, and its importance in oral drug absorption},
	volume = {3},
	issn = {0928-0987},
	url = {https://journals.scholarsportal.info/details/09280987/v3i3-5/247_eeoteuiiioda.xml},
	number = {3-5},
	urldate = {2024-09-02},
	journal = {European Journal of Pharmaceutical Sciences},
	author = {Fagerholm, Urban and Lennernäs, Hans},
	year = {1995},
	note = {Publisher: Elsevier Science},
	pages = {247--253},
}

@article{cone2009barrier,
  title={Barrier properties of mucus},
  author={Cone, Richard A},
  journal={Advanced drug delivery reviews},
  volume={61},
  number={2},
  pages={75--85},
  year={2009},
  publisher={Elsevier}
}

@article{swidsinski_comparative_2007,
	title = {Comparative study of the intestinal mucus barrier in normal and inflamed colon},
	volume = {56},
	copyright = {Copyright 2007 by Gut},
	issn = {0017-5749, 1468-3288},
	url = {https://gut.bmj.com/content/56/3/343},
	doi = {10.1136/gut.2006.098160},
	language = {en},
	number = {3},
	urldate = {2024-09-02},
	journal = {Gut},
	author = {Swidsinski, Alexander and Loening-Baucke, Vera and Theissig, Franz and Engelhardt, Holger and Bengmark, Stig and Koch, Stefan and Lochs, Herbert and Dörffel, Yvonne},
	month = mar,
	year = {2007},
	pmid = {16908512},
	note = {Publisher: BMJ Publishing Group
Section: Colonic inflammation},
	pages = {343--350},
}

@article{hilton_morphology_1902,
	title = {The morphology and development of intestinal folds and {Villi} in vertibrates},
	volume = {1},
	copyright = {Copyright © 1902 Wiley-Liss, Inc.},
	issn = {1553-0795},
	url = {https://onlinelibrary.wiley.com/doi/abs/10.1002/aja.1000010406},
	doi = {10.1002/aja.1000010406},
	language = {en},
	number = {4},
	urldate = {2024-10-17},
	journal = {American Journal of Anatomy},
	author = {Hilton, William A.},
	year = {1902},
	note = {\_eprint: https://onlinelibrary.wiley.com/doi/pdf/10.1002/aja.1000010406},
	pages = {459--505},
}

@article{skrzypek_mechanisms_2018,
	title = {Mechanisms involved in the development of the small intestine mucosal layer in postnatal piglets},
	volume = {69},
	issn = {1899-1505},
	doi = {10.26402/jpp.2018.1.14},
	language = {eng},
	number = {1},
	journal = {J Physiol Pharmacol},
	author = {Skrzypek, T. H. and Kazimierczak, W. and Skrzypek, H. and Valverde Piedra, J. L. and Godlewski, M. M. and Zabielski, R.},
	month = feb,
	year = {2018},
	pmid = {29769429},
	pages = {127--138},
}

@article{roberts_mucosa_1974,
	title = {The {Mucosa} of the {Small} {Intestine} of the {Horse}: {A} {Microscopical} {Study} of {Specimens} {Obtained} through a {Small} {Intestinal} {Fistula}},
	volume = {6},
	copyright = {© 1974 EVJ Ltd},
	issn = {2042-3306},
	shorttitle = {The {Mucosa} of the {Small} {Intestine} of the {Horse}},
	url = {https://onlinelibrary.wiley.com/doi/abs/10.1111/j.2042-3306.1974.tb03933.x},
	doi = {10.1111/j.2042-3306.1974.tb03933.x},
	language = {en},
	number = {2},
	urldate = {2024-10-22},
	journal = {Equine Veterinary Journal},
	author = {Roberts, M. C. and Hill, F. W. G.},
	year = {1974},
	note = {\_eprint: https://onlinelibrary.wiley.com/doi/pdf/10.1111/j.2042-3306.1974.tb03933.x},
	pages = {74--80},
}

@article{marsh_study_1969,
	title = {A study of the small intestinal mucosa using the scanning electron microscope},
	volume = {10},
	issn = {0017-5749},
	url = {https://gut.bmj.com/lookup/doi/10.1136/gut.10.11.940},
	doi = {10.1136/gut.10.11.940},
	language = {en},
	number = {11},
	urldate = {2024-10-22},
	journal = {Gut},
	author = {Marsh, M. N. and Swift, J. A.},
	month = nov,
	year = {1969},
	pages = {940--949},
}

@article{abbas_internal_1989,
	title = {Internal structure of the intestinal villus: morphological and morphometric observations at different levels of the mouse villus},
	volume = {162},
	issn = {0021-8782},
	shorttitle = {Internal structure of the intestinal villus},
	language = {eng},
	journal = {J Anat},
	author = {Abbas, B. and Hayes, T. L. and Wilson, D. J. and Carr, K. E.},
	month = feb,
	year = {1989},
	pmid = {2808121},
	pmcid = {PMC1256453},
	pages = {263--273},
}

@article{samanya_histological_2002,
	title = {Histological alterations of intestinal villi in chickens fed dried \textit{{Bacillus} subtilis} var. \textit{natto}},
	volume = {133},
	issn = {1095-6433},
	url = {https://www.sciencedirect.com/science/article/pii/S1095643302001216},
	doi = {10.1016/S1095-6433(02)00121-6},
	number = {1},
	urldate = {2024-10-22},
	journal = {Comparative Biochemistry and Physiology Part A: Molecular \& Integrative Physiology},
	author = {Samanya, Mongkol and Yamauchi, Koh-en},
	month = sep,
	year = {2002},
	pages = {95--104},
}

@article{seyyedin_histomorphometric_2017,
	title = {Histomorphometric study of the effect of methionine on small intestine parameters in rat: an applied histologic study},
	volume = {76},
	copyright = {In sending the manuscript the author(s) confirm(s) that (s)he has (they have) not previously submitted it to another journal (exept for abstracts of no more than 400 words) or published it elsewhere. The author(s) also agree(s), if and when the manuscript is accepted for publication, to automatic and free transfer of copyright to the Publisher allowing for the publication and distribution of the material submitted in all available forms and fields of exploitation. A transfer of copyright is an element of Author's Statement form that should be signed by the corresponding author and uploaded into the manuscript system at the time of manuscript submission.   The author(s) accept(s) that the manuscript will not be published elsewhere in any language without the written consent of the copyright holder, i.e. the Publisher. All manuscripts submitted should be accompanied by Author Statement document also confirming that the publication has been approved by all co-authors (if any), as well as (if required) by the responsible authorities at the institution where the work has been carried out. Additionally, the author(s) confirm(s) that (s)he is (they are) familiar with and will observe the 'Instruction to Authors' included in 'Folia Morphologica' and also that all sources of financial support have been fully disclosed. Materials previously published should be accompanied by written consent for reprinting from the relevant Publishers. In the case of photographs of identifiable persons, their written consent should also be provided. Any potential conflict of interest will be dealt with by the local court specific to the Publisher. Legal relations between the Publisher and the author( s) are in accordance with Polish law and with international conventions binding on Poland.    The form of Author Statement in availale at the end of Authors Guidelines section for this journal (above).},
	issn = {1644-3284},
	shorttitle = {Histomorphometric study of the effect of methionine on small intestine parameters in rat},
	url = {https://journals.viamedica.pl/folia_morphologica/article/view/FM.a2017.0044},
	doi = {10.5603/FM.a2017.0044},
	language = {en},
	number = {4},
	urldate = {2024-10-22},
	journal = {Folia Morphologica},
	author = {Seyyedin, S. and Nazem, M. N.},
	year = {2017},
	note = {Number: 4},
	pages = {620--629},
}

@article{goodarzi_histology_2021,
	title = {Histology of the small intestine in the common pheasant (): {A} scanning electron microscopy, histochemical, immunohistochemical, and stereological study},
	volume = {84},
	issn = {1097-0029},
	shorttitle = {Histology of the small intestine in the common pheasant ()},
	url = {https://onlinelibrary.wiley.com/doi/abs/10.1002/jemt.23794},
	doi = {10.1002/jemt.23794},
	language = {en},
	number = {10},
	urldate = {2024-10-22},
	journal = {Microscopy Research and Technique},
	author = {Goodarzi, Nader and Akbari Bazm, Mohsen and Poladi, Sadra and Rashidi, Fatemeh and Mahmoudi, Bahareh and Abumandour, Mohamed M. A.},
	year = {2021},
	note = {\_eprint: https://onlinelibrary.wiley.com/doi/pdf/10.1002/jemt.23794},
	pages = {2388--2398},
}

@article{fujimiya_serotonin-containing_1991,
	title = {Serotonin-containing epithelial cells in rat duodenum. {I}. {Quantitative} morphometric study of the distribution density},
	volume = {95},
	issn = {1432-119X},
	url = {https://doi.org/10.1007/BF00744992},
	doi = {10.1007/BF00744992},
	language = {en},
	number = {3},
	urldate = {2024-10-22},
	journal = {Histochemistry},
	author = {Fujimiya, M. and Maeda, T. and Kimura, H.},
	month = may,
	year = {1991},
	pages = {217--224},
}

@article{kimura2002gastrointestinal,
  title={Gastrointestinal transit and drug absorption},
  author={Kimura, Toshikiro and Higaki, Kazutaka},
  journal={Biological and Pharmaceutical Bulletin},
  volume={25},
  number={2},
  pages={149--164},
  year={2002},
  publisher={The Pharmaceutical Society of Japan}
}

@article{qi2008automated,
  title={Automated quantification of colonic crypt morphology using integrated microscopy and optical coherence tomography},
  author={Qi, Xin and Pan, Yinsheng and Hu, Zhilin and Kang, Wei and Willis, Joseph E and Olowe, Kayode and Sivak Jr, Michael V and Rollins, Andrew M},
  journal={Journal of biomedical optics},
  volume={13},
  number={5},
  pages={054055--054055},
  year={2008},
  publisher={Society of Photo-Optical Instrumentation Engineers}
}

@article{ma2007immmunohistochemical,
  title={Immmunohistochemical study of the blood and lymphatic vasculature and the innervation of mouse gut and gut-associated lymphoid tissue},
  author={Ma, B and Von Wasielewski, R and Lindenmaier, W and Dittmar, KEJ},
  journal={Anatomia, histologia, embryologia},
  volume={36},
  number={1},
  pages={62--74},
  year={2007},
  publisher={Wiley Online Library}
}

@article{shyer2013villification,
  title={Villification: how the gut gets its villi},
  author={Shyer, Amy E and Tallinen, Tuomas and Nerurkar, Nandan L and Wei, Zhiyan and Gil, Eun Seok and Kaplan, David L and Tabin, Clifford J and Mahadevan, L},
  journal={Science},
  volume={342},
  number={6155},
  pages={212--218},
  year={2013},
  publisher={American Association for the Advancement of Science}
}

@article{farkas2015cryosectioning,
  title={Cryosectioning method for microdissection of murine colonic mucosa},
  author={Farkas, Attila E and Gerner-Smidt, Christian and Lili, Loukia and Nusrat, Asma and Capaldo, Christopher T},
  journal={Journal of visualized experiments: JoVE},
  number={101},
  pages={53112},
  year={2015}
}

@article{Rohan2025,
   title={Hydrodynamics in a villi-patterned channel due to pendular-wave activity},
    author={Vernekar, Rohan and Ahmad, Faisal and Garic, Martin and Martín, Dácil Idaira Yánez and Loverdo, Claude and Tanguy, Stéphane and de Loubens, Clément},
    journal={arXiv preprint arXiv:2503.08852},
   year={2025}
}

@article{watson2005interferon,
  title={Interferon-$\gamma$ selectively increases epithelial permeability to large molecules by activating different populations of paracellular pores},
  author={Watson, Christopher J and Hoare, Catherine J and Garrod, David R and Carlson, Gordon L and Warhurst, Geoffrey},
  journal={Journal of cell science},
  volume={118},
  number={22},
  pages={5221--5230},
  year={2005},
  publisher={Company of Biologists}
}

@article{shalimar2013mechanism,
  title={Mechanism of villous atrophy in celiac disease: role of apoptosis and epithelial regeneration},
  author={Shalimar, DM and Das, Prasenjit and Sreenivas, Vishnubhatla and Gupta, Siddhartha Datta and Panda, Subrat K and Makharia, Govind K},
  journal={Archives of Pathology and Laboratory Medicine},
  volume={137},
  number={9},
  pages={1262--1269},
  year={2013},
  publisher={the College of American Pathologists}
}

@article{winne1989effect,
  title={Effect of villosity and distension on the absorptive and secretory flux in the small intestine},
  author={Winne, Dietrich},
  journal={Journal of theoretical biology},
  volume={139},
  number={2},
  pages={155--186},
  year={1989},
  publisher={Elsevier}
}

@article{oliver1998surface,
  title={What surface of the intestinal epithelium is effectively available to permeating drugs?},
  author={Oliver, Ruth E and Jones, Alan F and Rowland, Malcolm},
  journal={Journal of pharmaceutical sciences},
  volume={87},
  number={5},
  pages={634--639},
  year={1998},
  publisher={Elsevier}
}

@article{lentle2013mucosal,
  title={Mucosal microfolds augment mixing at the wall of the distal ileum of the brushtail possum},
  author={Lentle, RG and Janssen, PWM and DeLoubens, C and Lim, YF and Hulls, C and Chambers, P},
  journal={Neurogastroenterology \& Motility},
  volume={25},
  number={11},
  pages={881--e700},
  year={2013},
  publisher={Wiley Online Library}
}

@article{strugala2003colonic,
  title={Colonic mucin: methods of measuring mucus thickness},
  author={Strugala, Vicki and Allen, Adrian and Dettmar, Peter W and Pearson, Jeffrey P},
  journal={Proceedings of the Nutrition Society},
  volume={62},
  number={1},
  pages={237--243},
  year={2003},
  publisher={Cambridge University Press}
}

@article{johansson2011composition,
  title={Composition and functional role of the mucus layers in the intestine},
  author={Johansson, Malin EV and Ambort, Daniel and Pelaseyed, Thaher and Sch{\"u}tte, Andr{\'e} and Gustafsson, Jenny K and Ermund, Anna and Subramani, Durai B and Holm{\'e}n-Larsson, Jessica M and Thomsson, Kristina A and Bergstr{\"o}m, Joakim H and others},
  journal={Cellular and molecular life sciences},
  volume={68},
  pages={3635--3641},
  year={2011},
  publisher={Springer}
}

@article{puthumana2022steady,
  title={Steady streaming flow induced by active biological microstructures; application to small intestine villi},
  author={Puthumana Melepattu, Midhun and de Loubens, Cl{\'e}ment},
  journal={Physics of Fluids},
  volume={34},
  number={6},
  year={2022},
  publisher={AIP Publishing}
}

@article{wang2010multiscale,
  title={A multiscale lattice Boltzmann model of macro-to micro-scale transport, with applications to gut function},
  author={Wang, Yanxing and Brasseur, James G and Banco, Gino G and Webb, Andrew G and Ailiani, Amit C and Neuberger, Thomas},
  journal={Philosophical Transactions of the Royal Society A: Mathematical, Physical and Engineering Sciences},
  volume={368},
  number={1921},
  pages={2863--2880},
  year={2010},
  publisher={The Royal Society Publishing}
}

@article{bashkatov2003glucose,
  title={Glucose and mannitol diffusion in human dura mater},
  author={Bashkatov, Alexey N and Genina, Elina A and Sinichkin, Yuri P and Kochubey, Vyacheslav I and Lakodina, Nina A and Tuchin, Valery V},
  journal={Biophysical journal},
  volume={85},
  number={5},
  pages={3310--3318},
  year={2003},
  publisher={Elsevier}
}

@article{lennernaas1998human,
  title={Human intestinal permeability},
  author={Lennerna{\"a}s, Hans},
  journal={Journal of pharmaceutical sciences},
  volume={87},
  number={4},
  pages={403--410},
  year={1998},
  publisher={Elsevier}
}

@article{wang2017three,
  title={Three-dimensional mechanisms of macro-to-micro-scale transport and absorption enhancement by gut villi motions},
  author={Wang, Yanxing and Brasseur, James G},
  journal={Physical Review E},
  volume={95},
  number={6},
  pages={062412},
  year={2017},
  publisher={APS}
}

@article{vertzoni2019impact,
  title={Impact of regional differences along the gastrointestinal tract of healthy adults on oral drug absorption: An UNGAP review},
  author={Vertzoni, Maria and Augustijns, Patrick and Grimm, Michael and Koziolek, Mirko and Lemmens, Glenn and Parrott, Neil and Pentafragka, Christina and Reppas, Christos and Rubbens, Jari and Van Den Abeele, Jens and others},
  journal={European journal of pharmaceutical sciences},
  volume={134},
  pages={153--175},
  year={2019},
  publisher={Elsevier}
}

@article{holzheimer1989influence,
  title={Influence of distension on absorption and villous structure in rat jejunum},
  author={Holzheimer, G and Winne, D},
  journal={American Journal of Physiology-Gastrointestinal and Liver Physiology},
  volume={256},
  number={1},
  pages={G188--G197},
  year={1989},
  publisher={American Physiological Society Bethesda, MD}
}

\end{document}